\documentclass[a4paper,fleqn]{cas-dc}

\usepackage[authoryear,longnamesfirst]{natbib}
\usepackage{amssymb}
\usepackage{epsfig} 
\usepackage{amsmath} 
\usepackage{graphicx}
\usepackage{color}
\usepackage{lipsum}
\usepackage{float}
\usepackage{xcolor}
\usepackage{amssymb}
\usepackage{multirow}
\usepackage{eqnarray}
\usepackage{enumitem}

\def\tsc#1{\csdef{#1}{\textsc{\lowercase{#1}}\xspace}}
\tsc{WGM}
\tsc{QE}
\begin{document}
\let\WriteBookmarks\relax
\def\floatpagepagefraction{1}
\def\textpagefraction{.001}

\shorttitle{Thermal properties of the core of magnetar}    
\shortauthors{T. Sarkar, S. Yadav, M. Sinha}  

\title [mode = title]{Thermal properties of the core of magnetar}  

%\tnotemark[<tnote number>] 

%\tnotetext[<tnote number>]{<tnote text>} 


\author[1]{Trisha Sarkar}%[<options>]

%\cormark[1]

\fnmark[1]

\ead{sarkar.2@iitj.ac.in}

%\ead[url]{<URL>}

%\credit{<Credit authorship details>}

\affiliation[a]{organization={Indian Institute of Technology},
            addressline={Jodhpur}, 
            city={Rajasthan},
%          citysep={}, % Uncomment if no comma needed between city and postcode
            postcode={342030}, 
            country={India}}

\author[2]{Shalu Yadav}%[<options>]

% Footnote of the second author
\fnmark[2]

% Email id of the second author
\ead{yadav.6@iitj.ac.in}
\affiliation[a]{organization={Indian Institute of Technology},
            addressline={Jodhpur}, 
            city={Rajasthan},
%          citysep={}, % Uncomment if no comma needed between city and postcode
            postcode={342030}, 
            country={India}}

\author[3]{Monika Sinha}%[<options>]

% Footnote of the second author
\fnmark[3]

% Email id of the second author
\ead{ms@iitj.ac.in}
\affiliation[a]{organization={Indian Institute of Technology},
            addressline={Jodhpur}, 
            city={Rajasthan},
%          citysep={}, % Uncomment if no comma needed between city and postcode
            postcode={342030}, 
            country={India}}




%\cortext[1]{Corresponding author}



\begin{abstract}
During very early age of neutron stars, the core cools down faster compared to the crust creating a large thermal gradient in the interior of the star. During $10-100$ years, a cooling wave propagates from the core to the crust causing the interior of the star to thermalize. During this duration thermal properties of the core material is of great importance to understand the dynamics of the interior of the star. The heat capacity and thermal conductivity of the core depends on the behaviour of matter inside the core. We investigate these two properties in case of magnetars. Due to presence of large magnetic field, the proton superconductivity is quenched partially inside the magnetars depending upon the comparative values of upper critical field and the strength of the magnetic field present. This produces non-uniformity in the behaviour of matter throughout the star. Moreover, such non-uniformity arises from the variation of nature of the pairing and values of the pairing gap energy. We find that the heat capacity is substantially reduced due to the presence of superfluidity. On the other hand, the thermal conductivity of neutron is enhanced due to proton superconductivity and gets reduced due to neutron superfluidity. Hence, the variation of the thermal properties due to superfluidity in presence of magnetic field is different at different radius inside the star. However, in all the cases the %minimum
maximum variation is of the order one. This affects the thermal relaxation time of the star and eventually its the thermal evolution.
\end{abstract}


\maketitle

% Main text
\section{Introduction}\label{intro}

Neutron stars (NS) are unique astrophysical objects born in supernova explosions with initial temperature $\sim~10^{11}$ K. Just after its birth a large temperature gradient is set up inside the star. However, within a few days its surface temperature goes down to $\sim~10^9$ K with the inner region of the star remaining hotter compared to its surface. The crust and the core cool down independently with different cooling mechanisms. Due to the lack of sufficient thermal conduction in the interior of the star, the thermalization inside the star is not able to take place at this stage. The cooling is faster around the neutron drip density of the crust while the inner crust cools comparatively at a slower rate , as a consequence of which the inner crust remains hotter in comparison with the outer crust. The core of the star cools down quickly by neutrino emission, while the crust continues to be hotter as compared the core. This results in heat flow from the outer to inner region of the star. The entire phenomena can be pictured as the propagation of a cooling wave from the core towards the surface \citep{Gnedin:2000me} which occurs within a short period of time, $10-100$ years, known as the era of thermal relaxation.

At a latter stage of thermal relaxation, the surface temperature drops down very fast  
due to the large value of thermal conductivity and the thermalization between the inner and the outer core. This results in the formation of a nearly isothermal region within the entire core, which may even extend to some parts of the inner crust. In this context the heat capacity and thermal conductivity are two of the most crucial ingredients to understand and construct the cooling model of NS, which are able to estimate the total loss of thermal energy of the star. 
It is to be noted that both of these quantities are completely dependent on the state of the interior matter compositions. For the detailed review on the thermal properties in the interior of a NS, see the work by  \citet{2001PhR...354....1Y} and the references mentioned in it.

The direct observation of NS cooling is detected for $\sim330$ years old NS Cassiopeia A. Analyzing nine years of data ($2000-2009$) obtained by NASA's \emph{Chandra} satellite its temperature is found to decline from $2.12\pm0.01\times10^6$ K to $2.04\pm0.01\times10^6$ K ($\sim5.4\sigma$) \cite{2010ApJ...719L.167H}. Before this the observation of thermal radiation emitted from NS was only limited to the estimation of the  temperature at any individual time \emph{e.g.} Crab pulsar \cite{2004ApJ...601.1050W}, Vela pulsar \cite{2001ApJ...552L.129P}, Geminga pulsar \cite{1997ApJ...477..905H} and so on. The observed cooling rate of Cassiopeia A is much faster than the standard cooling calculated from modified URCA process. One possible explanation of such fast cooling could possibly be the coexistence of neutron superfluidity and proton superconductivity in the stellar interior \cite{2011PhRvL.106h1101P}. In this work we consider the transition of normal neutrons and protons to superfluid and superconducting states respectively. The neutrons may reside in superfluid state pairing via triplet ($^3P_{2}$) channel at higher density of the inner core, while the singlet pairing ($^1S_0$) of neutrons are active in the lower density regions, in the vicinity of the crust \citep{1970PhRvL..24..775H}. The proton may form $^1S_0$ pairing even in the inner part of the core, due to its lower density fraction as compared to the neutrons. The presence of superfluidity/superconductivity substantially affects the thermal properties, heat capacity and thermal conductivity 
\citep{2000A&A...353.1129D, 2007MNRAS.381.1143C, 2014APS..APR.C8006S, 2014PhRvC..90e5803M,2012arXiv1201.5602P,2017PhRvC..95b5806C}.
 
The magnetars belong to a special class of NSs which possess extremely high surface magnetic field ($\gtrsim10^{13}$ G). Several astrophysical observation such as soft gamma repeaters and anomalous X-ray pulsars are related to such a distinctive stellar body \citep{2008AIPC..983..227W, 2020ApJ...904L..19B, 2020ATel14092....1C, 2020ApJ...894..159R}, in which the structural properties and behaviour of the highly dense matter are affected extraordinarily \citep{2015PhRvC..91c5805S}. Consequently the thermal properties of the matter inside such stars are also expected to exhibit diverse characteristics. Due to the presence of large magnetic field the proton superconductivity in magnetars is subdued significantly  \citep{2015PhRvC..91c5805S} which results in astounding repercussions in its thermal properties. In this work we analyse and estimate the heat capacity and thermal conductivity of dense matter inside the core of the magnetar, considering the possibilities of its matter configuration to be both normal and superfluid/superconducting. We present a comparative analysis of these two quantities at different times during its thermal evolution.

The plan of the work is as follows.
In section \ref{sec:formalism} we describe the complete model of the stars including its superfluid/superconducting pairing gap profiles both in presence and in absence of the magnetic field. We also illustrate the heat capacity and thermal conductivity of different particle species in this section and describe the concerned results. In section \ref{mag_sf} the effect of magnetic field on these two quantities in addition to superfluidity/superconductivity is explained. Finally in section \ref{dis} we conclude and briefly discuss our results.

\section{Model}\label{sec:formalism}
%%%%%%%%%%%%%%%%%%%%%%%%%%%%%%%%%%%%%%%%

We consider magnetars with sufficiently strong surface magnetic field ($B\sim10^{15}$ G). The field inside the core is assumed to be at least one order of magnitude greater than its surface value. In the core, the matter is considered to be composed of only nucleons and  electrons ($npe$), governed by \textcolor{black}{DDME2} equation of state (EoS) \citep{2005PhRvC..71b4312L} \textcolor{black}{which satisfies several astrophysical observables}.  
The interior matter is charge neutral and in $ \beta $-equilibrium. We consider two model magnetars having two extreme mass limits, $2.3~M_{\odot}$ and $1.4~M_{\odot}$. The radii of the massive ($M=2.3~M_{\odot}$) and canonical mass star ($M=1.4~M_{\odot}$) are \textcolor{black}{$12.72$ and $12.30$ km} respectively. 

We consider two magnetic field profile inside the model magnetars which are discussed as follows.

(a) Exponential profile:\hspace{2mm} $B{_{exp}}(n_{b}/n_{0})=B_{s}+B_{c}\{1-\beta~ e^{(n_{b}/n_{0})^{\gamma}} \}$ \citep{PhysRevLett.79.2176}.
Here $B_{c}$ and $B_{s}$ are the magnetic field at the core and the surface of the star respectively. $\beta$ and $\gamma$ are two arbitrary parameters which are selected in such a way that the desired magnetic field on the surface can be obtained. In this work we have considered \textcolor{black}{$\beta=0.04$ and $\gamma=3.22$}, to obtain $B_{s}\sim10^{15}$ G and $B_{c}\sim10^{16}$ G.

(b) Universal profile:\hspace{2mm}
$B{_{uni}}(x)=B_{c} (1-1.6x^{2}-x^{4}+4.2 x^{6}-2.4 x^{8})$ \citep{2019PhRvC..99e5811C}.
Here $x=r/R$, r and R are the radial distance from the centre and the radius of the star respectively.

\begin{table*}[t!]
\begin{center}
\caption{Different ranges of radii ($R$) expressed in km in which $^1S_0$ and $^3P_2$ pairing of nucleons are active, following the lower curves of $\Delta_n$ and $\Delta_p$ as shown in fig. \ref{del} for the two stars, both in presence and absence of magnetic field.}
\begin{tabular}{|p{0.9cm}|p{2.5cm}|p{2.5cm}|p{2.5cm}|p{2.5cm}|p{2.5cm}|} 
 \hline
 Mass\par($M_{\odot}$) &        \hspace{5mm}$\Delta_n(^1S_0)$ & \hspace{5mm}$\Delta_n(^3P_2)$ & \hspace{5mm}$\Delta_p(^1S_0)$ \newline \hspace*{5mm}($B=0$) & \hspace{5mm}$\Delta_p(^1S_0)$ \newline \hspace*{5.4mm}($B_{uni}$) &  \hspace{5mm}$\Delta_p(^1S_0)$ \newline \hspace*{5.4mm}($B_{exp}$)  \\ \hline\hline
 1.4 & \textcolor{black}{12.3 - 11.8} & \textcolor{black}{11.8 - 8.48}   &  \textcolor{black}{11.8-3.13}  & \textcolor{black}{11.8 - 8.71}  & \textcolor{black}{11.8- 9.27} \\  \hline
 2.3 & \textcolor{black}{12.72 - 12.5} & \textcolor{black}{12.5 - 11} & \textcolor{black}{12.5-9.4} & \textcolor{black}{12.5-10.85} & \textcolor{black}{12.5-10.98} \\
  \hline
\end{tabular}
\label{table:1}
\end{center}
\end{table*}

\begin{table*}[t!]
\begin{center}
\caption{Same as Table \ref{table:1} following the upper curves of $\Delta_n$ and $\Delta_p$}
\begin{tabular}{|p{0.9cm}|p{2.5cm}|p{2.5cm}|p{2.5cm}|p{2.5cm}|p{2.5cm}|} 
 \hline
 Mass\par($M_{\odot}$) &        \hspace{5mm}$\Delta_n(^1S_0)$ & \hspace{5mm}$\Delta_n(^3P_2)$ & \hspace{5mm}$\Delta_p(^1S_0)$ \newline \hspace*{5mm}($B=0$) & \hspace{5mm}$\Delta_p(^1S_0)$ \newline \hspace*{5.4mm}($B_{uni}$) &  \hspace{5mm}$\Delta_p(^1S_0)$ \newline \hspace*{5.4mm}($B_{exp}$)  \\
 \hline\hline
 1.4 & \textcolor{black}{12.30 - 11.8} & \textcolor{black}{11.8 - 0.8}  &  \textcolor{black}{11.8 - 0.8} & \textcolor{black}{11.8 - 6.45} & \textcolor{black}{11.8 - 7.22} \\ 
 \hline
 2.3 & \textcolor{black}{12.72-12.5} & \textcolor{black}{12.5-0.8} & \textcolor{black}{12.5-5.69} & \textcolor{black}{12.5-9.44} & \textcolor{black}{12.5-9.86} \\
  \hline
\end{tabular}
\label{table:2}
\end{center}
\end{table*}

\begin{center}
\begin{figure*}[ht]
\includegraphics[scale=0.31]{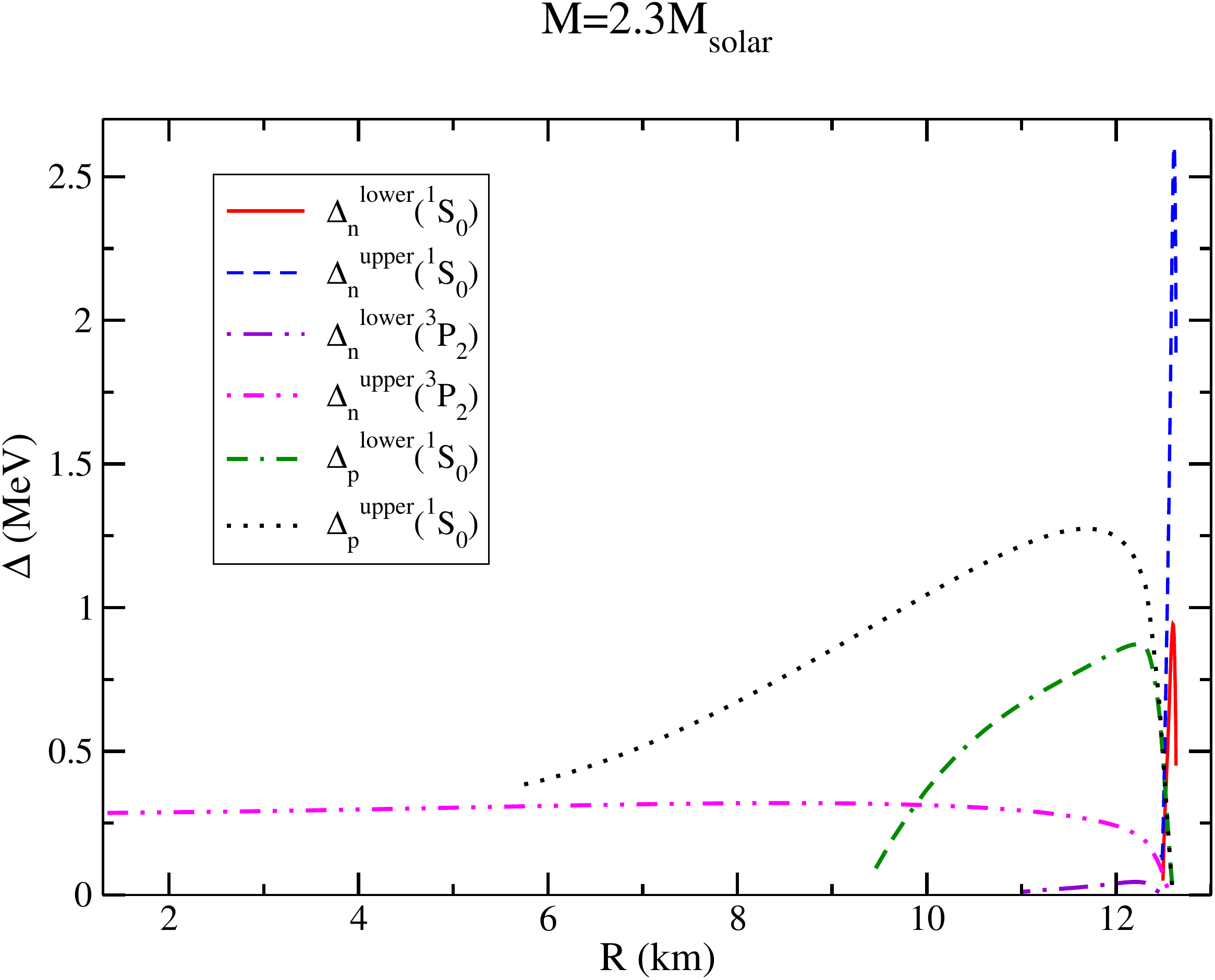}\hspace*{1mm}\includegraphics[scale=0.31]{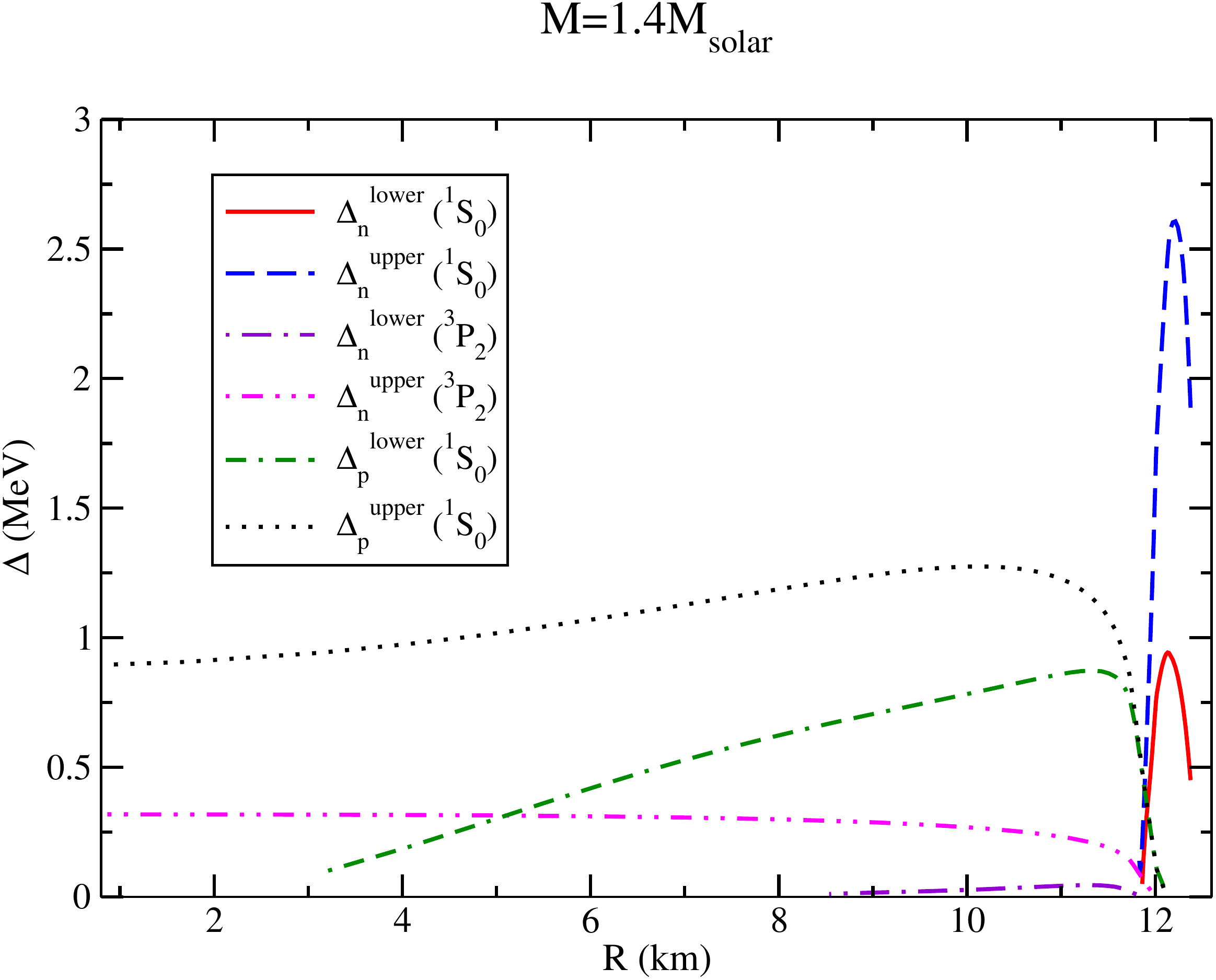}\\
%\hspace*{-5mm}
\includegraphics[scale=0.31]{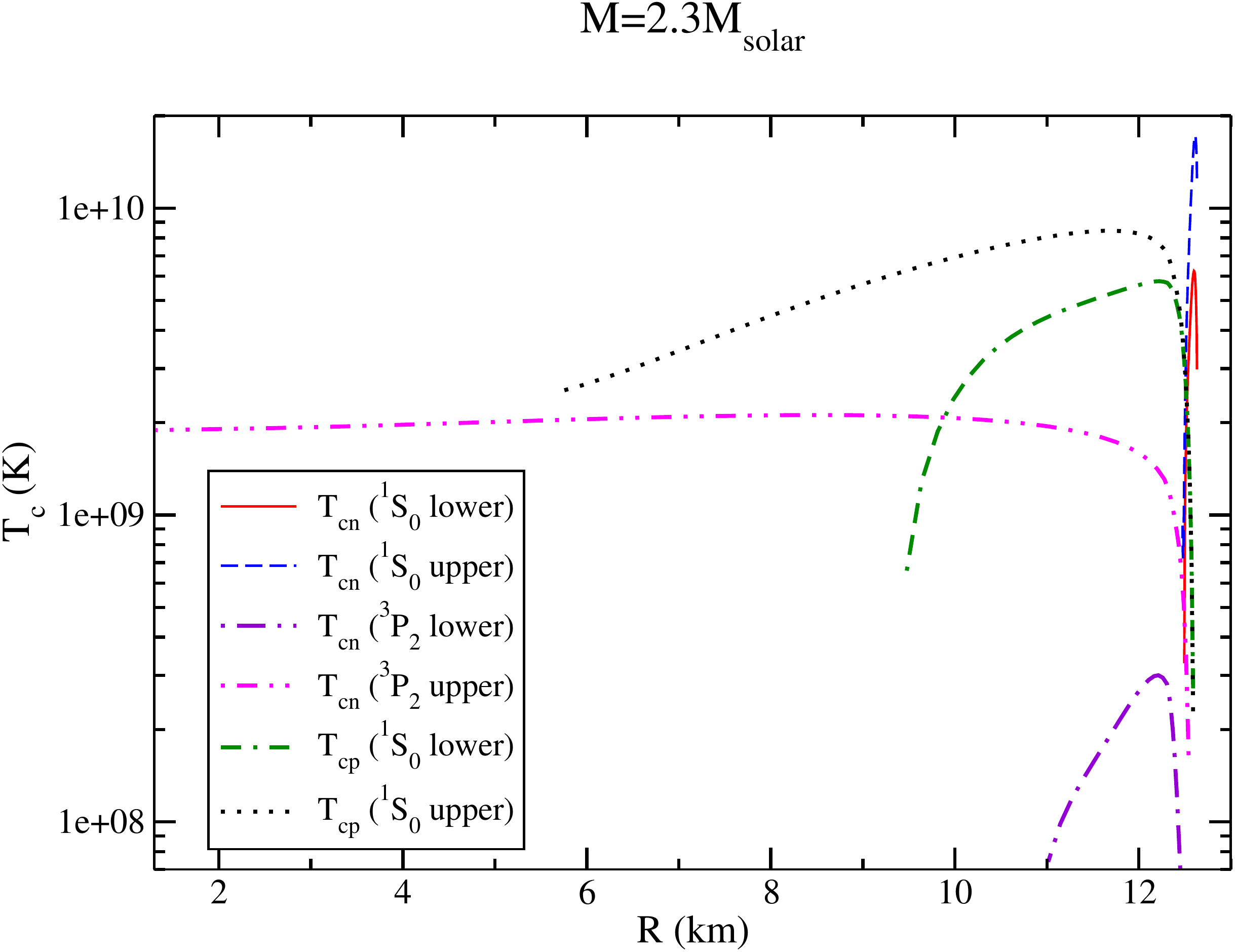}\hspace*{1mm}\includegraphics[scale=0.31]{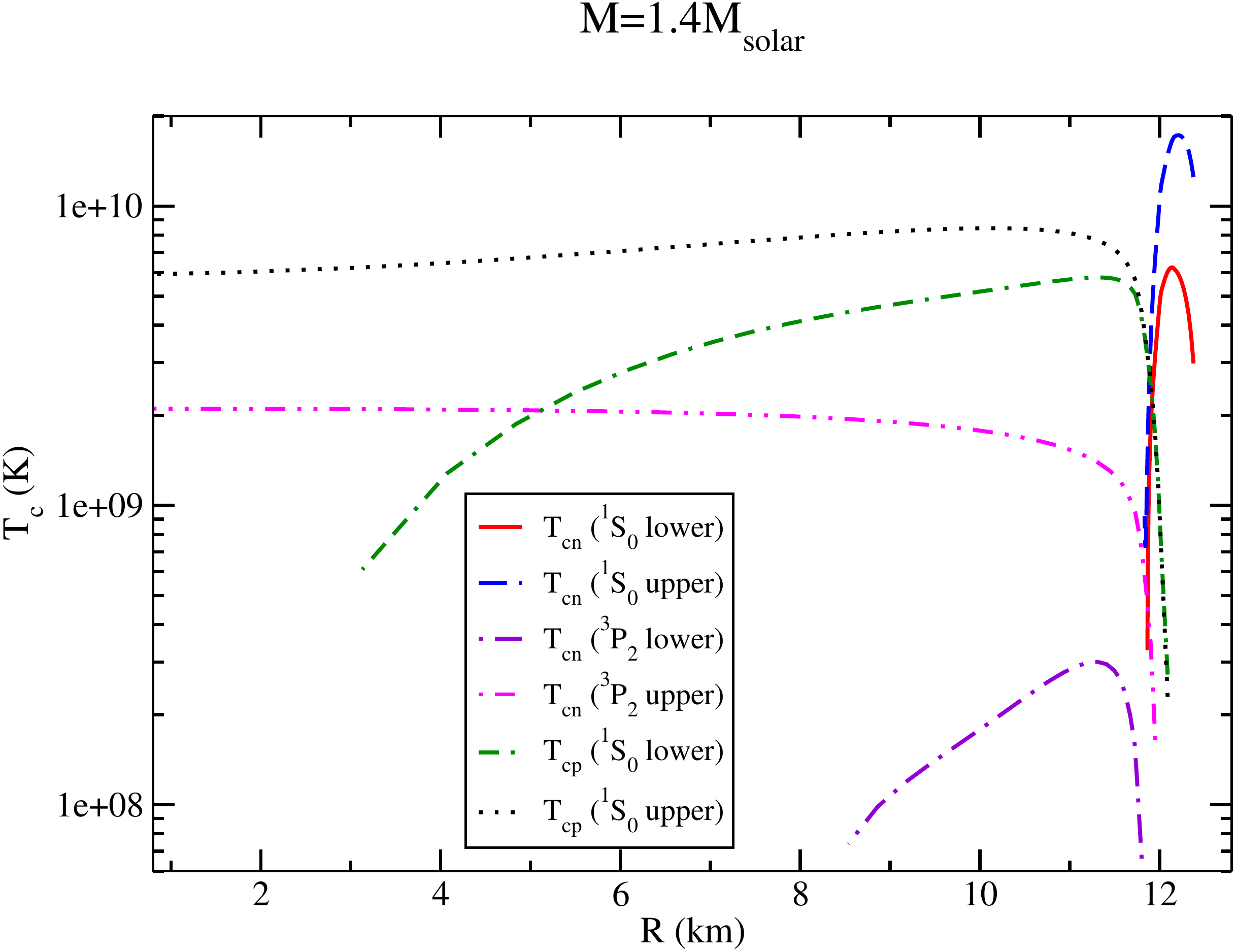}\\
%\hspace*{-5mm}
\includegraphics[scale=0.31]{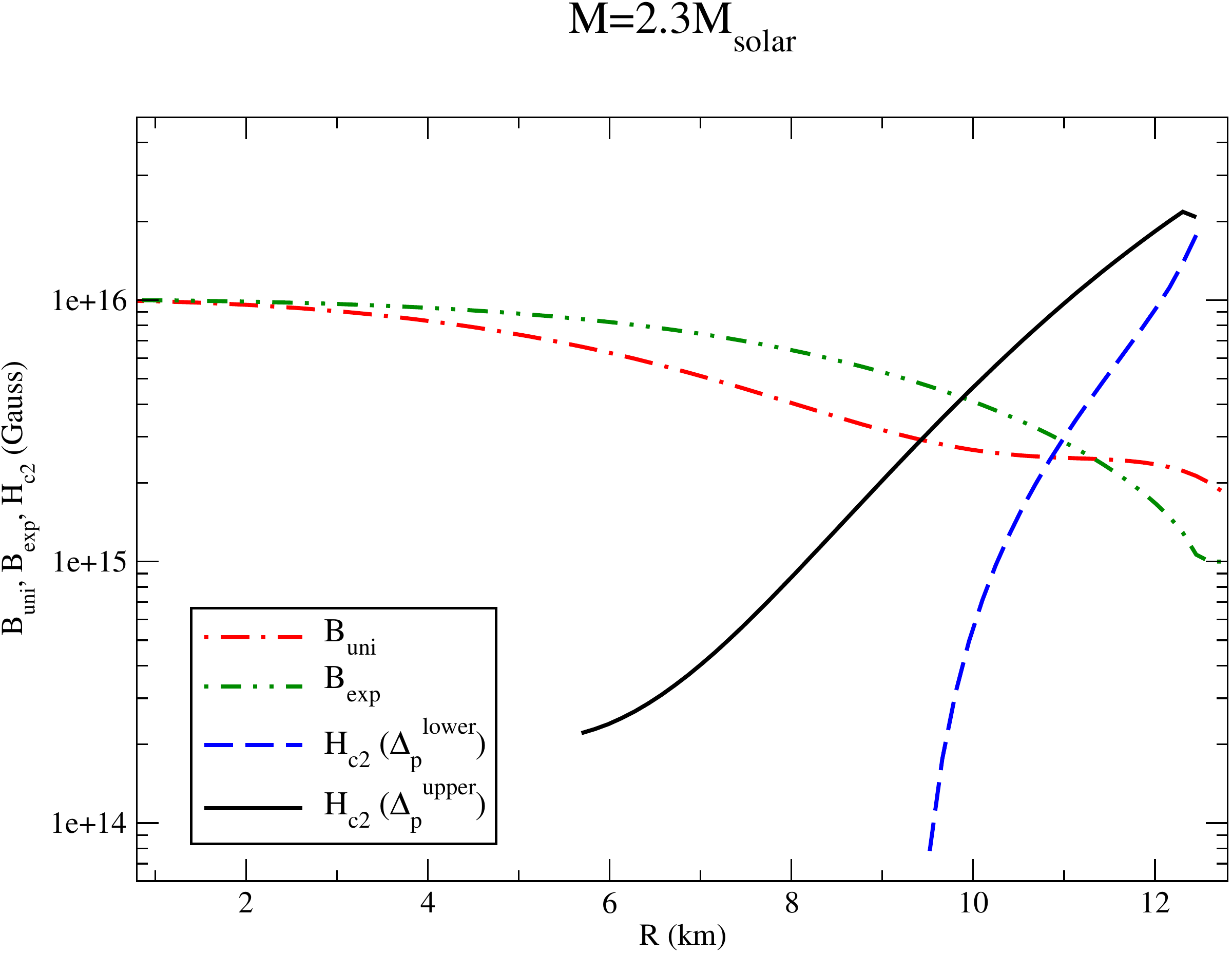}\hspace*{1mm}\includegraphics[scale=0.31]{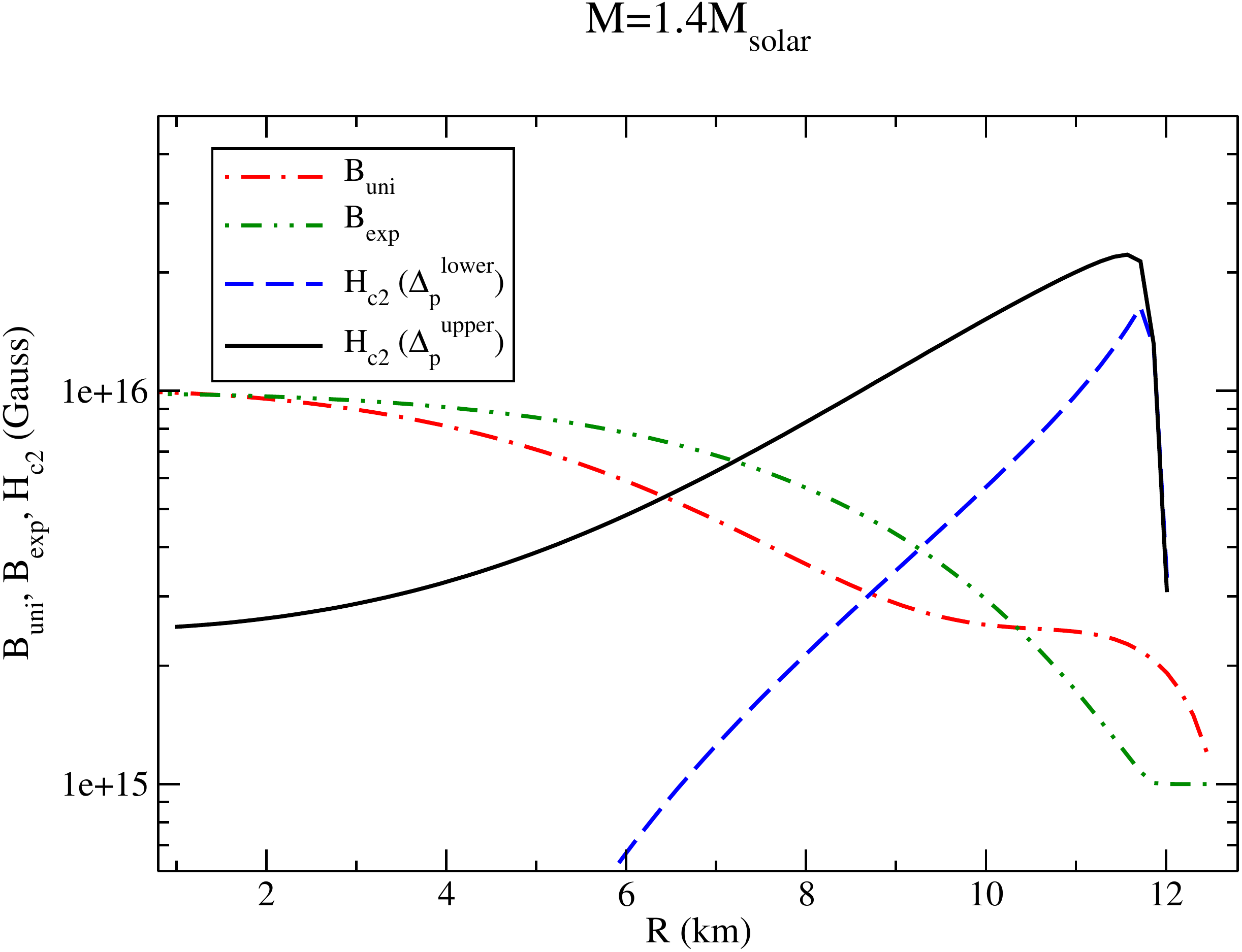} 
\caption{Top panel: Variation of the energy gap of neutron ($\Delta_n$) and proton ($\Delta_p$) with radius ($R$) for different pairing types; middle panel: Variation of the critical temperature of neutrons ($T_{cn}$) and protons ($T_{cp}$) with radius ($R$) for different pairing interaction channel; bottom panel: Variation of the critical magnetic field ($H_{c2}$) and magnetic field profiles  with radius ($R$). Left panels are for star with mass $M\sim 2.3M_\odot$ and right panels for the star of mass $M\sim 1.4M_\odot$, \textcolor{black}{the interior matter governed by DDME2 EoS}. $H_{c2}$ is plotted for the upper ($\Delta_{p}^{upper}$) and lower ($\Delta_{p}^{lower}$) curve of proton pairing.}
\label{del}
\end{figure*}
%\FloatBarrier
\end{center}

In generally, immediately after the birth, the initial temperature of the star is $\sim 10^{10} - 10^{11}$ K and gradually it cools down. The matter inside the core is  highly dense, and neutrons (protons) may form Cooper pairs with strong interaction leading to superfluidity (superconductivity). The concept of superfluidity in NS is well established \cite{1959NucPh..13..655M} and have been studied extensively in many previous works \citep{2019EPJA...55..167S, 2019arXiv190609641S}. Inside NS the superfluidity of the baryons is considered to be of BCS type. The neutrons form singlet pairing ($^{1}S_{0}$) near the crust at lower density regions where $\rho\lesssim\rho_{0}$ ($\rho_{0}\sim 0.17 fm^{-3}$ is the nuclear saturation density), while at the higher density regions of the core the pairing is formed via triplet channel ($^{3}P_{2}$), since the singlet state pairing becomes repulsive at larger densities \citep{1966ApJ...145..834W}. Although at higher densities both $^{1}D_{0}$ and $^{3}P_{2}$ pairing channels are dominant, $^{3}P_{2}$ is the most attractive among them \citep{2019arXiv190609641S}. The triplet pairing is originally understood by $^{3}P_{2}~-~^{3}F_{2}$ channel. However, because of the smallness of $^{3}F_{2}$ phase, only $^{3}P_{2}$ wave is considered for simplicity \citep{1992NuPhA.536..349B}. The protons form $^{1}S_{0}$ state pairing even inside the deeper layers of the core, due to its smaller number density fraction compared to the neutrons \citep{2004ARA&A..42..169Y}. There are several phenomenological models for the superfluid energy gap ($\Delta$) in both $^1S_0$ and $^3P_2$ channels for the protons and neutrons, depending on the nature of the strong nuclear interactions. From the overall picture of different models, two extreme bounds of energy gap for both of the neutrons ($\Delta_n$) and protons ($\Delta_p$) are considered \cite{2019EPJA...55..167S}. For these two extreme cases the variation of $\Delta_n$ and $\Delta_p$ inside the two model magnetars are shown in the left and right panels respectively, of the top row of fig. \ref{del}. It can be pointed out comparing the left and right panels that the proton superconductivity is very prominent near the surface of both the stars, although for the massive star the region containing the proton pairs extends up to a smaller depth as compared to the canonical mass star. 

The neutron superfluidity in $^1S_0$ state is also restricted in a narrow region near the surface of the magnetars%, although its strength is much higher in the canonical mass star
. It is also evident from fig. \ref{del} that for the massive star the neutron superfluidity is confined within some specific portion of the star. \textcolor{black}{In case of the canonical mass star, in absence of magnetic field, the neutron superfluidity remains active throughout almost the entire star for the upper boundaries of the pairing energy gap, while for the lower boundaries it is only effective in a small radial zone inside the star.} %Further, the neutron pairing in $^3P_2$ channel is stronger in the massive star as compared to the canonical mass star. 
The critical temperatures ($T_c$) corresponding to the various scenarios of both $\Delta_n$ and $\Delta_p$ variation inside the stars are shown in the middle row of the same figure.

In the magnetars, the protons are considered to be type-II superconductors, which causes the magnetic field to penetrate by constructing the array of quantized fluxtubes in the limit, $B \leq H_{c2}$, $H_{c2}$ being the upper critical field. With increase in magnetic field, the fluxtubes become more and more densely packed. Finally when the magnetic field exceeds $H_{c2}$ $i.e.$ $B\ge H_{c2}$ the fluxtubes are contacted with each other \citep{2014MNRAS.437..424L} and this leads to the destruction of the proton superconductivity. 
Depending on the variation of $H_{c2}$ and the magnetic field of the star, the proton superconductivity is quenched at those specific regions of the star, where the magnetic field is larger than $H_{c2}$ \citep{2015PhRvC..91c5805S}. In terms of the coherence length $\zeta_{p}$ the upper critical field $H_{c2}$ is given by,
\begin{equation}\label{eqn23}
H_{c2}=\frac{\phi_{0}}{2\pi\zeta_{p}^{2}},
\end{equation}
where $\phi_{0}$ is the quantum of flux at $B \leq H_{c2}$. 
The lower critical field $H_{c1}$ is insignificant in case of NSs due to their large electrical conductivity, unlike ordinary type-II superconductors constructed at laboratories \citep{2014MNRAS.437..424L, 2017JApA...38...43C, 2018ASSL..457..401H}. 

The variation of $H_{c2}$ for two extreme profiles of $\Delta_p$ and the profiles of the magnetic field inside the magnetars are illustrated in the bottom row of fig. \ref{del}. 
It is clearly evident from the plots that in presence of the magnetic field, the region where proton pairing is functional, is narrowed down and shifted more towards the surface. In case of the %canonical mass star the universal profile of magnetic field results in greater suppression of proton pairing as compared to the exponential profile, while for 
the massive star, the suppression is nearly similar for the two kinds of magnetic field profiles\textcolor{black}{, while for the canonical mass star the suppression appears to be occurring a little further with the consideration of the exponential profile of the magnetic field in comparison with the universal one. }.

The notion of different pairing interaction of the nucleons at the different radial regions of the model stars, both in presence and absence of magnetic field, is illustrated in table \ref{table:1} and \ref{table:2}.

During the thermal relaxation stage the effective surface temperature of the star does not reflect the thermal state of its core, since the core and crust remain thermally decoupled. At the end of this period the temperature is dropped drastically when the cooling wave coming from the core hits the surface of the star, initiating the process of neutrino cooling. At this stage the core becomes nearly isothermal and in case of the model stars, the temperature is even reduced to as low as \textcolor{black}{$\sim10^8$ K}. As the overall critical temperature of neutrons ($T_{cn}$) and protons ($T_{cp}$) in the pairing channels is much greater than \textcolor{black}{$\sim10^8$} K, as shown in the middle row of fig. \ref{del}, for both of the massive and canonical mass stars, it can be concluded that the neutrons (protons) inside the stars are converted to superfluid (superconducting) phase, soon after the thermal relaxation period is ended. \textcolor{black}{However, it is to be noted that in case of the lower boundary of the triplet pairing channel of the neutrons, $T_{cn}$ is quite low, even lower than $10^8$ K at some parts of the stars, as compared to the other pairing channels.} It is needless to say that in the NS interior superfluid neutrons and superconducting protons can coexist. The thermal relaxation period lasts \textcolor{black}{around $\sim 100$ years for the massive star, while it extends a little more in case of the canonical mass star.} The duration of the thermal relaxation period impacts the thermal evolution of the magnetars remarkably. Due to the shorter thermal relaxation time, the massive star starts to cool down faster than the canonical mass star, because of which the massive star becomes superfluid at an earlier stage, as compared to the canonical mass star. Once the stellar matter is converted from normal to superfluid/superconducting, both the heat capacity and the thermal conductivity are changed exceptionally.
\textcolor{black}{In our model stars, the baryons start forming superfluid pairs at the later stage of  the thermal relaxation process which might be responsible for slowing down their overall cooling rates. }

\section{Thermal properties in absence of magnetic field}
\subsection{Heat capacity}
The total heat capacity of the core of a NS is taken as the sum of the partial heat capacity of each component particle $i.e.$ $C_v=\sum\limits_{i=n,p,e}C_i$, since the core is consisted of strongly degenerate Fermions. Here the partial heat capacity of the $i$-th particle species is given by \citep{1981SSRv...28..450L},
\begin{equation}\label{eq:bin}
    C_i = \frac{2\hbar}{(2\pi \hbar)^3}\int d^3k_i(\epsilon _i-\mu_i)\frac{\mathrm{d} f_i}{\mathrm{d} T},
\end{equation}
where $f_i = 1/(1+e^z)$ is Fermi Dirac distribution function with $z=({\epsilon_i - \mu_i})/k_B T$.  $\epsilon_i$ and $\mu_i$  are energy and chemical potential of $i$-th particles respectively, $k_B$ is Boltzmann constant and $T$ is the temperature. Inside the NS, the electrons form an ultra relativistic highly degenerate Fermi gas, while the nucleons form a non relativistic non-ideal Fermi liquid.
Solving eq. \eqref{eq:bin} for these two cases, one obtains the general expression of partial heat capacity for $i$-th particle as follows,
\begin{equation}
    C_{{i}{0}} = \frac{m_{i}^{*} k_{F_{i}} k_{B}^2 T}{3\hbar^3},
    \label{eq:CwithoutB}
\end{equation}
where $i = n, p, e$ and $m_{i}^{*}$ and $k_{F_i}$ are effective mass and Fermi momentum of $i$-th species respectively. The subscript '$0$' implies the presence of normal nucleons.

\subsubsection{Effect of superfluidity/superconductivity}\label{cv_sf}
If any particle species ($i$) is considered to be superfluid then below its critical temperature ($T_{c}$) all the particles of that particular species transit to the ground state which reduces their free energy and consequently, the entropy. When the temperature reaches below $T_{c}$ the energy gap increases. In NS, only those nucleons having energy $|\epsilon_{i}-\mu_{i}|\lesssim k_{B}T$ contribute to the heat capacity. In presence of strong superfluidity $i.e.$ at $T<<T_{c}$, the energy gap $\Delta(T)$ exceeds $k_{B}T$, as a consequence of which the corresponding nucleons cannot contribute to the heat capacity for which it gets lowered \citep{1999PhyU...42..737Y}. The concept is realised mathematically by introducing a multiplicative reduction factor, denoted by $\mathcal{R}(T)$, so that the suppression of heat capacity in presence of superfluidity can be expressed as, $C=\sum_{i=n,p,e} C_{i0}\mathcal{R}(T)$, $C_{i0}$ is mentioned in eq. \eqref{eq:CwithoutB}.

\begin{center}
\begin{figure*}[ht]
\includegraphics[scale=0.31]{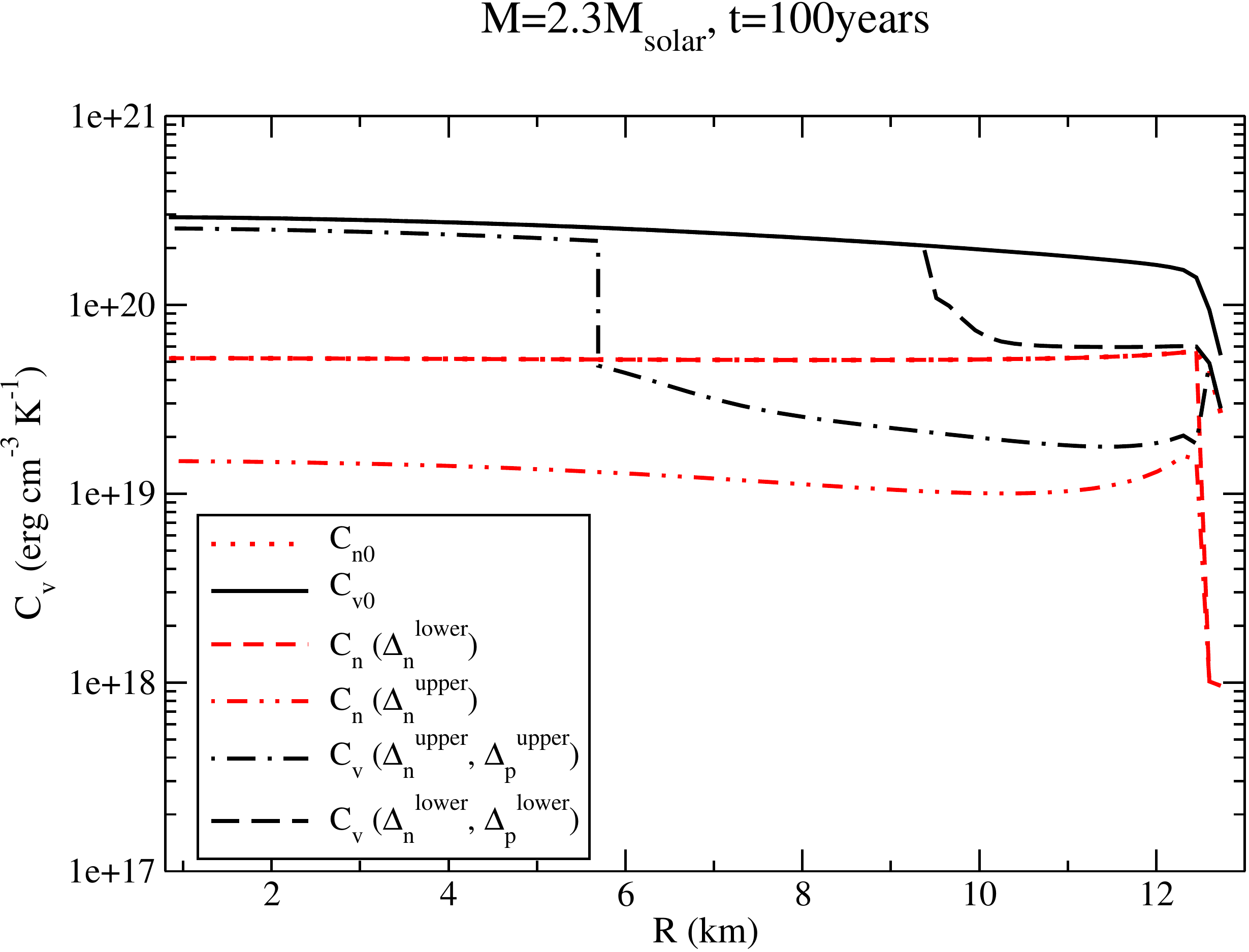}\hspace*{1mm}\includegraphics[scale=0.31]{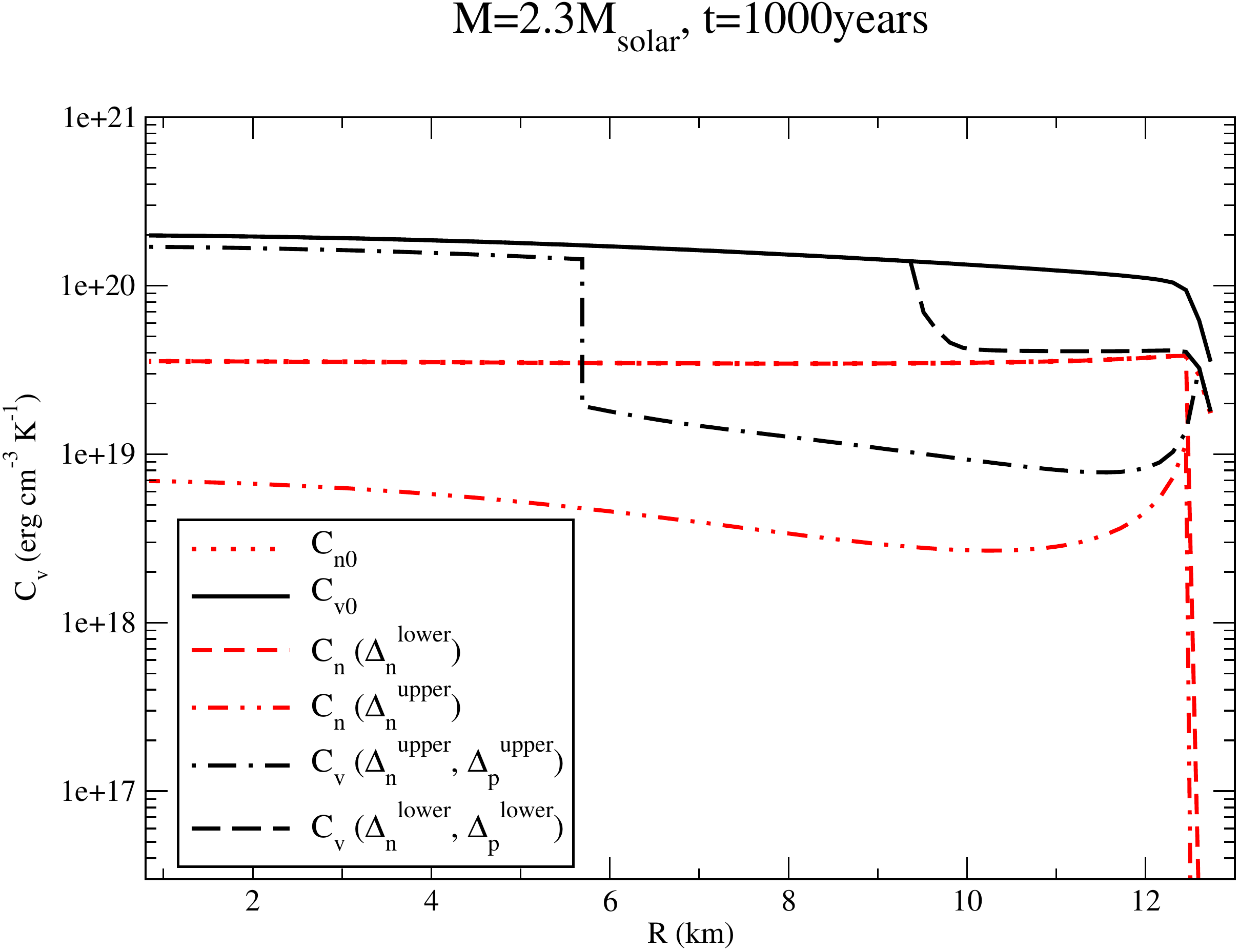}\\
\includegraphics[scale=0.31]{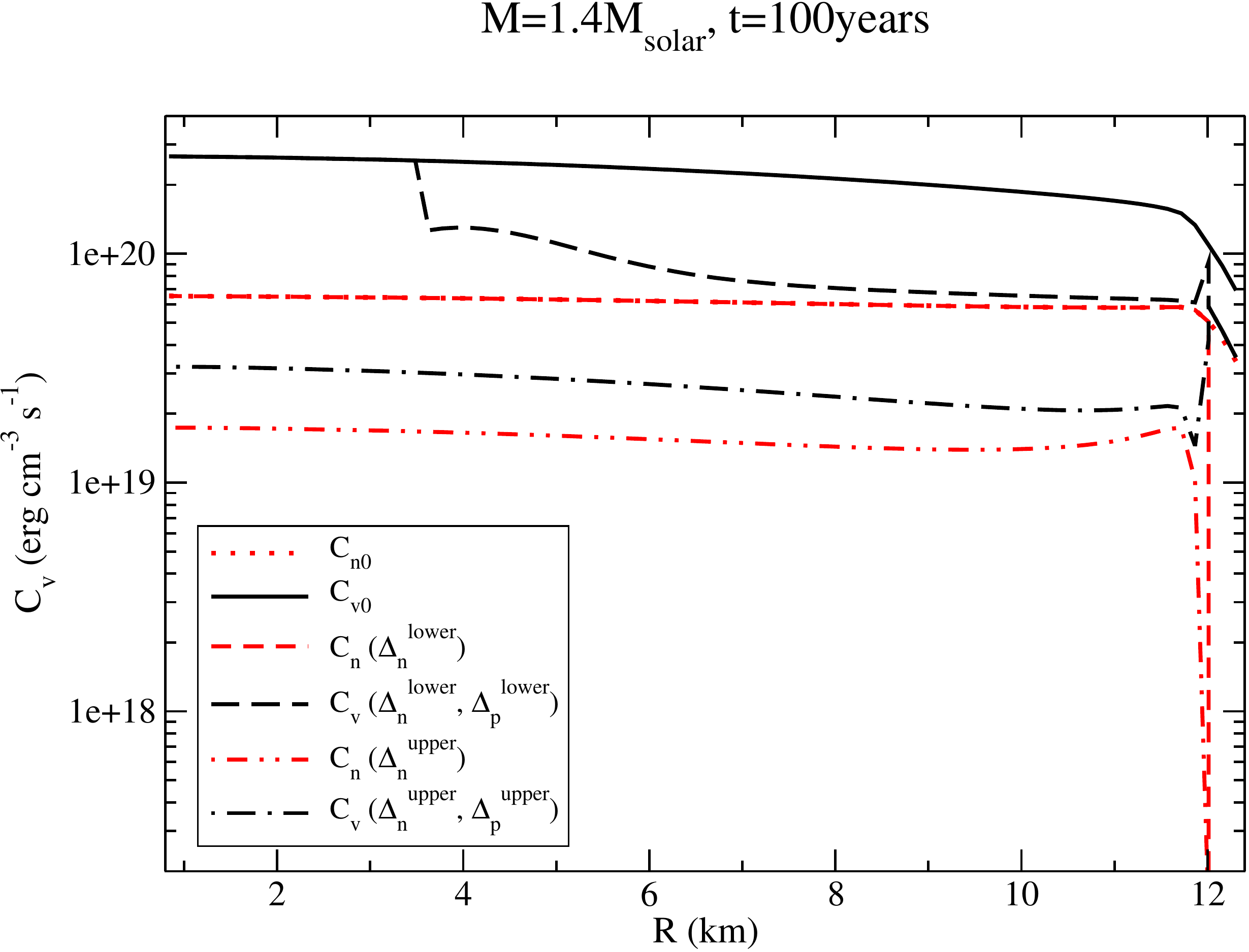}\hspace*{1mm}\includegraphics[scale=0.31]{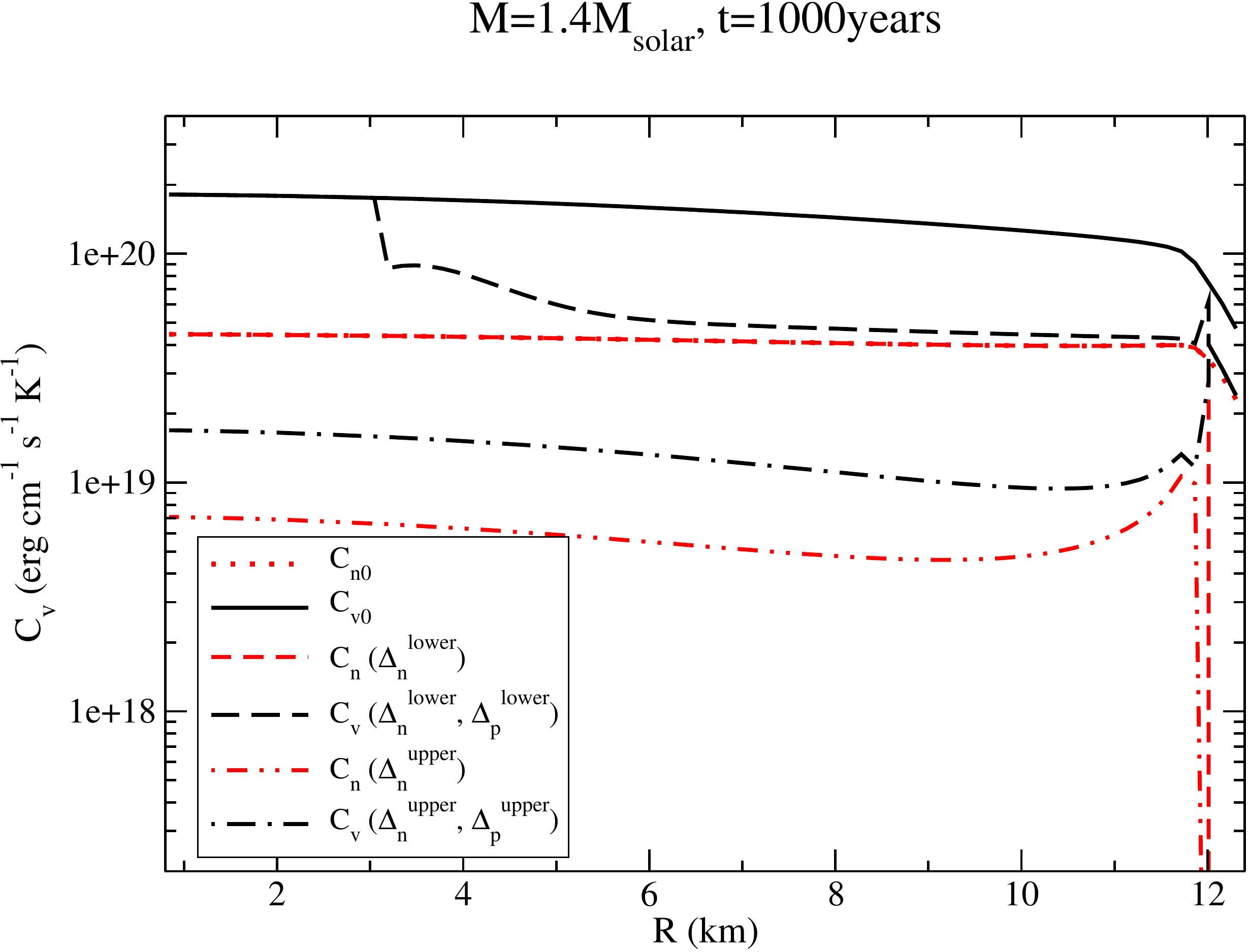}
\caption{Variation of heat capacity inside the star \textcolor{black}{for DDME2 EoS}. Solid line without superfluidity, long-dashed curves for total $C_v$ with lower $\Delta$ curve and dash-dotted curves for with upper $\Delta$ curve for total $C_v$. For neutron contribution to $C_v$ dotted curves are without superfluidity, short-dashed curve with lower $\Delta$ curve and dash-dot-dotted curves with upper $\Delta$ curves. Top panel: for the massive star $M\sim 2.3 M_\odot$; bottom panel: for $M\sim 1.4 M_\odot$.  Left panel: for $t=100$ years; right panel: for $t=1000$ years.}
\label{1.4lw1}
\end{figure*}
\end{center}

The reduction factor depends on the nature of the pairing energy gap of the nucleons, $\Delta_n$ and $\Delta_p$ \citep{1999PhyU...42..737Y} .
For $^1S_0$ pairing, $\Delta$ is isotropic, denoted by $\Delta=\delta_{0}$ and is independent of the orientation of the particle momenta with respect to the quantization axis. The dependence of $\delta_{0}$ with $T$ is obtained from the BCS theory which can be expressed as \citep{1994ARep...38..247L,1995A&A...297..717Y},
\begin{align}\label{eqn12}
y_{n}(^1S_0) \equiv \frac{\delta_{0}(T)}{k_{B}T} \equiv y_p(^1S_0) \nonumber \\  = \sqrt{1-T/T_{c}} \left( 1.456-\frac{0.157}{\sqrt{T/T_{c}}}+\frac{1.764}{T/T_{c}} \right) 
\end{align}
For $^3P_2$ pairing the energy gap becomes anisotropic \citep{1970PhRvL..24..775H} because of its dependence on the direction of particle momenta. However the energy gap can still be assumed to be isotropic if the orientation of the quantization axis is arbitrary instead of having a fixed direction. Therefore for triplet pairing the energy gap, denoted by $\delta_{1}$, is expressed as \citep{1994ARep...38..247L, 1995A&A...297..717Y, 2000A&A...357.1157H,  2001A&A...374..151B}, \begin{equation}\label{eqn13}
y_{n}(^3P_2) \equiv \frac{\delta_{1}(T)}{k_{B}T}=\sqrt{1-T/T_{c}} \left(0.7893+\frac{1.188}{T_{c}}\right).
\end{equation}
The quantities $y_{n}$ and $y_p$ mentioned in eqn. \eqref{eqn12} and \eqref{eqn13} are dimensionless. It is evident that for $^1S_0$ and $^3P_2$ pairing the reduction factors are different, that are expressed in \citep{1999PhyU...42..737Y} as algebraic expressions. All the reduction factors tend to be unity in the absence of superfluidity/superconductivity.

The variation of heat capacity with radius is shown in fig. \ref{1.4lw1} for the massive and the canonical mass stars at two different ages,  $t=100$ and $1000$ years. The reason for selecting these two specific times is that at $t\gtrsim 100$ years in both of the stars the thermal equilibrium is \textcolor{black}{about to be} established and the neutrons (protons) just start to convert into superfluid (superconducting), while at a much latter time, $t=1000$ years, the effect of the nucleon pairing is expected to be manifested much more prominently on the thermal properties. The total heat capacity of the matter with contribution from all the particle species is denoted by $C_{v(0)}$, while $C_{n(0)}$ represents the contribution from the neutrons only%, in absence of superfluid matter
. %In fig. \ref{1.4lw1} the solid and the dotted lines represent $C_{v0}$ and $C_{n0}$ respectively. 
The heat capacity of the protons and electrons are relatively much smaller, hence they are not shown separately in the figure. 

%%%%%%%%%%%%%CORRECT ONE%%%%%%%%%%%%%%%%%%%%%
%Near the crust, all the nucleons form $^1S_0$ pairing and the strength of neutron pairing dominates over proton pairing $i.e.$ $\Delta_n^{lower(upper)}(^1S_0)>\Delta_n^{lower(upper)}(^1S_0)$, in case of both the massive and canonical mass stars, as observed from fig. \ref{del}. \textcolor{blue}{In case of the massive star, the neutrons form $^1S_0$ pairing till $R\sim 12.59$ km and transit to $^3P_2$ pairing state after that, till $R\sim 10.98$ km ($R\sim 5.69$ km) for the lower (upper) bounds of $\Delta_p$.} It is to be noticed that the pairing strength in $^3P_2$ state is much weaker as compared to $^1S_0$ state. Therefore, the resultant heat capacity $C_v$ is highly suppressed near the surface region of the star $i.e.$ at \textcolor{blue}{$R\gtrsim 12.45$ km}, for both the lower and upper bounds of $\Delta_n$ and $\Delta_p$.
%%%%%%%%%%%%%%%%%%%%%%%%%%%%%%%%%%%%%%%%%%%%%%%

Near the crust, all the nucleons form $^1S_0$ pairing and the strength of neutron pairing dominates over proton pairing $i.e.$ $\Delta_n^{lower(upper)}(^1S_0)>\Delta_p^{lower(upper)}(^1S_0)$, in case of both the massive and canonical mass stars, as observed from fig. \ref{del}. In case of the massive star, the neutrons form $^1S_0$ pairing till \textcolor{black}{$R\sim 12.5$ km } and transit to $^3P_2$ pairing state after that, till \textcolor{black}{$R\sim 11$ km, for $\Delta_n^{lower}$, while in case of $\Delta_n^{upper}$ the neutron pairing exists till the inner core }. It is to be noticed that the pairing strength in $^3P_2$ state is much weaker as compared to $^1S_0$ state. \textcolor{black}{The lower bound of $^3P_2$ provides very small value of pairing gap energy inside the star. Any particular particle species ($i$) may not exist in the pairing state, if the temperature of the star remains greater than the critical temperature of the particles $i.e.$ $T>T_{ci}$, even though its pairing energy gap is non-zero. }

Therefore, the resultant heat capacity $C_v$ is highly suppressed near the surface region of the star $i.e.$ at \textcolor{black}{$R\gtrsim 12.5$ km}, for both $\Delta_n^{lower}$ and $\Delta_p^{upper}$. The protons are superconducting \textcolor{black}{within} the radial distance \textcolor{black}{$12.5\gtrsim R\gtrsim 9.4$ km ($12.5\gtrsim R\gtrsim 5.69$ km)}, corresponding to $\Delta_p^{lower(upper)}$. \textcolor{black}{The region of active proton superconductivity is altered in presence of magnetic field which is discussed in detail in section 4.2. }

Within the radial zone \textcolor{black}{$12.5~ \text{km}\gtrsim R \gtrsim10.98$ km, the pairing energy gaps of $^1S_0$ protons and $^3P_2$ neutrons are non-zero in case of their respective lower boundaries. However, since the internal temperature of the star varies within $T\sim (6-9)\times10^8$ K at $t=100$ years and $\sim(4-6)\times10^8$ K at $t=1000$ years, the critical temperature of $^3P_2$ neutrons for $\Delta_n^{lower}$ becomes smaller than the temperature of the star $(T_{cn}^{lower}<T)$, in these regions. As a consequence, the neutrons remain non-superfluid, despite of the non-zero value of $\Delta_n^{lower}$ for triplet state which is extremely small. Proton superconductivity is further extended till $R\sim 9.4$ km, above which both the baryons are normal. Unlike the triplet neutrons, the corresponding critical temperature $T_{cp}^{lower}$ is larger than $T$ which activates the proton superconductivity. This results in the order of $C_v$ to stay nearly constant at $\sim6\times10^{19}$ erg cm$^{-3}$ K$^{-1}$ within the range $12.5\gtrsim R\gtrsim 9.4$ km at $t=100$ years which is lower than the range of $C_{v0}$ in this region. At $R\sim 9.36$ km, $C_v$ is suddenly lifted to $2\times10^{20}$ to merge with $C_{v0}$ and is slightly increased to $3\times10^{20}$ erg cm$^{-3}$ K$^{-1}$ till the inner core. In this scenario, the neutron contribution to specific heat, $C_n$ remains the same as $C_{n0}$ at $R\lesssim 12.5$ km, as observed from the plot, due to the loss of the triplet pairing of the neutrons.} 

\textcolor{black}{In case of $\Delta_n^{upper}$ and $\Delta_p^{upper}$, the scenario is quite different. Here, at $R\gtrsim 12.5$ km neutrons are in $^1S_0$ pairing state and at $R\lesssim 12.5$ km neutron superfluidity is converted from singlet to triplet pairing state. The proton pairing is also turned on at $R\sim 12.5$ km. At $R\lesssim 5.69$ km the protons are normal, while the neutrons remain in the $^3P_2$ state till the inner core. Since $T_{cn}^{upper}$ for the $^3P_2$ neutrons remains higher than the stellar temperature due to the presence of sizable $\Delta_n^{upper}$, the $^3P_2$ neutrons remain intact nearly all throughout the star. The proton superconductivity also remains activated in the radial zone $12.5\gtrsim R \gtrsim 5.69$ km, as the condition $T_{cp}^{upper}>T$ remains satisfied. This causes $C_v$ to vary within $(2-5)\times 10^{19}$ erg cm$^{-3}$ K$^{-1}$ for $12.5\gtrsim R \gtrsim 5.69$ km and jump suddenly to $\sim 2.8\times 10^{20}$ erg cm$^{-3}$ K$^{-1}$ at $R\sim 5.69$ km, due to the conversion of singlet protons to normal ones and remain constant till the inner core.}
At $t=1000$ years both $C_v$ and $C_{v0}$ are reduced to some extent in all the regions of the star.

In case of the canonical mass star the neutrons form $^1S_0$ pairing in the narrow region around the crust \textcolor{black}{$R\gtrsim11.8$ km for $\Delta_n^{lower}$ which causes $C_v$ to have very small values near the surface of the star. For the $^3P_2$ neutrons, although $\Delta_n^{lower}$ is non-zero within the radial range $11.8\gtrsim R \gtrsim 8.48$ km, due to its very small value, $T_{cn}^{lower}<T$ with $T\sim (6-9)\times10^8$ K at $t=100$ years and $T\sim(4-6)\times10^8$ K at $t=1000$ years, which causes the neutrons to exist in normal state in this area of the star, similar to the scenario of the massive star. For the protons, although $\Delta_p^{lower}$ is non-zero inside the radial zone $R\sim 11.8$ to $R\sim 3.13$ km, the proton superconductivity is actually active inside the star from $R\sim 11.86$ km to $R\sim 3.63$ km as the conditions $T_{cp}^{lower}>T$ is no longer valid at $R\leq 3.63$ km. Therefore, $C_v$ varies from $6\times10^{19}$ to $\sim 1.5\times 10^{20}$ erg cm$^{-3}$ K$^{-1}$ within the radial range $12\gtrsim R \gtrsim 3.63$ km which is suddenly raised to $\sim 2.5\times10^{20}$ erg cm$^{-3}$ K$^{-1}$ at $R\sim 3.63$ km and remains nearly the same beyond that.}

\textcolor{black}{For the upper boundaries of the nucleon pairing energy, within the radial zone at $R\gtrsim 11.8$ km neutrons form $^1S_0$ pairing and reside in $^3P_2$ state in the rest of the star. The proton pairing is also active in the entire star starting from $R\sim 11.8$ km. Both $T_{cn}^{upper},~T_{cp}^{upper}$ in different pairing channels are greater than the stellar temperature in all the region of the star. The range of $C_v$ prevails within $(2-3)\times10^{19}$ erg cm$^{-3}$ K$^{-1}$ in almost through the entire star $i.e.$ $R\lesssim 11.8$ km.} Further, it is needless to mention that both $C_v$ and $C_{v0}$ are lower at $t=1000$ years as compared to $t=100$ years. 

The distinct features of $C_v$ in the different radial regions of the massive star at different time are illustrated in table 3. The entire star is divided into different radial zones depending on the nature of nucleon pairing at different zones which are different for the lower and upper boundaries of the pairing energy gap. Although the variation of $C_{v0}$ is continuous throughout the star, its value is mentioned separately in the distinct zones to point out how much $C_v$ is lesser than $C_{v0}$ in superfluid/superconducting nuclear matter. 

%\begin{widetext}
\begin{center}
\begin{table*}[h!]\label{t}
\hspace*{-0.5cm}
\begin{tabular}{|c|c|c||c|c|}
\hline
\multicolumn{5}{|c|}{$M=2.3~M_{\odot}$} \\
\hline
 Time$\rightarrow$ & \multicolumn{2}{c||}{$t=100$ yrs } & \multicolumn{2}{c|}{$t=1000$ yrs }\\
  \hline
 Radial zone$\rightarrow$  & \textcolor{black}{$12.5-9.4$} km & \textcolor{black}{$\leq 9.4$} km &  \textcolor{black}{$12.5-9.4$} km  & \textcolor{black}{$\leq 9.4$} km \\
  \hline
   $C_{v0}$ & \textcolor{black}{$5\times10^{19}-2\times10^{20}$} & \textcolor{black}{$(2-3)\times10^{20}$} & \textcolor{black}{$3\times10^{19}-1.5\times10^{20}$} & \textcolor{black}{$(1.5-2)\times10^{20}$} \\
   \hline 
   $C_v$ ($\Delta_n^{lower}, \Delta_p^{lower}$)  & \textcolor{black}{$6\times10^{19}$} & \textcolor{black}{$(2-3)\times10^{20}$} & \textcolor{black}{$4\times10^{19}$} & \textcolor{black}{$(1.5-2)\times10^{20}$} \\
  \hline
 \hline
 \end{tabular}
% \end{table*}
% \end{center}
% \begin{center}
% \begin{table*}
\begin{tabular}
{|p{2.4cm}|p{2.5cm}|p{2.5cm}||p{2.5cm}|p{2.5cm}|} 
\hline
 Time$\rightarrow$ & \multicolumn{2}{c||}{$t=100$ yrs} & \multicolumn{2}{c|}{$t=1000$ yrs}\\
  \hline
 Radial zone$\rightarrow$  & \textcolor{black}{$12.5-5.69$} km & \textcolor{black}{$\leq 5.69$} km & \textcolor{black}{$12.5-5.69$} km & \textcolor{black}{$\leq 5.69$} km  \\
  \cline{1-5}
  $C_{v0}$ & \textcolor{black}{$5\times10^{19}-2.8\times10^{20}$}  & \textcolor{black}{$(2.8-3)\times10^{20}$} & \textcolor{black}{$3\times10^{19}-1.8\times10^{20}$} & \textcolor{black}{$2\times10^{20}$} \\
  \hline
  $C_v$($\Delta_n^{upper}, \Delta_p^{upper} $) & \textcolor{black}{$(2-5)\times10^{19}$} & \textcolor{black}{$2.8\times10^{20}$} & \textcolor{black}{$8\times10^{18}-2\times10^{19}$} &  \textcolor{black}{$1.8\times10^{20}$}    \\
 \hline
 \end{tabular}
\caption{Estimation of the resultant heat capacity of all the particle species, $C_{v}=\sum_{i=n,p,e}C_{i)}$, in the unit of erg cm$^{-3}$ K$^{-1}$ in the different radial zones of the massive star ($M=2.3M_{\odot}$), in presence of normal ($C_{v0}$) and superfluid/superconducting nucleons ($C_v$), considering the lower and upper boundaries of their pairing energies ($\Delta_n$, $\Delta_p$) at the two different ages, $t=100$ and $t=1000$ years. Different radial zones are selected accordingly as the existence of the nucleon pairing, as mentioned in Tables 1 and 2.  }
\end{table*}
\end{center}
%\end{widetext}

%\begin{widetext}
\begin{center}
\begin{table*}[h!]
\begin{tabular}{|c|c|c||c|c|} 
\hline
\multicolumn{5}{|c|}{$M=1.4~M_{\odot}$} \\
\hline
 Time$\rightarrow$ & \multicolumn{2}{c||}{$t=100$ yrs} & \multicolumn{2}{c|}{$t=1000$ yrs}\\
  \hline
Radial zone$\rightarrow$  &  \textcolor{black}{$11.8-3.63$} km & \textcolor{black}{$\leq 3.63$} km & \textcolor{black}{$11.8-3.63$} km & \textcolor{black}{$\leq 3.63$} km \\
 \hline
 $C_{v0}$ & \textcolor{black}{$7\times10^{19}-2.8\times10^{20}$} & \textcolor{black}{$2.8\times10^{20}$} & \textcolor{black}{$6\times10^{19}-1.8\times10^{20}$} & \textcolor{black}{$1.8\times10^{20}$}  \\
 \hline
  $C_v(\Delta_n^{lower},\Delta_p^{lower})$ & \textcolor{black}{$6\times10^{19}-1.5\times10^{20}$} & \textcolor{black}{$(2.5-2.8)\times10^{20}$} & \textcolor{black}{$(4-9)\times10^{19}$} & \textcolor{black}{$1.8\times10^{20}$}  \\
 \hline\hline
 \end{tabular}
 \begin{tabular}{|p{2.4cm}|p{4cm}||p{4cm}|}
  Time$\rightarrow$ & $t=100$ yrs & $t=1000$ yrs\\
  \hline
  Radial zone$\rightarrow$ &  \textcolor{black}{$\leq 11.8$} km & \textcolor{black}{$\leq 11.8$} km \\
  \cline{1-3}
  $C_{v0}$ & \textcolor{black}{$8\times10^{19}-2.8\times10^{20}$} & \textcolor{black}{$6\times10^{19}-1.8\times10^{20}$} \\
  \hline
 $C_v(\Delta_n^{upper},\Delta_p^{upper})$ & \textcolor{black}{$(2-3)\times10^{19}$} & \textcolor{black}{$(1-1.8)\times10^{19}$} \\
 \hline
\end{tabular}
\caption{Same as table 3, for the canonical mass star ($M=1.4M_{\odot}$)}\label{tt}
\end{table*}
 \end{center}
%\end{widetext}

\subsection{Thermal conductivity}\label{kappa}
Neutrons are the most significant heat carriers in the neutron star core being the most abundant particles, while the protons have much smaller number density. Therefore, the contribution of neutrons is much larger to the thermal conductivity as compared to the protons. The leptons also play a crucial role in heat conduction along with neutrons due to their high mobility and large mean free path. In the envelope and outer crust, the heat conduction by electrons is much significant where electron-ion scattering takes place. In the inner crust and core free neutrons start to appear. Heat conduction by neutrons occur by neutron-neutron ($nn$) and neutron-proton ($np$) collisions via strong interaction, while the electrons undergo Coulomb collision with protons ($ep$) and other electrons ($ee$). Neutron-electron ($ne$) scattering is neglected in this work. 
It is to be mentioned that the thermal conductivity for neutrons and electrons can be studied independently \citep{2015SSRv..191..239P}. Theoretical development of thermal conductivity in dense matter with $npe$ configuration, along with other transport coefficients are analysed in previous works \cite{1976ApJ...206..218F, 1979ApJ...230..847F}. The key ingredient of thermal conductivity calculation is the evaluation of nucleon-nucleon scattering which occurs via the strong interaction. The problem is discussed with different approaches and nuclear potentials in several literatures  \citep{1979ApJ...230..847F, 1993NuPhA.555..128W, 
1994PhLB..338..111S, 2001A&A...374..151B, 2020PhRvD.102f3010S}. 

As mentioned earlier, due to plenty of abundance, neutrons contribute more compared to protons to the thermal conductivity inside the stellar core, in which the nucleons form a strongly degenerate non-relativistic Fermi liquid while the electrons form a strongly degenerate ultra-relativistic gas. In the limit of strong degeneracy as the inter-particle interaction is taken into account, the Fermi liquid is considered to be composed of $quasiparticles$ \cite{Landau:1956zuh}. Since neutrons and electrons are the only significant heat carriers, they are treated as quasiparticles while the protons are considered to be scatterers. The calculation of thermal conductivity of Fermi liquid was done previously in the low temperature limit \cite{1959RPPh...22..329A,1970AnPhy..56....1S}. The quasiparticles are treated as a weakly interacting dilute gas having non-equilibrium probability distribution function $F_c(\mathbf{r},\mathbf{v},t)$ where $c=n,e$. 
The dynamics of the quasiparticles is determined by Boltzmann transport equation, given by
\begin{equation}\label{eqn1}
\left(\frac{\partial}{\partial t}+\mathbf{v}_c.\mathbf{\nabla}_r+\mathbf{g}_{c}.\mathbf{\nabla}_{v}\right) F_c(\mathbf{r},\mathbf{v},t)=J_{ci},
\end{equation}
where $\mathbf{v}_c$ is the velocity of the heat carrier ($c$), $\mathbf{g}_{c}$ is the force  per unit mass exerted by it. The collisions occurring among different particle species restore the non-equilibrium to equilibrium. Here $\mathbf{\nabla}_r$ and $\mathbf{\nabla}_v$ denote partial differentiation in configuration and momentum space respectively. $J_{ci}$ is the collision integral for collision occurring between the quasiparticle heat carrier $c$ and $i$-th particle species. To estimate the thermal conductivity of the neutrons, the total contribution to the collision integral, $J_{ci}$ is represented by, $J_n=J_{nn}+J_{np}$, while in case of the electrons it is expressed as, $J_e=J_{ep}+J_{ee}$.
Since the protons are not treated as quasiparticles, %treated as the scatterers instead of heat carriers, they are not treated as quasiparticles. Hence protons 
they are described by equilibrium Fermi-Dirac (FD) distribution function, given by \citep{2001A&A...374..151B}
\begin{equation}\label{eqn2}
F_{p}=f_{p}=[1+exp (\epsilon_{p}-\mu_{p})]^{-1},
\end{equation}
where $\epsilon_{p}$ and $\mu_{p}$ are the energy and chemical potential of the proton respectively. The distribution function for neutron and electron $F_{c}$ is expressed as \citep{1979ApJ...230..847F, 1995NuPhA.582..697G,2001A&A...374..151B}
\begin{equation}\label{eqn3}
F_{c}=f_{c}-\phi_{c} \frac{\partial f_{c}}{\partial \epsilon_{c}}.
\end{equation}
Here $f_{c}$ ($c=e,n$) is the equilibrium FD distribution function, similar to eqn. \eqref{eqn2}. The small deviation from the equilibrium is introduced in terms of the factor $\phi_c$, which is represented by \citep{ziman1960electrons}
\begin{equation}\label{eqn4}
\phi_{c}=-\tau_{c} (\epsilon_{c}-\mu_{c}) \frac{\mathbf{v}_{c}.\mathbf{\nabla}{T}}{T},
\end{equation}
where $\tau_{c}$ is the effective relaxation time corresponding to the collision of the quasiparticle heat carrier ($c$) and $T$ is the temperature. Now for the scattering process $c+i\rightarrow c^{'}+i^{'}$, the collision integral has the general form given by the following expression
%\begin{widetext}
\begin{center}
\begin{eqnarray}\label{eqn6}
J_{ci}=\frac{1}{(2\pi\hbar)^{9}}\int d^{3}\mathbf{k}_{c}^{'}d^{3}\mathbf{k}_{i}d^{3}\mathbf{k}_{i}^{'} \sum_{\sigma_{c}^{'},\sigma_{i},\sigma_{i}^{'}} \mathcal{P}_{ci} \nonumber \\
\times [F_{c}^{'}F_{i}^{'}(1-F_{c})(1-F_{i})-F_{c}F_{i}(1-F_{c}^{'})(1-F_{i}^{'})],
\end{eqnarray}
\end{center}
%\end{widetext}
where $\mathcal{P}_{ci}$ is the differential transition probability. The unprimed (primed) quantities correspond to the particles before (after) collision and
$\sigma_{i}^{(')}$ is the spin of the $i^{(')}$-th particle.
Since the particles remain in strongly degenerate state at the core, only those particles near the vicinity of the Fermi surface, $\hbar k_{F} |k-k_{F}| \lesssim k_{B} T$ contribute to the collision integral significantly, given by eq. \eqref{eqn6}. Using eqn. \eqref{eqn2}, \eqref{eqn3}, \eqref{eqn4}, \eqref{eqn6} in eq. \eqref{eqn1}, we get
%\begin{widetext}
\begin{eqnarray}\label{eqn8}
\mathbf{v}_{c}\cdot\frac{\partial f_{c}}{\partial T} \mathbf{\nabla} T= 
\frac{\tau_{c} \nabla{T}}{(2\pi)^{9}T^{2}}\sum_{i}\sum_{\sigma_{c}^{'},\sigma_{i},\sigma_{i}^{'}}\int\int\int d\mathbf{k}_{c}^{'} d\mathbf{k}_{i}^{'} d\mathbf{k}_{i}  \nonumber \\
\times  f_{c}f_{i}(1-f_{c}^{'})(1-f_{i}^{'}) \mathcal{P}_{ci}(\phi_{c}^{'}+\phi_{i}^{'}-\phi_{c}-\phi_{i}).
\end{eqnarray}
%\end{widetext}
Solving eq. \eqref{eqn8}, one arrives at the following expression
\begin{equation}\label{eqn9}
\tau_{c}=\frac{1}{\sum\limits_{i}\nu_{ci}},
\end{equation}
where $\nu_{ci}$ denotes the collision frequency corresponding to the collision  occurring between the heat carrier ($c$) and $i$-th particle species ($i=n,p$ if $c=n$ ; $i=e,p$ for $c=e$).Thus, in case of neutrons, $\tau_n=(\nu_{nn}+\nu_{np})^{-1}$ and for electrons, $\tau_e=(\nu_{ee}+\nu_{ep})^{-1}$.

Thermal conductivity $\kappa_{c}$ of a particular heat carrier $c$ can be expressed by Fourier's heat conduction equation as
\begin{equation}\label{eqn5}
\mathbf{q_{c}}= -\kappa_{c}\mathbf{\nabla}T
=\frac{2}{(2\pi\hbar)^{3}} \int d^{3} \mathbf{k}_{c} (\epsilon_{c}-\mu_{c})\mathbf{v}_{c}F_{c},
\end{equation}
where $\mathbf{q_{c}}$ is the heat flux. Using eqs. \eqref{eqn3} and \eqref{eqn4} into \eqref{eqn8} and comparing with eq. \eqref{eqn5}, one can obtain the expression of thermal conductivity as
\begin{eqnarray}\label{eqn7}
\kappa_{c0}=\frac{\pi^{2}Tn_{c}\tau_{c}}{3m_{c}^{*}} \nonumber \\
=\frac{\pi^{2}Tn_{c}}{3m_{c}^{*}\sum\limits_{i}\nu_{ci}}
\end{eqnarray}
The subscript '$0$' implies that both the nucleons are normal. $n_{c}$ is the number density and $m_{c}^{*}$ is its effective mass of the heat carrier ($c$). The collision frequency of the baryon and lepton are calculated independently.

\paragraph{Baryon:}

The expression of collision frequency for the neutrons ($c=n$) in the dense matter is given by \citep{ 2001A&A...374..151B}
\begin{equation}\label{eqn10}
\nu_{ni}=\frac{64m_{n}^{*}m_{i}^{*_{2}}k_{B}^{2}T^{2}S_{ni}}{5m_{n}^{2}\hbar^{3}},
\end{equation}
where $m^*_i$ is the effective mass of $i$-th species. The quantity $S_{ni}$ has the dimension of the cross section and are expressed as analytical fit expressions for $nn$ and $np$ scattering, given by \cite{2001A&A...374..151B} 

\begin{eqnarray} \label{eq}
S_{nn}= \frac{7.880(1-0.2241k_{Fn}+0.2006k_{Fn}^2}{k_{Fn}^2(1-0.1742k_{Fn})} 
\times \left(\frac{m_n}{m_n^*}\right)^2 \times & \nonumber \\ & \hspace*{-8cm}(0.4891+1.111u^2-0.2283u^3+0.01589k_{Fp}  \nonumber \\ & \hspace*{-10cm} -0.02099k_{Fp}^2+0.2773uk_{Fp}),\nonumber 
\end{eqnarray}
\begin{eqnarray}
S_{np} = \frac{0.8007k_{Fp}}{k_{Fn}^2}(1+31.28k_{Fp}-0.0004285k_{Fp}^2+ \nonumber \\ 26.85k_{Fn}+ 0.08012k_{Fn}^2)(1-0.5898k_{Fn}+ \nonumber \\ 0.2368k_{Fn}^2 +0.5838k_{Fp}^2+0.884k_{Fn}k_{Fp})^{-1}  \nonumber \\ \times \left(\frac{m_n}{m_p^*}\right)^2 (0.04377+1.1v^2+  0.118v^3  +0.1626k_{Fp} \nonumber \\ +  0.3871vk_{Fp}-0.299v^4) + \frac{0.383k_{Fp}^4}{k_{Fn}^{5.5}}(1+102k_{Fp}+\nonumber \\ 53.91k_{Fn}) (1-0.7087k_{Fn}+0.2537k_{Fn}^2+ \nonumber \\ 9.404k_{Fp}^2-1.589k_{Fn}k_{Fp})  
\times \left(\frac{m_n}{m_p^*}\right)^2 \nonumber \\ (0.0001313+1.248w^2 +0.2403w^3+0.3257k_{Fp}\nonumber \\ +0.5536wk_{Fp}-0.3237w^4+0.09786w^2k_{Fp}) \nonumber \\ 
 = S_{np,1}+S_{np,2}
\end{eqnarray}
%\end{widetext}
Here, $u=k_{Fn}-1.556$, $v=k_{Fn}-2.126$ and $w=k_{Fn}-2.116$, provided $k_{Fn}$ and $k_{Fp}$ are the respective Fermi momenta of neutron and proton. In eqn. \eqref{eq} both $S_{nn}$ and $S_{np}$ are expressed in the unit of $millibarn$. The fit expressions are consisted within the range, $1.1\leq k_{Fn} 2.6$ fm$^{-1}$ and $0.3\leq k_{Fp} \leq 1.2$ fm$^{-1}$, which are consisted with the two model stars considered in this analysis.

\paragraph{Lepton: } For the leptonic component, the collision frequencies corresponding to $ep$ and $ee$ collision are given by \cite{1995NuPhA.582..697G}
\begin{eqnarray}\label{eqn18}
\nu_{ep}=\frac{4\pi^{2}}{5\hbar}\left(\frac{e^{2}}{\hbar c}\right)^{2}\left(\frac{\hbar k_{F_{e}}}{ q_{0}}  \right)^{3} \frac{\mu_{e} m_{p}^{*2}k_{B}^{2} T^{2}\hbar^2}{k_{F_{e}}^2} \nonumber \\
\nu_{ee}=\frac{2\pi^{2}}{\hbar} \left( \frac{e^{2}}{\hbar c} \right)^{2}\left(\frac{\hbar k_{F_{e}}}{ q_{0}}\right)^{3}\frac{k_{B}^{2}T^{2}}{\mu_{e}},
\end{eqnarray}
where $q_{0}$ is the screening momentum transfer given by, $q_{0}^{2}=(4e^{2}/\pi)( m_e^*k_{F_{e}}+m_p^*k_{F_{p}})$ \citep{1995NuPhA.582..697G}. $k_{F_{e}}$ and $\mu_e$ are the Fermi momentum and chemical potential of the electron respectively. Using eqn. (15) and (17) in eqn. (14), the thermal conductivity of neutron and electron are estimated respectively, in normal nuclear matter inside the core of a magnetar.

%%%%%%%%%%%%%%%%%%%%%%%%%%%%%%%%%%%%%%%%%
\subsubsection{Effect of Superfluidity} 
%%%%%%%%%%%%%%%%%%%%%%%%%%%%%%%%%%%%%%%%%

The presence of nucleon superfluidity/superconductivity the thermal excitation of the particles is minimized and consequently the inter-particle interaction is subdued. As the dispersion relation of the particles is modified due to the formation of nucleon pairing, 
the collision integral is also revised accordingly due to the change in the available phase space of the particles. The electrons remain non-superconducting throughout the star during the entire thermal evolution period as their critical temperature is much lower than the usual neutron star temperature.

\paragraph{Baryon:}  Due to the modification in phase space the collision integral corresponding to $nn$ and $np$ collisions, $J_{ci}$ ($c=n$), mentioned in eq. \eqref{eqn6} is altered. The phenomena of suppression of the internucleon interactions due to the change in phase space is reflected in collision integral by the introduction of a reduction factor $\mathcal{R}_{ni}$; $J_{ni}^{sf}=J_{ni}\mathcal{R}_{ni}$. This in turn affects the quantity $S_{ni}$ in eqn. \eqref{eqn10}. Thus in presence of neutron (proton) superfluidity (superconductivity), eqn. \eqref{eqn10} is modified as
\begin{equation}\label{eqn14}
\nu_{ni}^{sf}=\frac{64m_n^*m_i^{*_{2}}k_B^2T^2S_{ni}^{sf}}{5m_n^2\hbar^3}.
\end{equation}
Here $S_{ni}^{sf}$ is the modified form of $S_{ni}$ due to the presence of neutron and proton pairing, which are represented as
%\begin{widetext}
\begin{eqnarray}
S_{nn}^{sf}=S_{nn}\mathcal{R}_{n2}(y_n)+\frac{3\times 14.57(1-0.0788k_{Fn}+0.0883k_{Fn}^2)}{k_{Fn}^{1.5}(1-0.1114k_{Fn})}\nonumber \\ \times \left(\frac{m_n}{m_n^*}\right)^2 (0.4583+0.892s^2-0.5497s^3-0.06205k_{Fp} \nonumber \\ +0.04022k_{Fp}^2+0.2122sk_{Fp}) (\mathcal{R}_{n1}(y_n)-\mathcal{R}_{n2}(y_n)), \nonumber 
\end{eqnarray}
%\end{widetext}
\begin{eqnarray}
S_{np}^{sf}=S_{np,2}\mathcal{R}_{p2}(y_n,y_p)+0.5S_{np,1}\times\{3\mathcal{R}_{p1}(y_n,y_p) \nonumber \\ -\mathcal{R}_{p2}(y_n,y_p)\}
\end{eqnarray}
Here $s=k_{Fn}-1.665$. The factors $\mathcal{R}_{n1}$, $\mathcal{R}_{n2}$, $\mathcal{R}_{p1}$ and $\mathcal{R}_{p2}$ are mentioned in the ref. \cite{2001A&A...374..151B}. $y_{n}$ and $y_p$ are the energy gap functions for neutrons and protons respectively, described in the section \ref{cv_sf}.
The relaxation time of collision is modified in presence of neutron and proton pairing accordingly which is expressed as
\begin{equation}\label{eqn144}
\tau_{n}^{sf}=\frac{\mathcal{R}(y_{n})}{\nu_{nn}^{sf}+\nu_{np}^{sf}}.
\end{equation}
$\mathcal{R}(y_{n})$ is another reduction factor different from  $\mathcal{R}_{ni}$ mentioned earlier.
$\mathcal{R}(y_n)$ arises from the kinematic correction which is expresse as
\begin{equation}
\mathcal{R}(y_n)=3\sqrt{2}\left(\frac{y_n}{\pi}\right)^{1.5}e^{-y_n}
\end{equation}  
In case of strong superfluidity ($T<<T_c$), $\mathcal{R}(y_n)<<1$ while at $T\geq T_{c}$, $\mathcal{R}(y_n)=1$ . Consequently the expression of thermal conductivity of neutron is changed as follows
\begin{equation}\label{eqn16}
\kappa_{n}^{sf}=\frac{\pi^{2}Tn_{n}\tau_{n}^{sf}\mathcal{R}^{2}(y_{n})}{3m_{n}^*}
\end{equation}

One crucial point to be pointed out here is that while the superfluid neutrons suppresses $\kappa_n$, it is enhanced by proton superconductivity, as proton pairing leads to the reduction in collision frequency between neutron and proton ($\nu_{np}^{sf}<\nu_{np}$). 

\begin{figure*}[ht]
\centering
\includegraphics[scale=0.3]{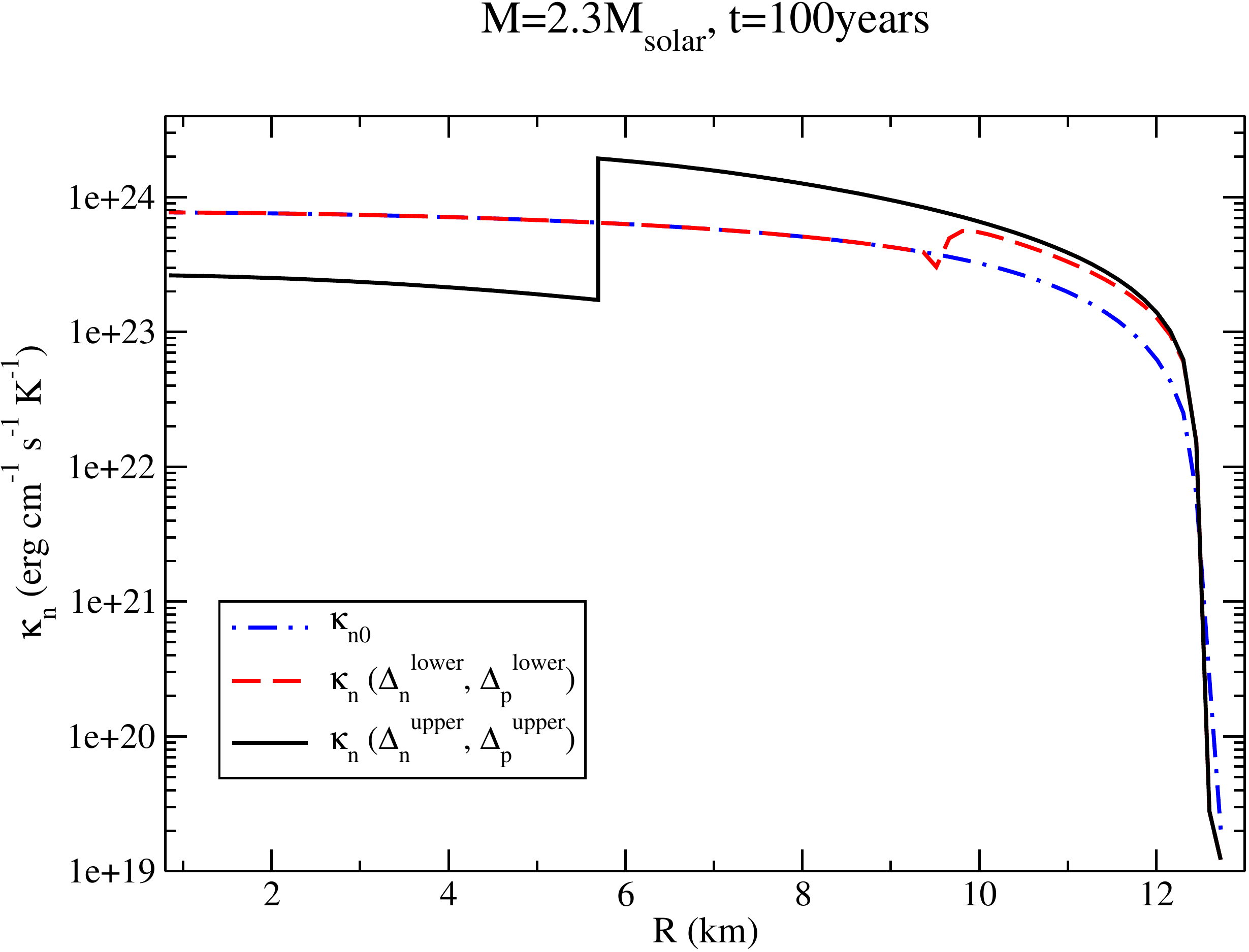}\hspace*{2mm}\includegraphics[scale=0.3]{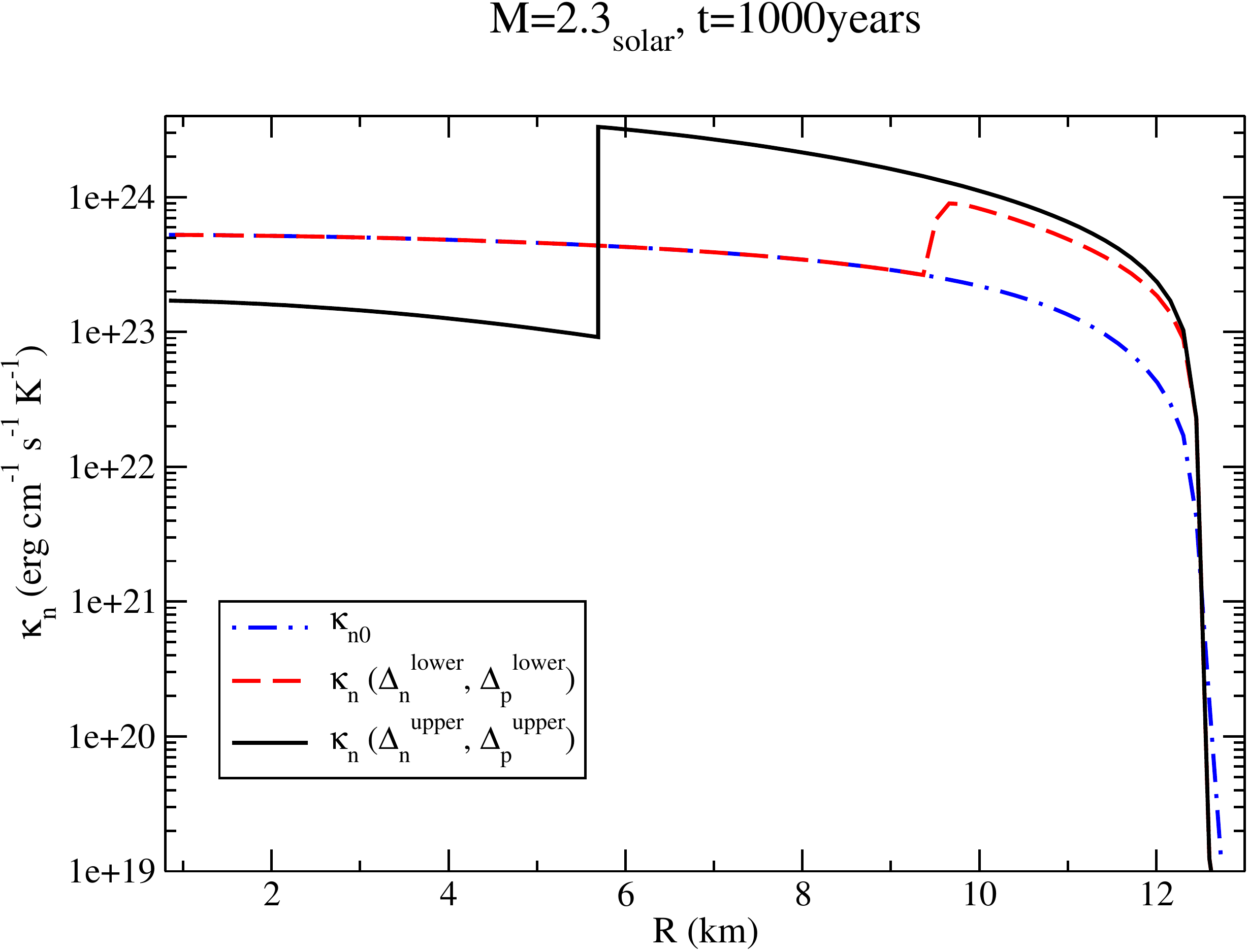}
\vspace{0.5cm}
\includegraphics[scale=0.3]{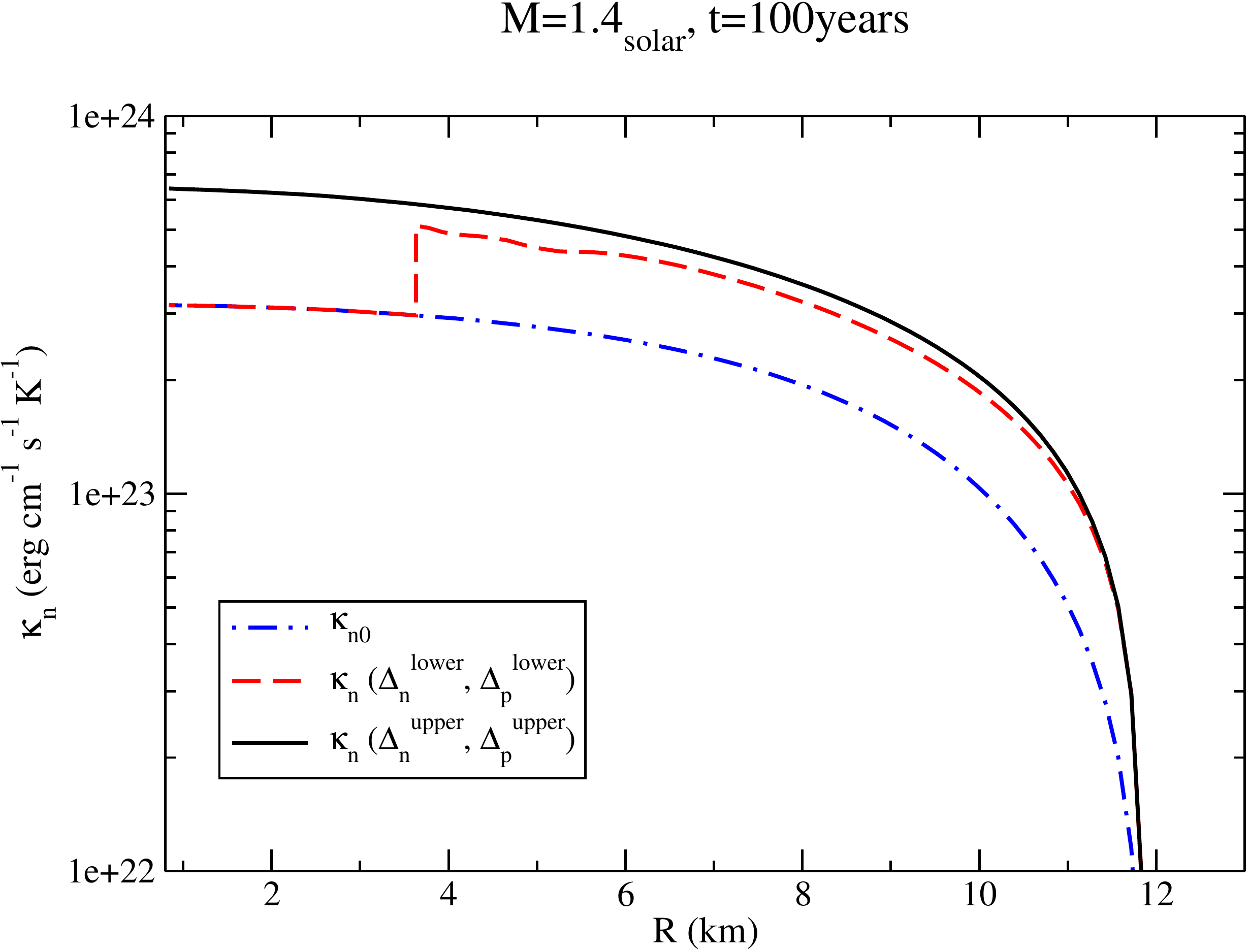}\hspace*{2mm}\includegraphics[scale=0.3]{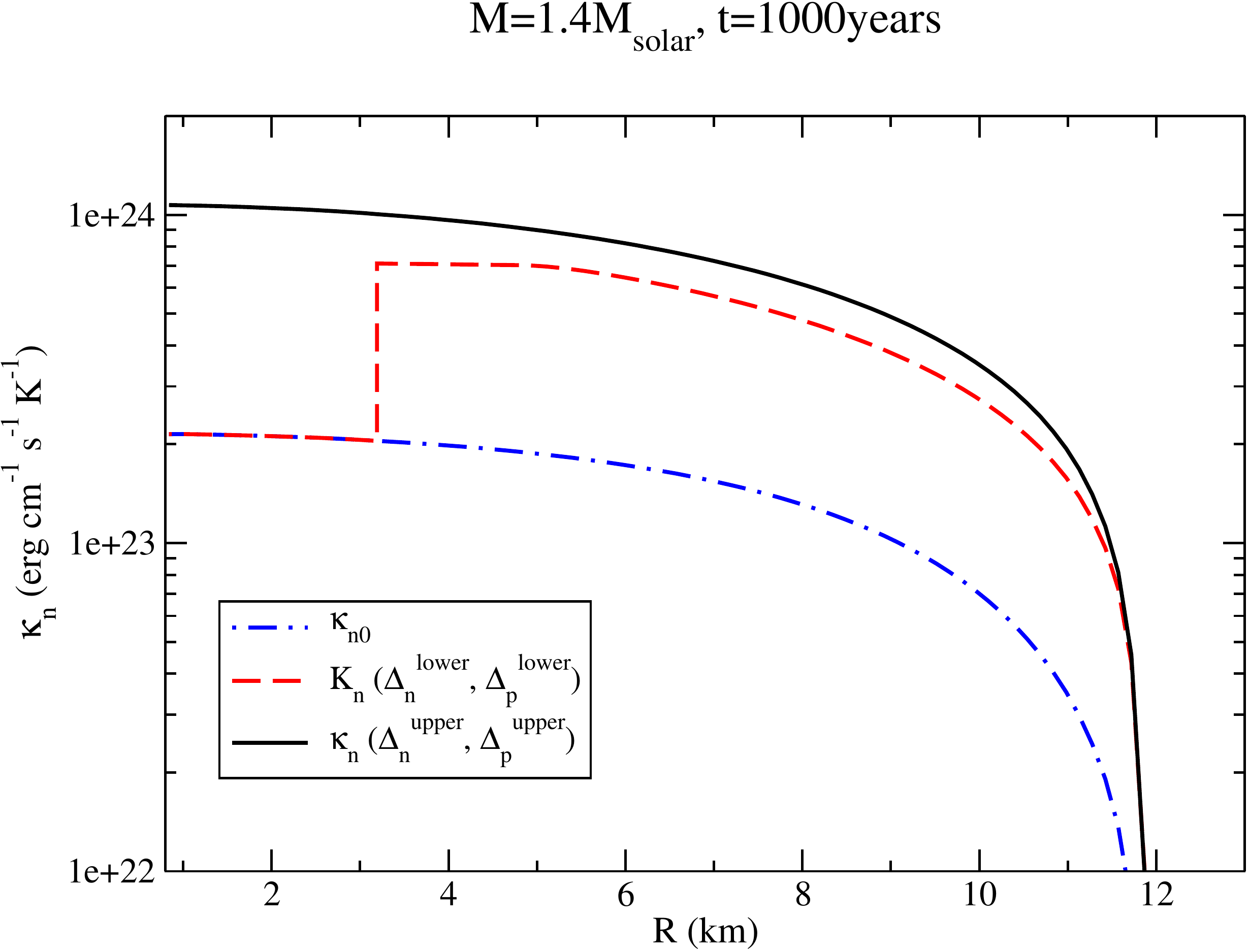}
\caption{Variation of neutron thermal conductivity with radius ($R$) in normal matter ($\kappa_{n0}$) and in presence of nucleon superfluidity/superconductivity ($\kappa_n$) for the lower and upper bounds of $\Delta_n$ and $\Delta_p$ .  
Top row: for the massive star ($M=2.3M\odot$); bottom row: for the canonical mass star ($M=1.4M_\odot$). Left panels: at $t=100$ years; right panels: at $t=1000$ years.}
\label{kn}
\end{figure*}

The scenario for neutron thermal conductivity $\kappa_n$ in absence of the magnetic field is depicted in the fig. \ref{kn}. The left and right panels of the upper row correspond to the massive star at two different times $t=100$ and $t=1000$ years respectively, while those in the lower row represent the results for the canonical mass star. 
In all the panels, the variation of $\kappa_n$ for the upper limits of $\Delta_n$ and $\Delta_p$ are presented by the solid curve, while the dashed curve represents the contribution in case of their lower bounds. The scenario of normal matter is expressed by the dot-dashed curves.

\textcolor{black}{In case of the massive star, $\kappa_n$ is very small at \textcolor{black}{$R\gtrsim12.5$ km } where the neutrons exist in $^1S_0$ pairing state.
The protons start forming $^1S_0$ pairing at $R\sim 12.5$ till $R\sim 9.4$ km ($R\sim 5.69$ km), corresponding to $\Delta_p^{lower(upper)}$. The pair formation of neutrons via $^3P_2$ channel is only possible for $\Delta_n^{upper}$ within the radial zone $R\lesssim 12.5$ km up to the inner core, while such pairing is not possible for $\Delta_n^{lower}$ within the region $12.5\lesssim R \lesssim 11$ km because of extremely low value of energy gap, as discussed in section 3.1.1. Overall, $\kappa_n$ exceeds its normal counterpart $\kappa_{n0}$,
because of the dominance of strong proton superconductivity in this region of the star.}

\textcolor{black}{Thus, at $t=100$ years in the radial zone $12.5\gtrsim R \gtrsim 9.4$ km, $\kappa_n$ varies within the range $10^{21}-6\times 10^{23}$ erg cm$^{-1}$ s$^{-1}$ K$^{-1}$. At $R\sim9.4$ km $\kappa_n$ is suddenly dropped to $3.5\times10^{23}$
erg cm$^{-1}$ s$^{-1}$ K$^{-1}$ and overlaps with $\kappa_{n0}$, increasing up to $8\times10^{23}$ erg cm$^{-1}$ s$^{-1}$ K$^{-1}$ towards the centre of the star. 
}
\textcolor{black}{In case of the upper boundaries of $\Delta_n$ and $\Delta_p$ the scenario is somewhat different. Here, at $R\sim 5.69$ km the superconducting protons become normal, while the neutrons continue to exist in $^3P_2$ pairing state till the inner core. This results in a sudden reduction in $\kappa_n$ and falls below $\kappa_{n0}$ at $R\sim 5.69$ km from $2\times10^{24}$ to $1.5\times10^{23}$ erg cm$^{-1}$ s$^{-1}$ K$^{-1}$ which increases slightly to $\sim 2.5\times10^{23}$ erg cm$^{-1}$ s$^{-1}$ K$^{-1}$ towards the deeper layers inside the star.  }

Comparing the left and right panels in the upper row of fig. 3 it is observed that $\kappa_{n0}$ is reduced slightly at $t=1000$ years in comparison with $t=100$ years. The distinction between $\kappa_n$ and $\kappa_{n0}$ at different radial regions of the star becomes greater at $t=1000$ years, as compared to $t=100$ years $i.e.$ in the region where $\kappa_n>\kappa_{n0}$, $\kappa_n$ exhibits slight enhancement with time, while in the region where $\kappa_n<\kappa_{n0}$, $\kappa_n$ is reduced with time.

\textcolor{black}{Now for the canonical mass star, at $t=100$ years $\kappa_n$ is very low ($10^{16(15)}$ erg cm$^{-1}$ s$^{-1}$ K$^{-1}$), around the surface, $R\gtrsim 11.8$ km in which the neutrons exist in the singlet pairing state .}

\textcolor{black}{Within the range $11.8\gtrsim R \gtrsim 8.48$ km, the neutrons are unable to form pairing in $^3P_2$ channel due to very small value of $\Delta_n^{lower}$, while protons are superconducting in this region which extends further till $R\sim 3.63$ km. This causes $\kappa_n$ to vary within $\sim 10^{22}-5\times 10^{23}$ erg cm$^{-1}$ s$^{-1}$ K$^{-1}$. At $R\sim 3.63$ km, as the superconducting protons are converted to the normal ones, $\kappa_n$ experiences sudden drop to $3\times 10^{23}$ erg cm$^{-1}$ s$^{-1}$ K$^{-1}$ and is superposed with $\kappa_{n0}$. For the upper bounds of $\Delta_n$ and $\Delta_p$, at $R\lesssim 11.8$ km neutron superfluidity is converted from $^1S_0$ to $^3P_2$ pairing state and both the nucleons remain in their respective pairing state till the inner core. This leads to the variation of $\kappa_n$ residing within the range $\sim 10^{22}-6.5\times10^{23}$ erg cm$^{-1}$ s$^{-1}$ K$^{-1}$, which is higher than $\kappa_{n0}$ as proton superconductivity in $^1S_0$ channel is stronger than $^3P_2$ neutron pairing.} The thermal evolution in the canonical mass star occurs in a similar fashion as the massive star.

%Now in case of the canonical mass star, at $t=100$ years $\kappa_n$ is found to be varying in the range $10^{23}-10^{24}$ erg cm$^{-1}$ s$^{-1}$ K$^{-1}$ which exceeds $\kappa_{n0}$ within the sector $12.45~\text{km}\lesssim R \lesssim 7.21$ km due to the stronger proton pairing, provided that both $\Delta_n$ and $\Delta_p$ follow their respective lower limits. At $R\gtrsim12.45$ km $\kappa_n$ is very low, because of the strong singlet pairing of the neutron and protons in this region. At $R\sim7.21$ km, $\kappa_n$ is suddenly dropped to $5\times10^{22}$ erg cm$^{-1}$ s$^{-1}$ K$^{-1}$ from $\sim 10^{24}$ erg cm$^{-1}$ s$^{-1}$ K$^{-1}$, which is caused due to the transition of the superconducting protons to normal ones. The value of $\kappa_n$ remains lower than $\kappa_{n0}$ starting from $R\sim 7.21$ km till the inner core, as the neutrons exist in $^3P_2$ pairing state nearly all throughout the star ($R\lesssim 12.45$ km). However, the scenario is completely different in case of the upper bounds of $\Delta_n$ and $\Delta_p$. Here, since the neutrons and protons remain in $^3P_2$ and $^1S_0$ pairing states respectively in the major portion of the star $i.e.$ $R\lesssim 11.5$ km, $\kappa_n$ stays higher than $\kappa_{n0}$ in this region. The thermal evolution in the canonical mass star occurs in a similar fashion as the massive star.

%\begin{widetext}

\begin{center}
\begin{table*}[h!]\label{t}
\hspace*{-0.7cm}
\begin{tabular}{|c|c|c||c|c|}
\hline
\multicolumn{5}{|c|}{$M=2.3~M_{\odot}$} \\
\hline
 Time$\rightarrow$ & \multicolumn{2}{c||}{$t=100$ yrs } & \multicolumn{2}{c|}{$t=1000$ yrs }\\
  \hline
   Radial zone$\rightarrow$ & $\textcolor{black}{12.5-9.4}$ km & $\textcolor{black}{\leq 9.4}$ km & $\textcolor{black}{12.5-9.4}$ km  & $\textcolor{black}{\leq 9.4}$ km \\
  \hline
   $\kappa_{n0}$ & \textcolor{black}{$10^{19}-3.5\times10^{23}$}  & \textcolor{black}{$(3.5-8)\times10^{23}$} & \textcolor{black}{$10^{18}-3\times10^{23}$} & \textcolor{black}{$(3-5)\times10^{23}$} 
   \\
\hline
$\kappa_n(\Delta_n^{lower}, \Delta_p^{lower}$) & \textcolor{black}{$10^{19}-6\times10^{23}$} & \textcolor{black}{$(3.5-8)\times10^{23}$} & \textcolor{black}{$10^{18}-9\times10^{23}$} & \textcolor{black}{$(3-5)\times10^{23}$}\\ 
 \hline
$\kappa_{e0}$ & \textcolor{black}{$3.5\times10^{23}-5\times10^{23}$} & \textcolor{black}{$5\times10^{23}$} & \textcolor{black}{$1.5\times10^{22}-3.5\times10^{23}$} & \textcolor{black}{$3.5\times10^{23}$}
\\
\hline
$\kappa_e$($\Delta_p^{lower})$ & \textcolor{black}{$1.5\times10^{22}-10^{23}$} & \textcolor{black}{$5\times10^{23}$} & \textcolor{black}{$(1-7)\times10^{22}$} & \textcolor{black}{$3.5\times10^{23}$}
\\
\hline \hline
 \end{tabular}
 \end{table*}
 \end{center}
\begin{table*}\label{t}
\vspace*{-0.3cm}
\hspace*{-0.6cm} \begin{tabular}
{|p{2.1cm}|p{3.2cm}|p{3.2cm}||p{3.2cm}|p{3.2cm}|} 
\hline
%Time$\rightarrow$ & $t=100$ yrs & $t=1000$ yrs & $t=100$ yrs & $t=1000$ yrs\\
Time$\rightarrow$ & \multicolumn{2}{c||}{$t=100$ yrs } & \multicolumn{2}{c|}{$t=1000$ yrs }\\
\hline
Radial zone$\rightarrow$  & \textcolor{black}{$12.5-5.69$} km & \textcolor{black}{$\leq 5.69$} km & \textcolor{black}{$12.5-5.69$} km & \textcolor{black}{$\leq 5.69$} km \\
  \hline
 $\kappa_{n0}$ & \textcolor{black}{$10^{19}-6\times10^{23}$} &  \textcolor{black}{$(6-8)\times10^{23}$}  & \textcolor{black}{$10^{18}-4\times10^{23}$} & \textcolor{black}{$(4-5)\times10^{23}$}   \\
  \hline
  $\kappa_n(\Delta_n^{upper},\Delta_p^{upper} )$ & \textcolor{black}{$10^{19}-2\times10^{24}$} & \textcolor{black}{$(1.5-2.5)\times10^{23}$} & \textcolor{black}{$10^{18}-3\times10^{24}$} & \textcolor{black}{$(1-1.8)\times10^{23}$} \\
 \hline
 $\kappa_{e0}$ & \textcolor{black}{$(3.5-5)\times10^{23}$} & \textcolor{black}{$5\times10^{23}$} & \textcolor{black}{$2\times10^{22}-3.5\times10^{23}$} & \textcolor{black}{$3.5\times10^{23}$} \\
 \hline
  $\kappa_e(\Delta_p^{upper} )$ & \textcolor{black}{$10^{22}-1.5\times10^{23}$} & \textcolor{black}{$5\times10^{23}$} & \textcolor{black}{$10^{22}-10^{23}$} & \textcolor{black}{$3.5\times10^{23}$}  \\
  \hline
\end{tabular}
\caption{Estimation of the thermal conductivity, in the unit of erg cm$^{-1}$ s$^{-1}$ K$^{-1}$, for the neutrons ($\kappa_n$) and electrons ($\kappa_e$) in the different radial zones of the massive star ($M=2.3M_{\odot}$) governed by \textcolor{black}{DDME2 EoS}, in presence of normal and superfluid/superconducting nucleons, considering the lower and upper boundaries of their pairing energies ($\Delta_n$, $\Delta_p$) at the two different ages, $t=100$ and $t=1000$ years. Different radial zones are selected accordingly as the existence of pairing of the different particle species, as mentioned in table 1 and 2.  }
\end{table*}

\begin{center}
\begin{table*}[h!]\label{t}
\hspace*{-0.7cm}
\begin{tabular}{|c|c|c||c|c|}
\hline
\multicolumn{5}{|c|}{$M=1.4~M_{\odot}$} \\
\hline
 Time$\rightarrow$ & \multicolumn{2}{c||}{$t=100$ yrs } & \multicolumn{2}{c|}{$t=1000$ yrs }\\
\hline
Radial zone$\rightarrow$ & $\textcolor{black}{11.8-3.63}$ km & $\textcolor{black}{\leq 3.63}$ km & $\textcolor{black}{11.86-3.63}$ km & $\textcolor{black}{\leq 3.63}$ \\
\hline
$\kappa_{n0}$ & \textcolor{black}{$10^{22}-3\times10^{23}$} & \textcolor{black}{$(3-3.2)\times10^{23}$} & \textcolor{black}{$10^{22}-2\times10^{23}$} & \textcolor{black}{$(2-2.2)\times10^{23}$}
\\
\hline
$\kappa_n(\Delta_n^{lower}, \Delta_p^{lower}$) & \textcolor{black}{$10^{22}-5\times10^{23}$} & \textcolor{black}{$(3-3.2)\times10^{23}$} & \textcolor{black}{$10^{22}-7\times10^{23}$} & \textcolor{black}{$(2-2.2)\times10^{23}$}  \\ 
\hline
$\kappa_{e0}$ & \textcolor{black}{$(3-4)\times10^{23}$} & \textcolor{black}{$4\times10^{23}$} & \textcolor{black}{$(2-2.5)\times10^{23}$} & \textcolor{black}{$2.5\times10^{23}$}
\\
\hline
$\kappa_e(\Delta_p^{lower})$ & \textcolor{black}{$6\times10^{21}-10^{23}$} & \textcolor{black}{$4\times10^{23}$} & \textcolor{black}{$4\times10^{21}-8\times10^{22}$} & \textcolor{black}{$2.5\times10^{23}$} 
\\
\hline \hline
 \end{tabular}
 \end{table*}
 \end{center}
 \begin{table*}\label{tt}
\vspace*{-0.3cm}
\hspace*{-0.6cm} \begin{tabular}
{|p{2.2cm}|p{2.4cm}||p{2.4cm}|} 
\hline
Time$\rightarrow$ & \multicolumn{1}{c||}{$t=100$ yrs } & \multicolumn{1}{c|}{$t=1000$ yrs }\\
\hline
Radial zone$\rightarrow$  & \textcolor{black}{$\leq 11.8$} km & \textcolor{black}{$\leq 11.8$} km \\
  \hline
 $\kappa_{n0}$ & \textcolor{black}{$10^{22}-3.2\times10^{23}$}  & \textcolor{black}{$10^{22}-2.2\times10^{23}$}   \\
  \hline
  $\kappa_n(\Delta_n^{upper},\Delta_p^{upper})$  & \textcolor{black}{$10^{22}-6.5\times10^{23}$} & \textcolor{black}{$10^{22}-1.2\times10^{24}$} \\
 \hline
$\kappa_{e0}$  & \textcolor{black}{$(3-4)\times10^{23}$} & \textcolor{black}{$(2-2.5)\times10^{23}$} \\
 \hline
$\kappa_e(\Delta_p^{upper})$  & \textcolor{black}{$6\times10^{21}-5\times10^{22}$} & \textcolor{black}{$4\times10^{21}-4.5\times10^{22}$}  \\
\hline
\end{tabular}
\caption{Same as table 5, for the canonical mass star $(M=1.4M_\odot)$}
\end{table*}

\begin{figure*}[ht]
\centering
\includegraphics[scale=0.3]{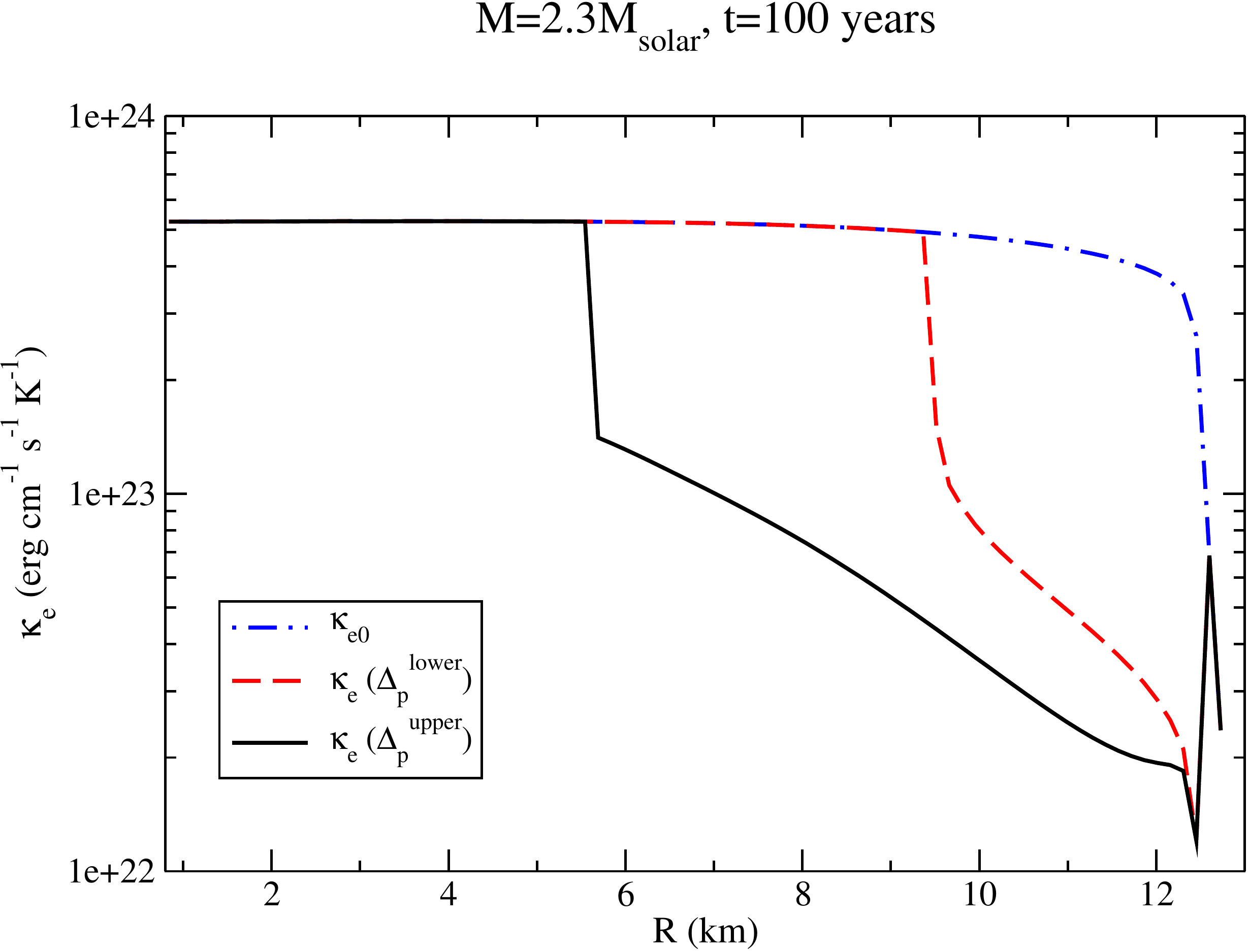}\hspace*{2mm}\includegraphics[scale=0.3]{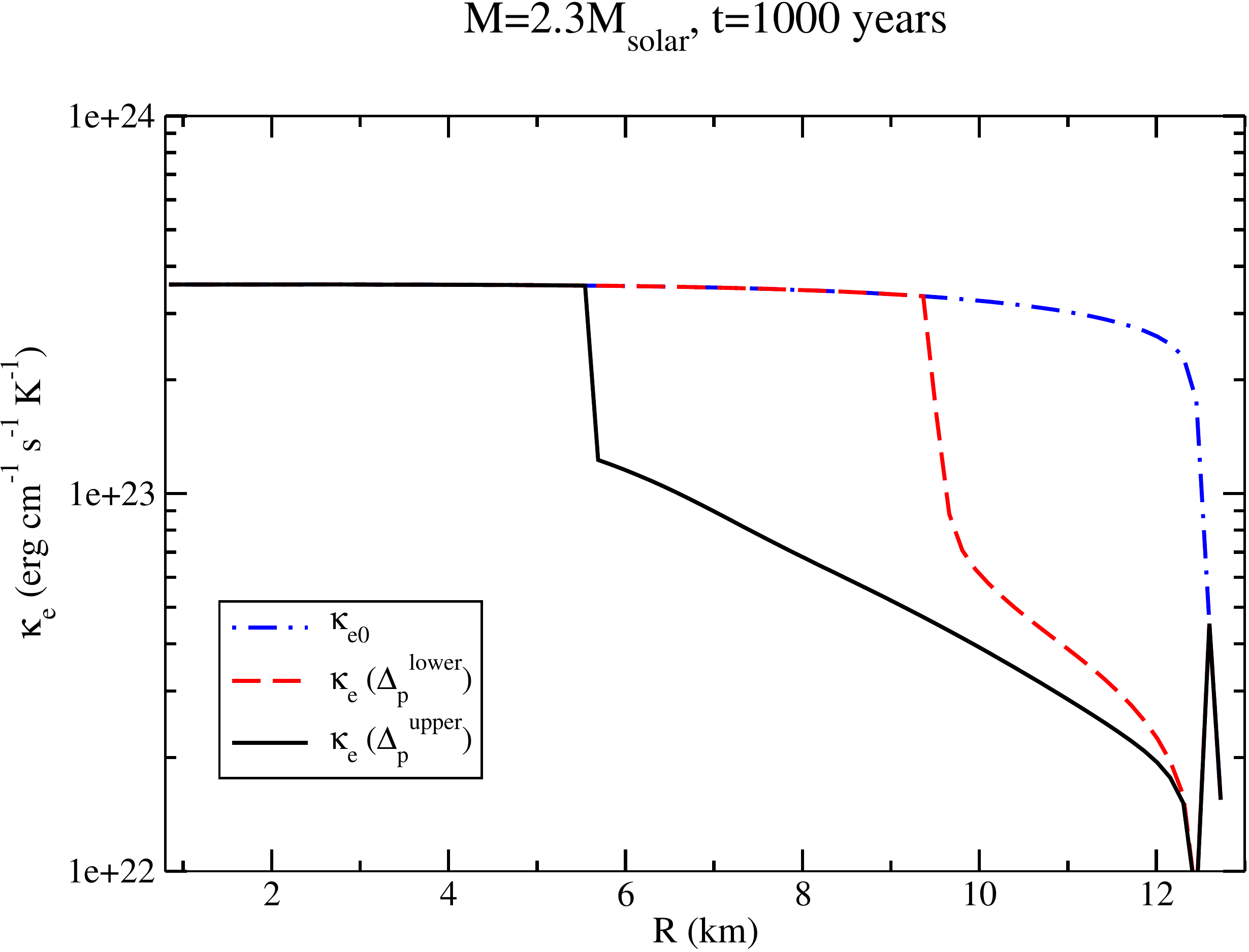}\\
\vspace{0.5cm}
\includegraphics[scale=0.3]{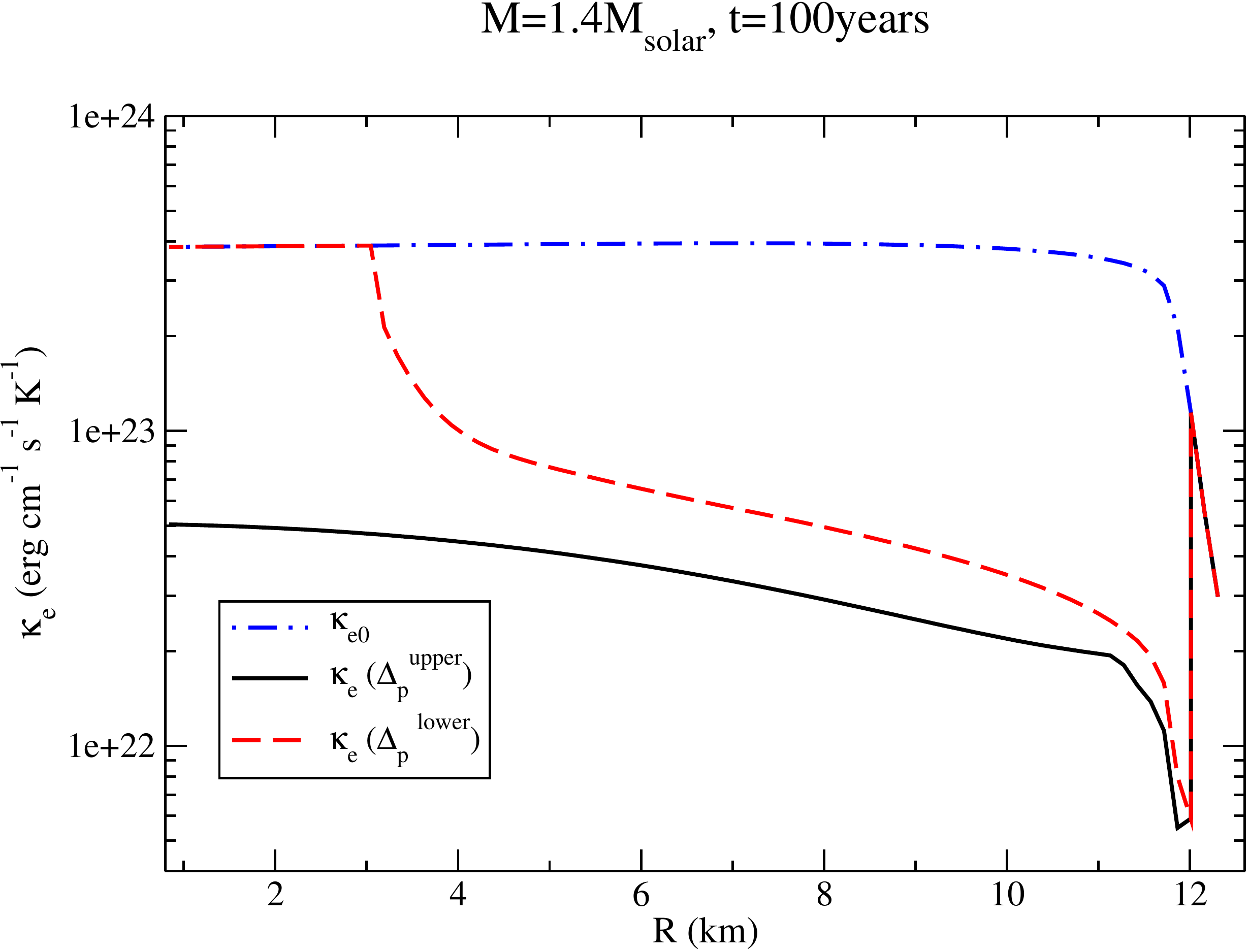}\hspace*{2mm}\includegraphics[scale=0.3]{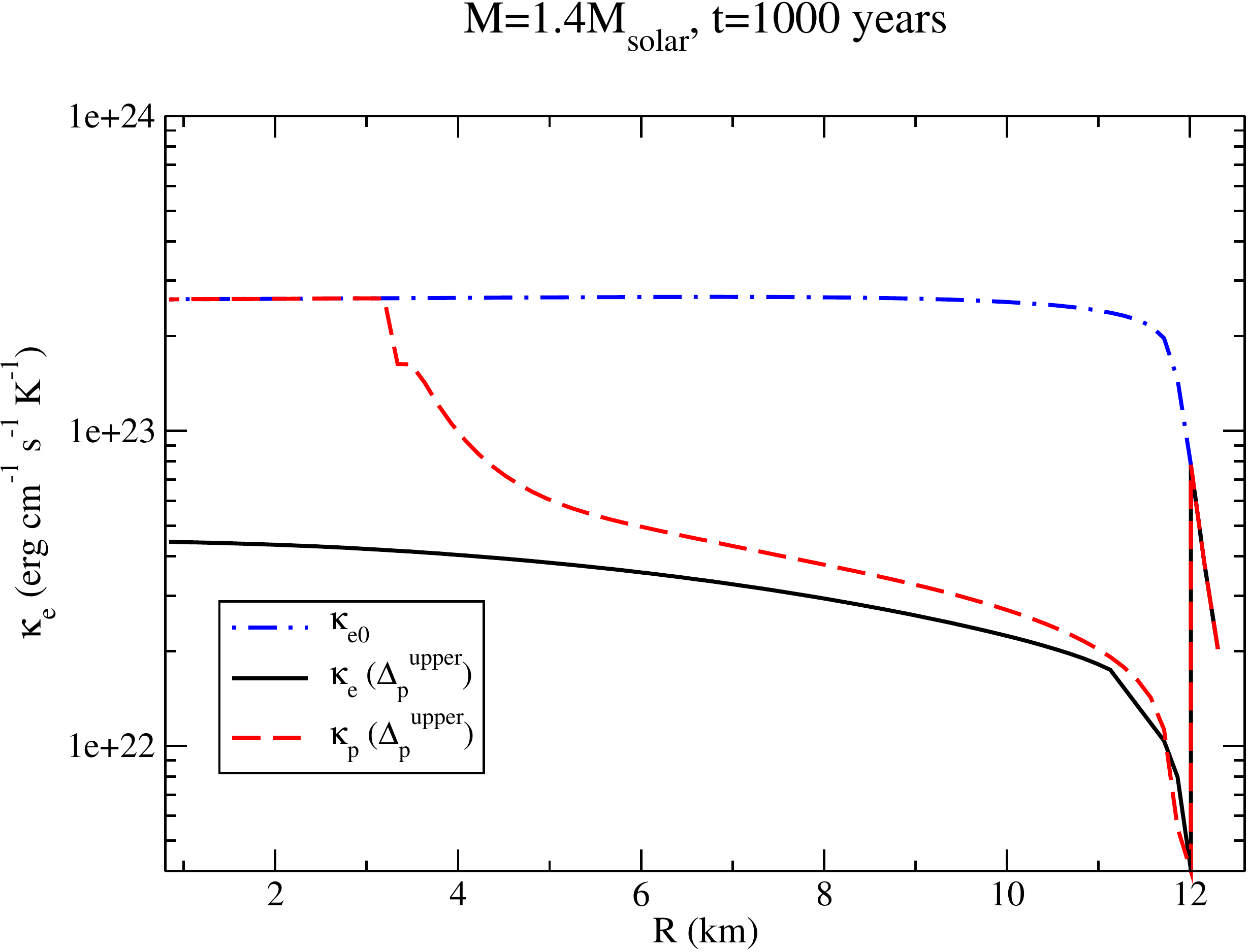}
\caption{Variation of electron thermal conductivity with radius ($R$) in normal matter ($\kappa_{e0}$) and in presence of superconducting protons ($\kappa_n$). Top left panel: for the massive star ($M\sim2.3M_\odot$) at $t=100$ years; top right panel: same as the left panel at $t=1000$ years; bottom panels: same as upper panels for the canonical mass star ($M\sim1.4M_\odot$). %The solid and dashed line correspond to superconducting protons having upper and lower bounds of $\Delta_p$ respectively, while the dot-dashed line shows the contribution for normal proton 
}
\label{lepton_kappa2.3}
\end{figure*}

\paragraph{Leptons:} When the protons become superconducting, the screening momentum $q_0$%given by \eqref{eqn18} 
is modified as, $q_0^{sf}=(4e^2/\pi)(m_e^*k_{Fe}+m_p^*k_{Fp}Z_p)$. The reduction factor $Z_p$ is multiplied with the second term due to proton superconductivity which results in overall enhancement of the screening momenta $i.e.$ $q_0^{sf}>q_0$. This is understood by the fact that when the proton pairing is formed, they are condensed to the ground state and the electric field screening mainly occurs due to the electrons. Both $ee$ and $ep$ collisions are affected by proton superconductivity. The corresponding collision frequencies are expressed as \cite{1995NuPhA.582..697G}
\begin{eqnarray}\label{23}
\nu_{ep}^{sf}=R(y_p)\frac{4\pi^{2}}{5\hbar}\left(\frac{e^{2}}{\hbar c}\right)^{2}\left(\frac{\hbar k_{F_{e}}}{q_{0}^{sf}}  \right)^{3} \frac{\mu_{e} m_{p}^{*2}k_{B}^{2} T^{2}}{p_{F_{e}}^2} \nonumber \\
\nu_{ee}^{sf}=\frac{2\pi^{2}}{\hbar} \left( \frac{e^{2}}{\hbar c} \right)^{2}\left(\frac{\hbar k_{F_{e}}}{q_{0}^{sf}}\right)^{3}\frac{k_{B}^{2}T^{2}}{\mu_{e}}
\end{eqnarray}
The additional reduction factor $R(y_p)$ comes from the modification in phase space due to proton pairing. Although both the screening momenta $q_0^{sf}$ and $R(y_p)$ affect the $ep$ collision, it is observed that the effect of $R(y_p)$ is more dominating which ultimately causes reduction in $\nu_{ep}$. On the other hand, as $q_0^{sf}>q_0$, $\nu_{ee}^{sf}$ becomes lower than $\nu_{ee}$. In presence of proton pairing the relaxation time is modified and is expressed as, $\tau^{sf}_e=1/(\nu_{ee}^{sf}+\nu_{ep}^{sf})$. %When protons are superconducting,
Since the contribution of $\nu_{ep}^{sf}$ becomes almost negligible and $\nu_{ee}^{sf}$ exceeds
$\nu_{ee}$, $\tau^{sf}_e<\tau_e$. Therefore, following \eqref{eqn7}, the thermal conductivity of the electrons undergoes depletion eventually, in presence of the proton superconductivity. 
Electrons remain normal all throughout the star. As the $en$ interaction is not taken into consideration, neutron superfluidity does not affect the electron thermal conductivity $\kappa_e$. 

\begin{figure*}[ht]
\centering
\includegraphics[scale=0.3]{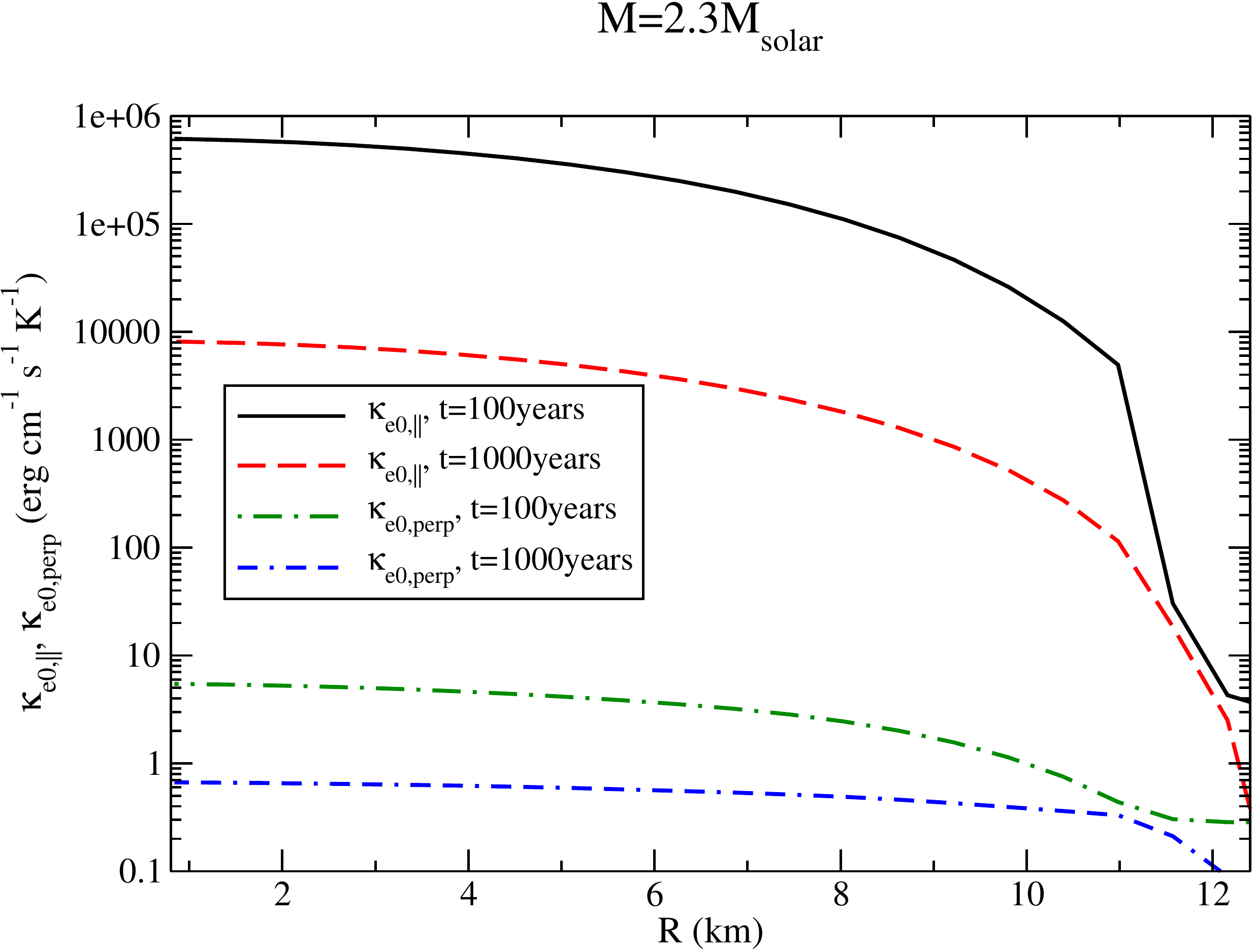}\hspace*{1cm}\includegraphics[scale=0.3]{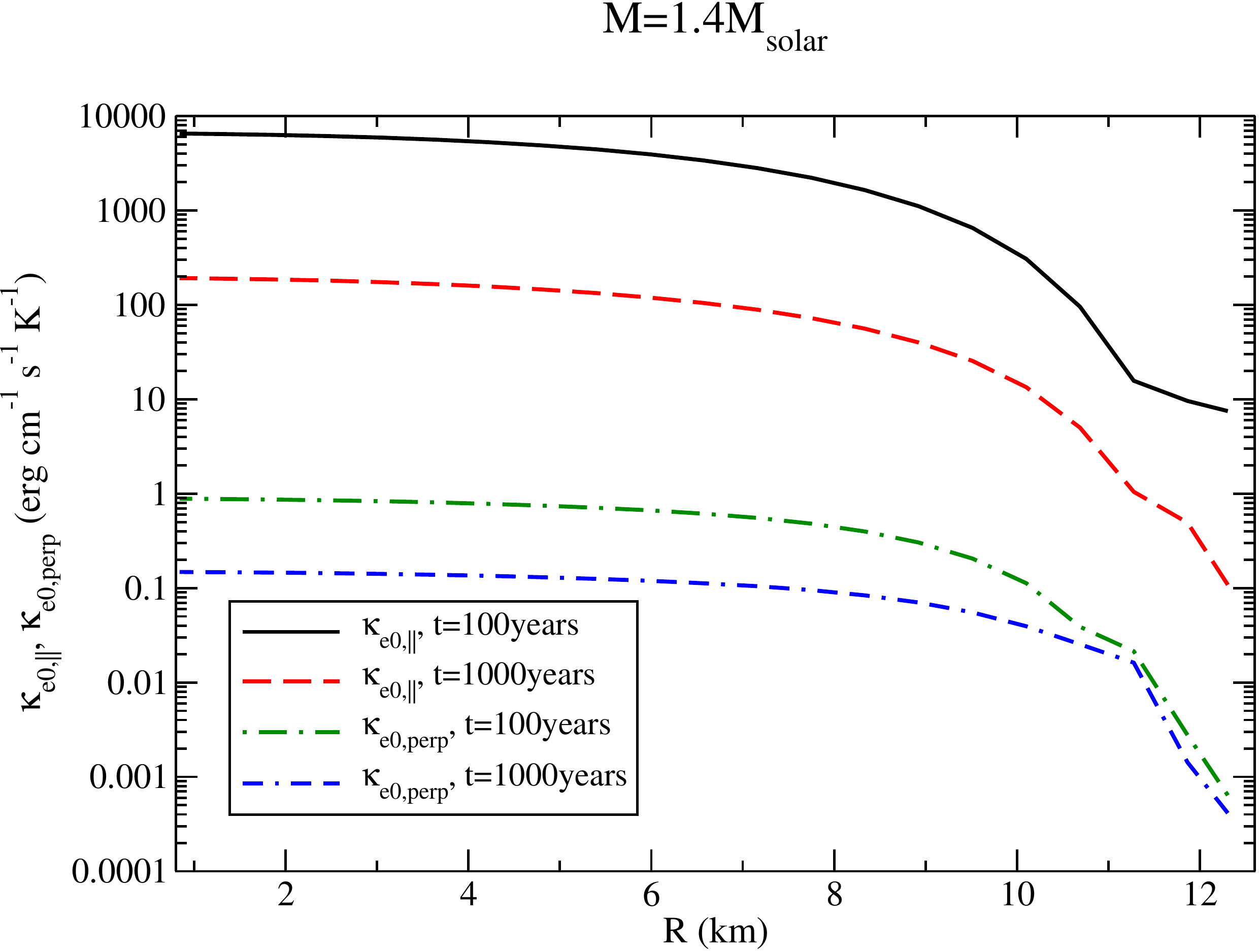}
\caption{Left panel: Variation of parallel ($||$) and perpendicular ($\perp$) components of the electron thermal conductivity with radius ($R$) in presence of magnetic field with universal profile for the star of mass $M=2.3M_\odot$, in the normal matter; right panel: Same for the star of mass $M=1.4M_\odot$.
}
\label{ke_mag}
\end{figure*}

The variation of $\kappa_e$ with radius is described in fig. \ref{lepton_kappa2.3}, similar to fig. 3, the upper and lower row corresponding to the massive and canonical mass stars, with the left and right panel exhibiting the results at two different times, $t=100$ and $t=1000$ years respectively. 

\textcolor{black}{As mentioned earlier, for the massive star the protons are normal within a very narrow region in the vicinity of the surface ($R>12.5$ km) and $\kappa_e$ remains in the range $\sim (2.5-7)\times 10^{22}$ erg cm$^{-1}$ s$^{-1}$ K$^{-1}$. The normal protons shift to $^1S_0$ superconducting state in the radial zone $12.5\gtrsim R \gtrsim 9.4$ km ($12.5 \gtrsim R \gtrsim 5.69$ km) for $\Delta_p^{\text{lower(upper)}}$ which causes $\kappa_e$ to be decreased suddenly at $R=12.5$ km from $\sim 7\times 10^{22}$ to $\sim 1.5\times 10^{22}$ erg cm$^{-1}$ s$^{-1}$ K$^{-1}$ at $t=100$ years and then increases gradually further towards the inner core of the star and reach at $\sim 10^{23}$ ($\sim 1.5\times 10^{23}$) erg cm$^{-1}$ s$^{-1}$ K$^{-1}$ at $R\sim 9.36$ km ($R\sim 5.69$ km). The proton superconductivity is absent beyond $R\sim 9.4$ km ($R\sim 5.69$ km) corresponding to $\Delta_p^{lower(upper)}$, for which $\kappa_e$ is drastically increased from $10^{23} (1.5\times10^{23})$ erg cm$^{-1}$ s$^{-1}$ K$^{-1}$, to overlap with $\kappa_{e0}$ and remains nearly constant at a value $\sim 5\times 10^{23}$ erg cm$^{-1}$ s$^{-1}$ K$^{-1}$ till the inner core.   }

\textcolor{black}{In case of the canonical mass star, the protons reside in $^1S_0$ pairing state within the range $11.8\gtrsim R \gtrsim3.63$ km, corresponding to $\Delta_p^{lower}$, while they remain superconducting approximately all through the entire star $i.e.$ in the region $R\lesssim 11.8$ km for $\Delta_p^{upper}$.
In the narrow radial zone around the surface $R\gtrsim 11.8$ km, for $t=100$ years the value of $\kappa_e$ dwells within $\sim 3\times10^{22}-10^{23}$ erg cm$^{-1}$ s$^{-1}$ K$^{-1}$. At $R\sim 11.8$ km $\kappa_e$ declines instantly from $\sim 10^{23}$ to $\sim 6\times 10^{21}$ erg cm$^{-1}$ s$^{-1}$ K$^{-1}$. Beyond that in case of $\Delta_p^{\text{lower}}$ , in the region $11.8 \gtrsim R \gtrsim 3.63$ km, $\kappa_e$ is confined within the range $\sim 6\times 10^{21}-10^{23}$, rising instantly to $\sim 4\times 10^{23}$ erg cm$^{-1}$ s$^{-1}$ K$^{-1}$ at $R\sim 3.63$ km, as at $R\lesssim 3.63$ km the $\Delta_p^{lower}$ becomes zero. $\kappa_e$ remains the same as $\kappa_{e0}$, nearly constant at $4\times 10^{23}$ erg cm$^{-1}$ s$^{-1}$ K$^{-1}$ till the inner core. However, in case of $\Delta_p^{\text{upper}}$ the protons exist in the singlet pairing state till the inner core, starting from $R\sim 11.8$ km. In this scenario $\kappa_e$ varies within the range $\sim 6\times 10^{21}-5\times 10^{22}$ erg cm$^{-1}$ s$^{-1}$ K$^{-1}$ which is lower than $\kappa_{e0}$ by an order approximately. At the latter time $t=1000$ years both $\kappa_e$ and $\kappa_{e0}$ are observed to be reduced a little as expected similar to the previous cases.}

%In case of the canonical mass star, the protons reside in $^1S_0$ pairing state within the range $12.45~ \text{km} \lesssim R \lesssim 7.21$ km corresponding to $\Delta_p^{lower}$, while they remain superconducting approximately all through the entire star $i.e.$ in the region $R\lesssim 13.03$ km, for $\Delta_p^{upper}$. In the first case, at $t=100$ years the value of $\kappa_e$ dwells within $\sim 1.5\times10^{22}-6\times10^{23}$ erg cm$^{-1}$ s$^{-1}$ K$^{-1}$ which is smaller than the value of $\kappa_{e0}$ in this region $i.e.$ $\sim 1.5\times10^{22}-6\times10^{23}$ erg cm$^{-1}$ s$^{-1}$ K$^{-1}$. At $R\sim7.21$ km, $\kappa_e$ rises drastically to $\sim 4\times 10^{24}$ erg cm$^{-1}$ s$^{-1}$ K$^{-1}$, concuring with $\kappa_{e0}$. Also, in the $R\gtrsim12.45$ km, as the protons are normal, there is a sudden jump in $\kappa_e$ from $\sim1.5\times10^{22}$ to $\sim 5\times10^{23}$ $i.e.$ $\kappa_e$ coincides with $\kappa_{e0}$. At the same time, in case of $\Delta_p^{upper}$, $\kappa_e$ remains within the range $\sim 10^{23}-10^{24}$ erg cm$^{-1}$ s$^{-1}$ K$^{-1}$ in the region $R\lesssim13.03$ km,. In the region $R\gtrsim13.03$ km, as protons are no superconducting, $\kappa_e$ overlaps with $\kappa_{e0}$. With time both $\kappa_e$ and $\kappa_{e0}$ are slightly reduced for both the stars, although their order of magnitude do not change. The variation of $\kappa_e$ and $\kappa_{e0}$ at two different times are displayed in table 5 and 6 for the massive and the canonical mass star respectively, alongwith $\kappa_n$ and $\kappa_{n0}$. 

\section{Thermal properties in presence of magnetic field}
\subsection{In normal matter}\label{thermal_mag}

Till now the effect of magnetic field was not taken into account in our analysis. Since the magnetic field is associated with the motion of charged particles, its effect is discernible only in case of the electron thermal conductivity. It is to be noted that the magnetic field affects the electrons much more than the protons as the electron mass is much smaller than the mass of the proton. In presence of magnetic field the 
motion of the electrons along the field direction remains the same, while along the direction perpendicular to the magnetic field it is highly suppressed. As a consequence, the thermal conductivity of electrons becomes anisotropic. In this context it is to be mentioned that although the neutrons are affected by the magnetic field due to its non-zero spin, the trajectory of its motion remains unchanged in presence of the magnetic field. This implies that unlike the electrons, the thermal conductivity of neutrons are not decomposed into distinct components along the different directions.  \textcolor{black}{The heat capacity, given by eqn. (2) also remains unaffected in normal matter, since it is not associated with the dynamical motion of the particles.} However, in presence of neutron superfluidity and superconductivity, both the specific heat and neutron thermal conductivity are modified, along with the thermal conductivity of the electrons, which are discussed in the next section.  

The expression of the thermal conductivity tensor of the electrons in presence of magnetic field is given by \citep{1999A&A...351..787P}
\begin{equation}\label{eqn24}
\kappa_{e0,ij}=\int \frac{(\mu-\epsilon)^2}{T}\frac{\mathcal{N}_B(\epsilon)}{\epsilon}\tau_{ij}(\epsilon)( -\frac{\partial f_0}{\partial \epsilon} ) d \epsilon.
\end{equation}
Here $\mu$ and $\epsilon$ represent the chemical potential and energy of the integral respectively, while $T$ is the temperature. $f_0$ is the equilibrium Fermi-Dirac distribution function, given by eqn (7) as $f_p$ for the protons. \textcolor{black}{$\mathcal{N}_B$ is the baryon number density in presence of the magnetic field}.

\textcolor{black}{In this context it is to be mentioned that in presence of the magnetic field the electrons are distributed into several Landau levels along the direction transverse to the magnetic field. However, in our analysis, the number of Landau levels are quite large, even in the presence of the strong magnetic field $B_c=10^{16}$ G. Therefore, the effect of magnetic field quantization becomes negligible and the system can be treated classically, although the transverse motion of the electron is affected \cite{1999A&A...351..787P}.}

%For sufficiently high temperature, however, the field can be treated as non-quantizing (classical). The Electron conduction in magnetized neutron star envelopes non-quantizing magnetic field does not affect thermodynamic properties of matter, but hampers transverse transport and causes the Hall currents. In a weakly quantizing field, where electrons populate several Landau levels, the thermodynamic functions and kinetic coefficients oscillate with increasing density around their classical values.Strongly quantizing magnetic field confines most electrons to the ground Landau level. In this case, thermodynamic and kinetic properties of matter are very different from those in the classical regime.
In eqn. (24) $\tau_{ij}$ is the relaxation time corresponding to the electron scattering in the magnetic field. The subscript '$0$' implies the case of normal matter.
Due to the symmetry property of $\kappa_{ij}$, $\tau_{xx}=\tau_{yy}$ which are the transverse components of the electron conductivity and $\tau_{zz}$ is its longitudinal counterpart, provided that the magnetic field is pointed along the $z$-direction. From now onwards, we denote $\tau_{zz}$ and $\tau_{xx}$ or $\tau_{yy}$ as $\tau_{||}$ and $\tau_{\perp}$ respectively.
To determine the components of the relaxation time, we follow the formalism described in the ref. \citep{1999A&A...351..787P} in which the longitudinal and transverse components of the relaxation time are expressed as
\begin{eqnarray}\label{eqn25}
\tau_{||}=\tau_0, \\ \notag
\tau_{\perp}=\frac{\tau_0}{1+(\omega_g \tau_0)^2}. \\ \notag
\end{eqnarray} 
Here, $\omega_g=eB/\epsilon$ which is known as the gyrofrequency of electron. $\tau_0$ is the relaxation time in absence of magnetic field. Here the nucleon superfluidity/superconductivity effects are not taken into account.

In fig. \ref{ke_mag}, the two components of $\kappa_{e0}$ have been illustrated at two different times, $t=100$ years and $1000$ years considering the universal profile of magnetic field with the central magnetic field $B_c=10^{16}$ G. The left and right panel correspond to the massive and canonical mass star respectively. In both the panels, the solid and the dashed curve correspond to the parallel component $\kappa_{e0,||}$ at $t=100$ and $t=1000$ years respectively. The dot-dashed and dot double-dashed line represent the same for the perpendicular component $\kappa_{e0,\perp}$. 

It can be observed from the left panel, \textcolor{black}{in case of the massive star both $\kappa_{e0,||}$ and $\kappa_{e0,\perp}$ varies slowly throughout the star, falling rapidly near the surface of the star. The order of $\kappa_{e0,\perp}$ is much smaller compared to $\kappa_{e0,||}$ in case of both the stars. At $t=100$ yrs, in case of the massive star $\kappa_{e0,||}$ varies within $\sim 4-7\times10^{5}$ erg cm$^{-1}$ s$^{-1}$ K$^{-1}$ starting from its surface to the centre, while at the latter time $t=1000$ yrs, its order is reduced even further residing within the range, $1-10^{4}$ erg cm$^{-1}$ s$^{-1}$ K$^{-1}$. The order of $\kappa_{e,\perp}$ is nearly $10$ at the centre of the star at $t=100$ years which decreases to a much smaller value towards the surface. Its order is reduced even further at the latter time, $t=1000$ years. It is to be mentioned that in absence of the magnetic field in normal matter, $\kappa_{e0}$ remains nearly constant throughout the core and its order is much higher ($\sim 10^{23}$), which is observed to be diminished remarkably in presence of the magnetic field.}

\textcolor{black}{In case of the canonical mass star, the order of $\kappa_{e0,||}$ and $\kappa_{e0,\perp}$ are lower by at least two orders, as compared to the massive star. At $t=100$ years, $\kappa_{e,||}$ acquires a value of $\sim 7\times10^3$ erg cm$^{-1}$ s$^{-1}$ K$^{-1}$ at the inner core which is decreases towards the surface at $t=1000$ years. $\kappa_{e,\perp}$ is much lower as compared to $\kappa_{e,||}$, similar to the case of the massive star. %Unlike the case of the massive star, here both the components of thermal conductivity reduce at a faster rate towards the surface. 
In both the stars, the order of both $\kappa_{e0,||}$ and $\kappa_{e0,\perp}$ are reduced by one or two orders at the latter time.}

\begin{figure*}[ht]
\centering
\includegraphics[scale=0.3]{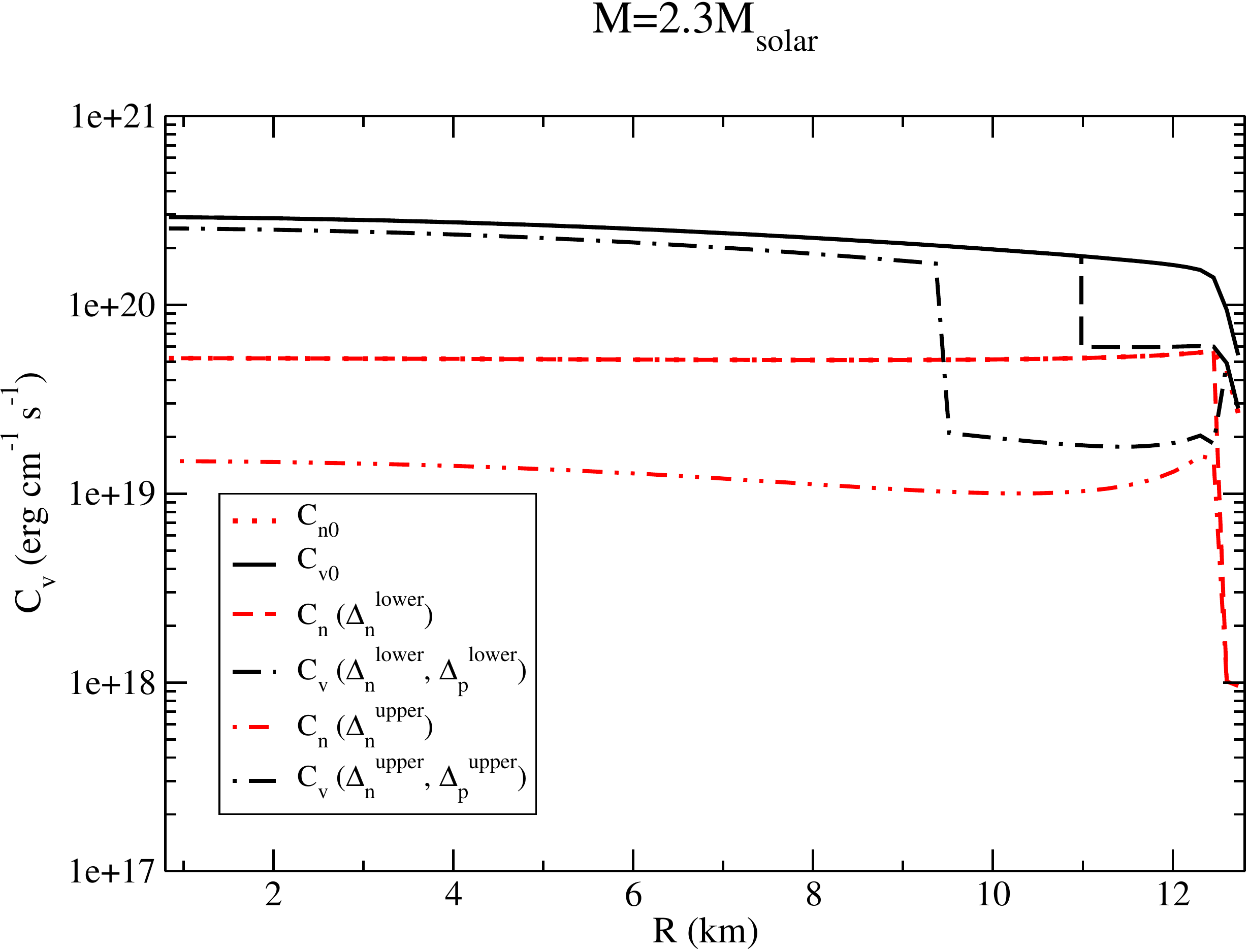}\hspace*{1cm}\includegraphics[scale=0.3]{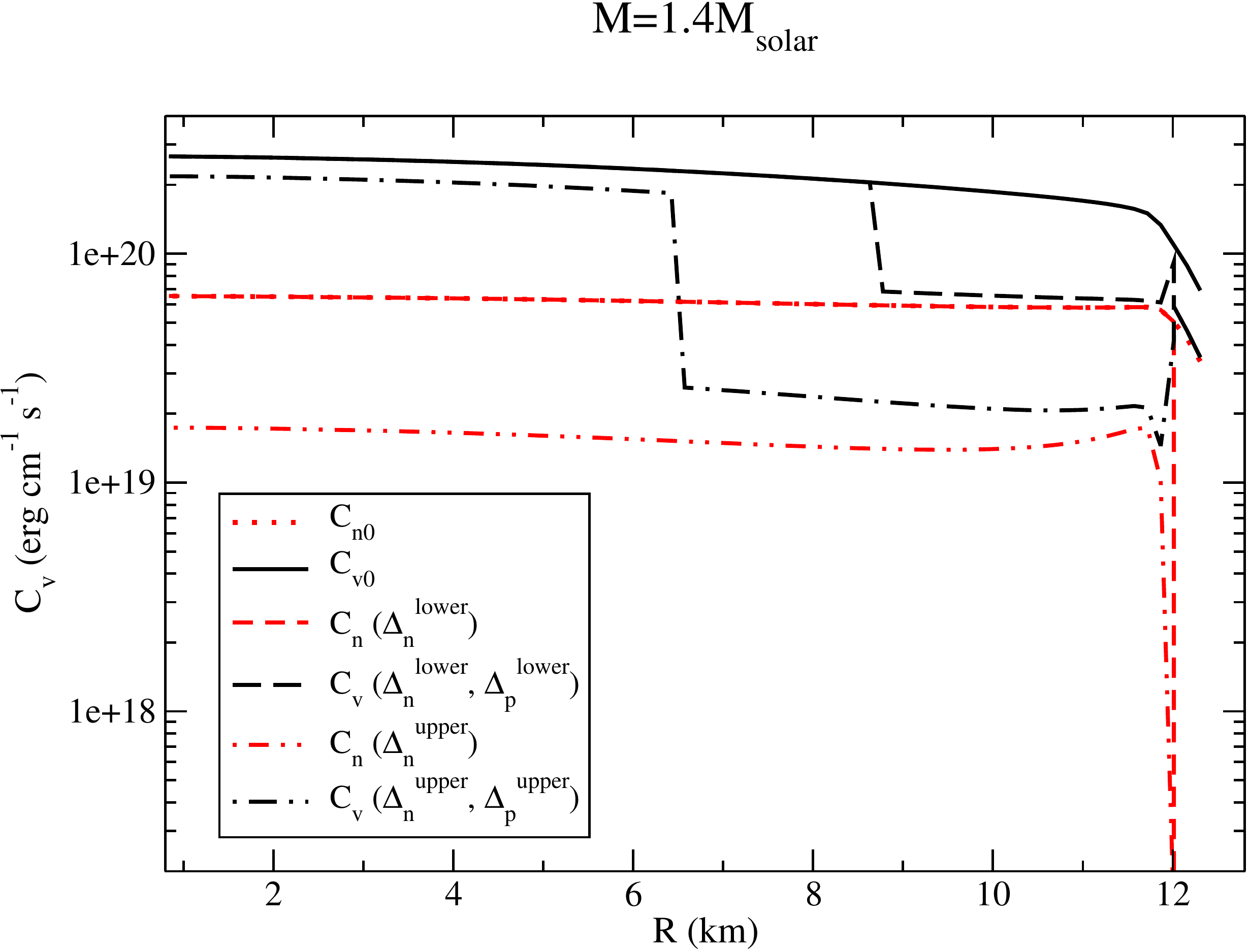}
\caption{Variation of $C_v$ inside the star in the presence of magnetic field following the universal profile ($B_{uni}$) at $t = 100 years$ for upper and lower delta. Left panel: For star of mass $M\sim1.4M_\odot$. Right panel: For $M\sim2.3M_\odot$. Solid lines and dotted lines corresponds to total heat capacity and partial heat capacity of neutrons with superfluidity respectively. Long dash-and small dash curves are for lower delta while long dash - dotted and dash - double dotted are for upper delta curves.
}
\label{B_1.4_2.3}
\end{figure*}

\begin{figure*}[ht]
\hspace*{-0.2cm}\includegraphics[scale=0.3]{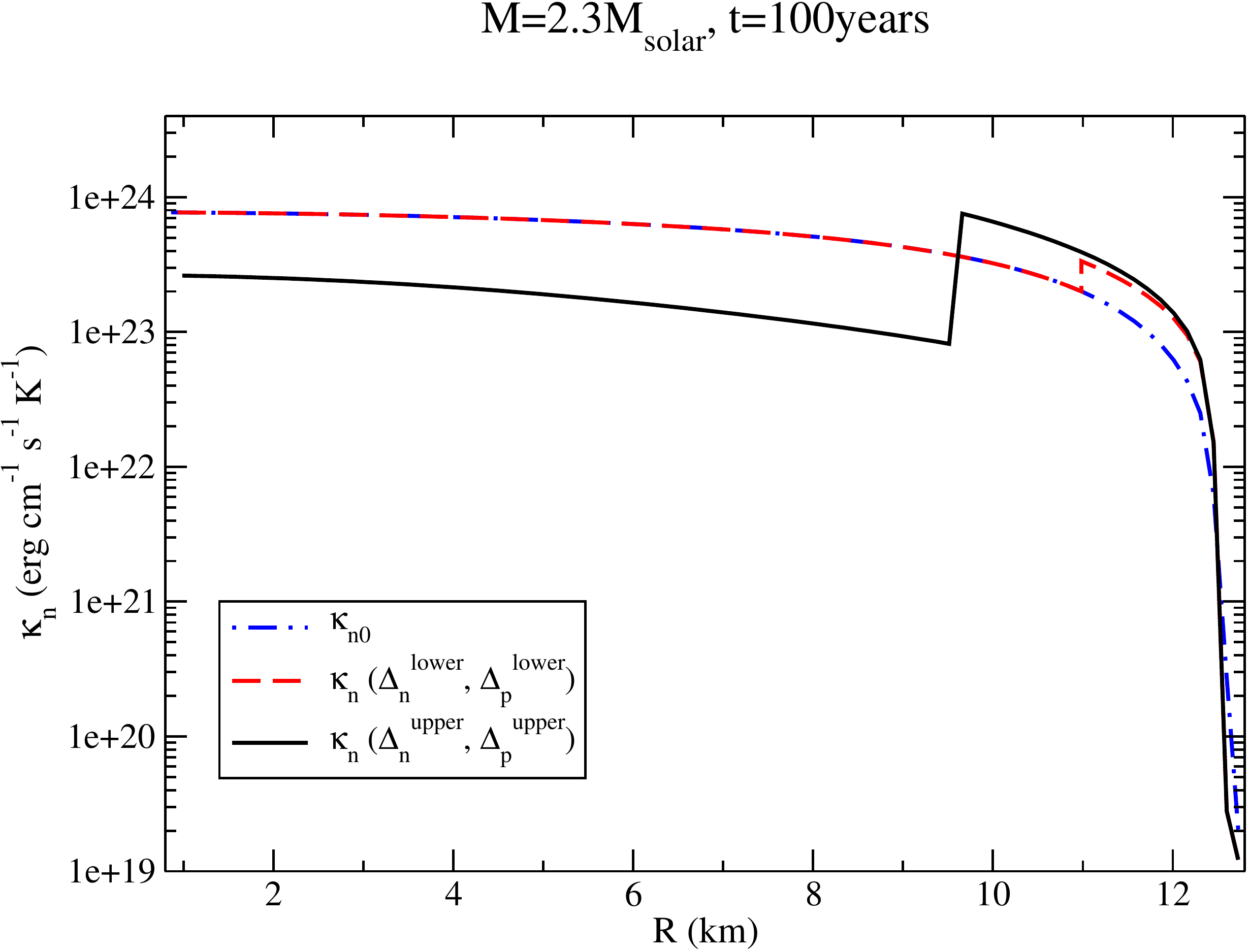}\hspace*{1cm}\includegraphics[scale=0.3]{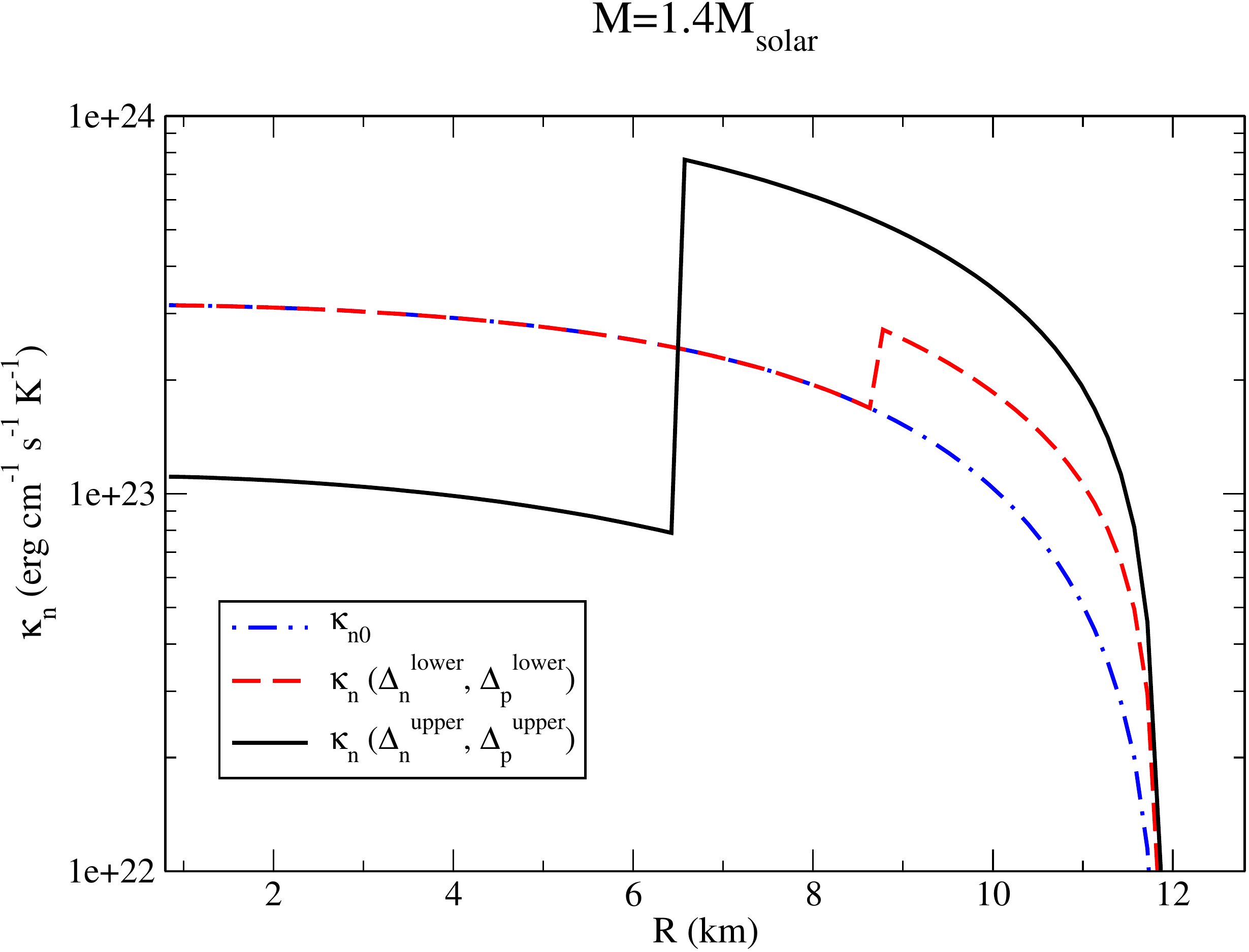}
\caption{Left panel: Variation of $\kappa_n$ in presence of magnetic field following the universal profile ($B_{uni}$) for the upper and lower bounds of $\Delta_n$ and $\Delta_p$ at $t=100$ years for the massive star ($M\sim2.3M_\odot$); right panel: Same for the low mass star ($M\sim1.4M_\odot$). The solid and the dashed line correspond to the nucleon pairing with pairing energy gap functions, $\Delta_n$ and $\Delta_p$ following upper and lower bounds respectively. The dot-dashed line is related to the case of non-superfluid/superconducting matter.}
\label{kn_mag}
\end{figure*}

\subsection{In superfluid/superconducting matter}\label{mag_sf}

\begin{figure*}[ht]
    \centering
    \includegraphics[scale=0.3]{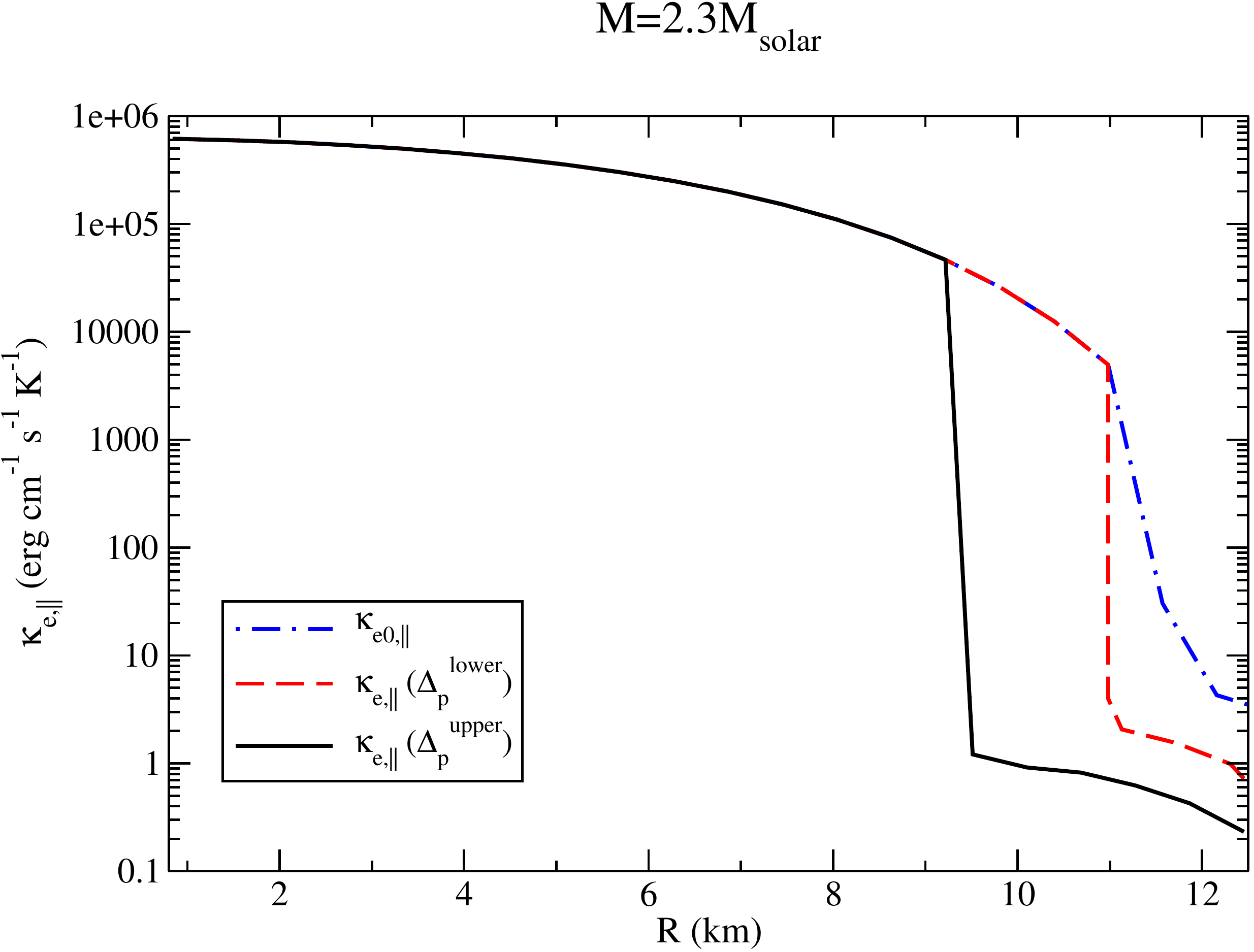}\hspace*{1mm} \includegraphics[scale=0.3]{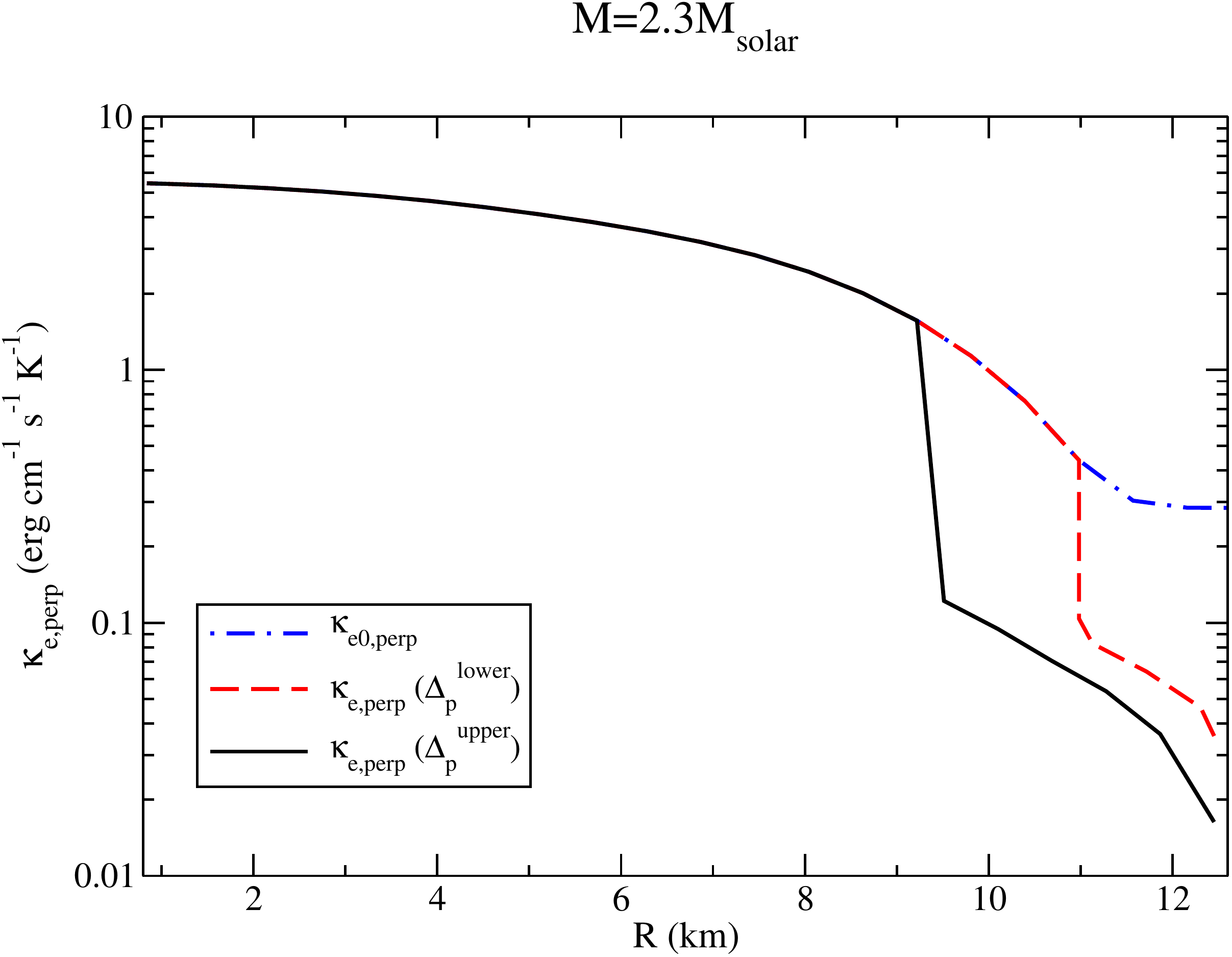}
    \caption{Left panel : Variation of parallel ($\kappa_{e,||}$) of electron thermal conductivity with radial distance inside the massive star ($M=2.3M_{\odot}$); right panel: Variation of parallel ($\kappa_{e,\perp}$) of electron thermal conductivity with radial distance inside the massive star}
    \label{mag_sf_2.3}
\end{figure*}

\begin{figure*}[ht]
    \centering
    \includegraphics[scale=0.3]{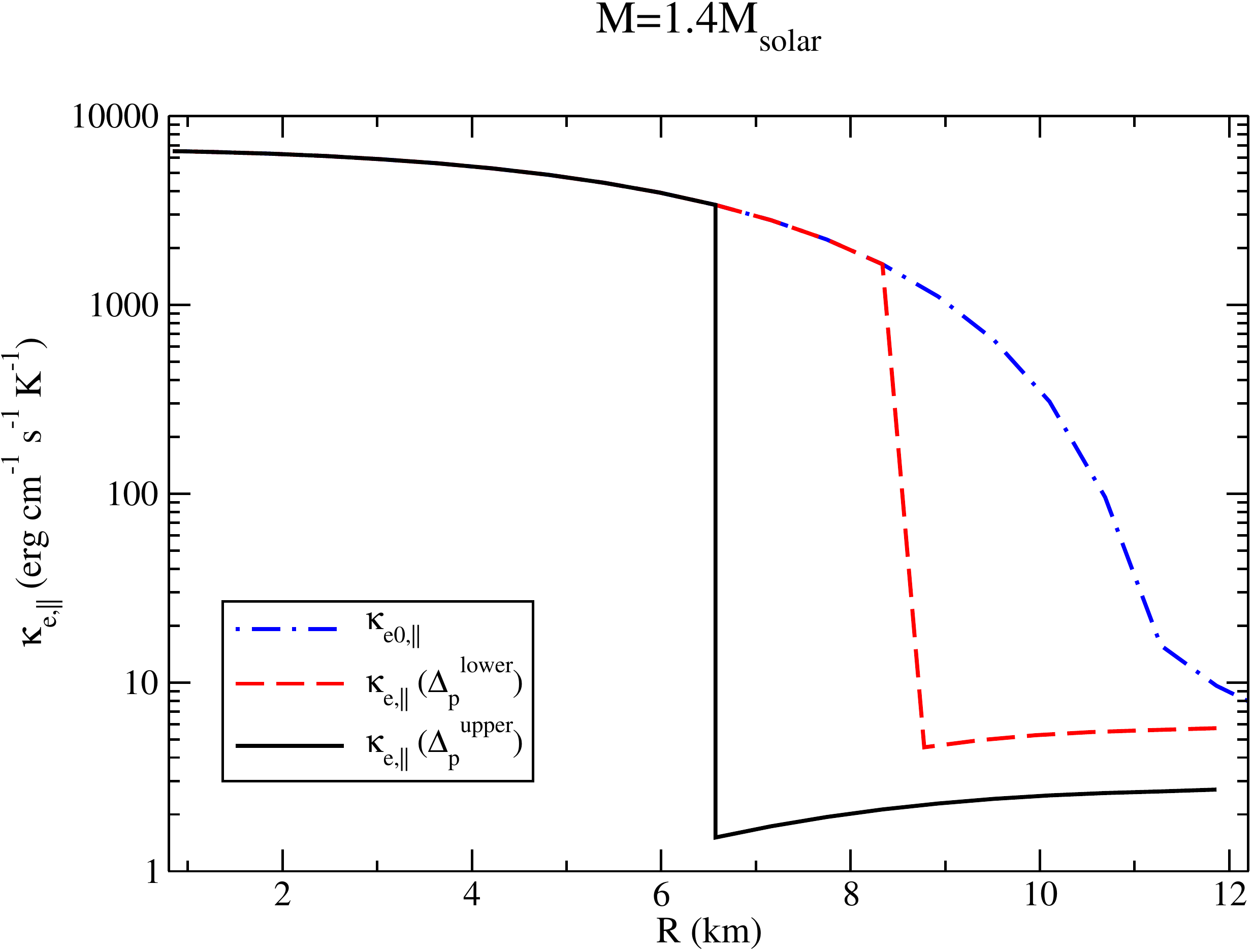}\hspace*{1mm} \includegraphics[scale=0.3]{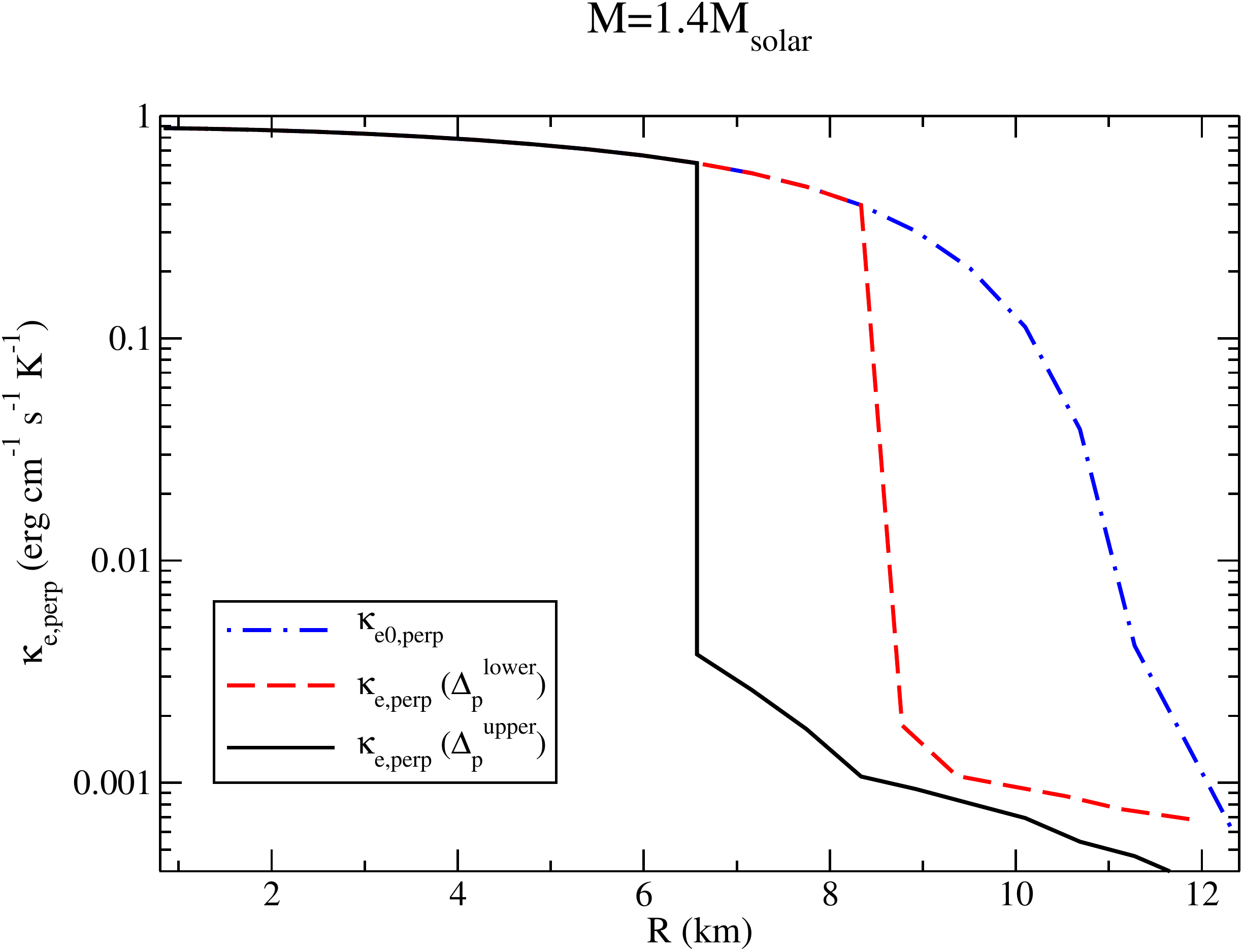}
    \caption{Same as fig. 8, for the canonical mass star ($M=1.4M_{\odot}$)}
    \label{mag_sf_1.4}
\end{figure*}

In this section the effect of magnetic field on the thermal properties is analysed considering the stellar matter to be consisted of superfluid neutrons and superconducting protons. 
The large magnetic field of the magnetar suppresses the proton superconductivity which causes the proton pairing to remain active in a comparatively narrower radial section towards the surface \citep{2015PhRvC..91c5805S}. In this region the frequency of $ep$ collision is also expected to be suppressed.
On the other hand, in case of the neutrons, the effect of magnetic field on superfluidity is still unknown \citep{2018ASSL..457..401H}. Therefore, in this work, we focus only on the effect of proton superconductivity in presence of magnetic field and analyze the thermal properties accordingly.
We present our results only for the universal profile of magnetic field, as the results do not differ significantly in the two kinds of field profiles. For example, as observed from the table 1 and 2, in the massive star the active region of proton superconductivity is suppressed to the radial range \textcolor{black}{$\sim12.5-10.85$ km ($\sim12.5-9.44$ km) and $\sim12.5-10.98$ km ($\sim12.5-9.86$ km)} for the universal and exponential profile respectively, in case of $\Delta_p^{lower(upper)}$. \textcolor{black}{Similarly, for the canonical mass star, the universal profile suspends the proton superconductivity in the radial zone $12\gtrsim R \gtrsim 8.71 (6.45)$ km for the universal profile and $12\gtrsim R \gtrsim 9.27 (7.22)$ km for the exponential profile of the magnetic field, in case of $\Delta_p^{lower(upper)}$. }

%in case of $\Delta_p^{lower}$ the quenching effect provided by the two profiles do not differ much. However, for $\Delta_p^{upper}$ the exponential profile does not cause any suppression to the proton superconductivity, while the universal profile results in strong suppression.

The combined effect of the magnetic field and proton superconductivity is illustrated in case of the heat capacity in fig. \ref{B_1.4_2.3}. The results are exhibited only at time $t=100$ years. For the massive star, at \textcolor{black}{$R\sim10.85$ km ($R\sim9.44$ km)} $C_v$ experiences a sudden jump from \textcolor{black}{$\sim6\times10^{19}$ ($\sim2\times10^{19}$) to $\sim 2\times10^{20}$ ($\sim 1.8\times10^{20}$)} erg cm$^{-3}$ K$^{-1}$, corresponding to $\Delta_p^{lower(upper)}$. The rest of the features are observed to be identical to the non-magnetic case shown in fig. 2.
Within the radial range \textcolor{black}{$10.85~\text{km}\gtrsim R \gtrsim 9.36$ km ($9.44~\text{km}\gtrsim R \gtrsim 5.69$ km)} in case of \textcolor{black}{$\Delta_p^{lower(upper)}$}, the order of $C_v$ is \textcolor{black}{$\sim2\times10^{20}$} (\textcolor{black}{$\sim1.8\times10^{20}$}) and \textcolor{black}{$\sim 6\times10^{19}$} (\textcolor{black}{$\sim (2-5)\times10^{19}$}) erg cm$^{-3}$ K$^{-1}$ respectively, in presence and absence of the magnetic field. %\textcolor{blue}{This implies the overall increase in $C_v$ in presence of the magnetic field if the protons present in the stellar matter are superconducting.} %The difference is a bit lower in case of $\Delta_p^{lower}$.

For the canonical mass star, at \textcolor{black}{$R\sim 8.71 (6.45)$} km  $C_v$ is changed promptly from \textcolor{black}{$\sim7\times10^{19}$ ($2\times10^{19}$) to $\sim2\times10^{20}$ ($1.5\times10^{20}$)} erg cm$^{-3}$ K$^{-1}$ in case of $\Delta_p^{lower(upper)}$. Also, within the radial zone \textcolor{black}{$8.71~\text{km}\lesssim R \lesssim 3.63$} km (\textcolor{black}{$R \lesssim 6.45$} km), corresponding to $\Delta_p^{lower(upper)}$, the value of $C_v$ is \textcolor{black}{$\sim 2.8\times10^{20}$($\sim 2\times10^{20}$)}  erg cm$^{-3}$ K$^{-1}$ in presence of the magnetic field, while it is somewhat lesser, \textcolor{black}{$\sim6\times10^{19}-1.5\times10^{20}$ ($\sim 3\times10^{19}$)}  erg cm$^{-3}$ K$^{-1}$, in absence of the magnetic field. \textcolor{black}{This implies the overall enhancement in the heat capacity in presence of the magnetic field, if the protons present in the stellar matter are superconducting.}

Fig. 7 depicts the variation of $\kappa_n$  in presence of both proton superconductivity and the magnetic field at time $t=100$ years. For the massive star, at \textcolor{black}{$R\sim10.85$ km ($R\sim 9.44$ km)} $\kappa_n$ is dropped from \textcolor{black}{$\sim3.5\times10^{23}$ ($\sim8\times10^{23}$) to $\sim2\times10^{23}$ ($\sim7\times10^{22}$)} erg cm$^{-1}$ s$^{-1}$ K$^{-1}$ corresponding to $\Delta_p^{lower(upper)}$ respectively. Now it can be pointed out that for $\Delta_p^{upper}$, within the radial range \textcolor{black}{$R\sim 9.4-5.69$ km} $\kappa_n$ is \textcolor{black}{lower by at least an order} in the scenario of magnetic case as compared to the non-magnetic case, as observed by comparing fig. 3 (left panel in the upper row ) and 7. Such distinction in the order of $\kappa_n$ is not significant corresponding to $\Delta_p^{lower}$ within the range \textcolor{black}{$R\sim 10.85-9.4$} km.    

Now in case of the canonical mass star, at \textcolor{black}{$R\sim8.71$} km there is a rapid reduction in $\kappa_n$ from \textcolor{black}{$\sim 3\times10^{23}$ erg cm$^{-1}$ s$^{-1}$ K$^{-1}$ to $\sim1.5\times10^{23}$ erg cm$^{-1}$ s$^{-1}$ K$^{-1}$ }for $\Delta_p^{lower}$, while for $\Delta_p^{upper}$ the jump occurs at \textcolor{black}{$R\sim6.45$} km from \textcolor{black}{$\sim8\times10^{23}$ erg cm$^{-1}$ s$^{-1}$ K$^{-1}$ to $\sim8\times10^{22}$ erg cm$^{-1}$ s$^{-1}$ K$^{-1}$}. Within the range \textcolor{black}{$R\sim8.71-3.63$ km there is no significant  difference in the order of $\kappa_n$ in presence of the magnetic field compared to the case of zero magnetic field for $\Delta_p^{lower}$},
while in the scenario of $\Delta_p^{upper}$ \textcolor{black}{the difference is about an order in the region $R\lesssim 6.45$ km.}

In fig. 8, the left and right panel exhibit the variation of $\kappa_{e,||}$ and $\kappa_{e,\perp}$ with radius for the massive star at the time $t=100$ years, while for the canonical mass star similar results are presented in fig. 9. In case of the massive star the value of $\kappa_{e,||}$ is raised instantaneously from \textcolor{black}{$\sim2$ ($\sim1$) erg cm$^{-1}$ s$^{-1}$ K$^{-1}$ to $\sim5\times10^3$ ($\sim4\times10^{4}$) erg cm$^{-1}$ s$^{-1}$ K$^{-1}$ at $R\sim10.85$ km ($R\sim9.4$ km)} and coincide with $\kappa_{e0,||}$, as proton superconductivity is limited in the region \textcolor{black}{$12.5~\text{km}\gtrsim R \gtrsim 10.85$ km ($12.5~\text{km}\gtrsim R \gtrsim 9.4$ km)}, corresponding to $\Delta_p^{lower(upper)}$ while, $\kappa_{e,\perp}$ also experiences abrupt elevation from \textcolor{black}{$\sim0.08$ ($\sim 0.1$) erg cm$^{-1}$ s$^{-1}$ K$^{-1}$ to $\sim 0.4$ ($\sim 1.5$) erg cm$^{-1}$ s$^{-1}$ K$^{-1}$}.

Similarly for the canonical mass star, $\kappa_{e,||}$ is suddenly increased from \textcolor{black}{$\sim 6 (\sim 2)$ to $2\times10^3$ ($4\times10^3$) erg cm$^{-1}$ s$^{-1}$ K$^{-1}$,} while there is a sudden rise in $\kappa_{e,\perp}$ to \textcolor{black}{$\sim 0.4$ ($\sim 0.8$) erg cm$^{-1}$ s$^{-1}$ K$^{-1}$ from $\sim 0.002$ ($\sim 0.004$) erg cm$^{-1}$ s$^{-1}$ K$^{-1}$ at $R\sim 8.71 (6.45)$ km.}
%\begin{widetext}
\begin{center}
\begin{table*}[h!]
\begin{tabular}{|c|c|c|} 
\hline
\multicolumn{3}{|c|}{$M=2.3~M_{\odot}, t=100$ yrs} \\
\hline
  \hline
 Radial zone$\rightarrow$ & \textcolor{black}{$12.5-10.85$} km & \textcolor{black}{$\leq 10.85$} km  \\
 \hline
 $\kappa_{e0,||}$ & \textcolor{black}{$4-7\times10^3$} & \textcolor{black}{$7\times10^3-6\times10^5$}   \\
 \hline
  $\kappa_{e,||}(\Delta_p^{lower})$ & \textcolor{black}{$0.2-2$} & \textcolor{black}{$7\times10^3-6\times10^5$}\\
\hline
$\kappa_{e0,\perp}$ & \textcolor{black}{$0.3-0.5$} & \textcolor{black}{$0.5-5.5$}  \\
\hline
  $\kappa_{e,\perp}(\Delta_p^{lower})$ & \textcolor{black}{$0.03-0.08$} & \textcolor{black}{$0.4-5.5$} \\
  \hline
 \hline
 Radial zone$\rightarrow$  & \textcolor{black}{$12.5-9.44$} km & \textcolor{black}{$\leq 9.44$} km   \\
  \hline
$\kappa_{e0,||}$ & \textcolor{black}{$4-4\times10^4$}  & \textcolor{black}{$4\times10^4-6\times10^5$}    \\
\hline
  $\kappa_{e,||}(\Delta_p^{upper})$ & \textcolor{black}{$0.2-1$} & \textcolor{black}{$4\times10^4-6\times10^5$} \\
 \hline
 $\kappa_{e0,\perp}$ & \textcolor{black}{$0.3-1.5$} & \textcolor{black}{$1.5-5.5$}  \\
 \hline
  $\kappa_{e,\perp}(\Delta_p^{upper})$ & \textcolor{black}{$0.01-0.15$} & \textcolor{black}{$1.5-5.5$}  \\
  \hline
\end{tabular}
\caption{Estimation of the two components of $\kappa_{e0}$ and $\kappa_e$, in presence of magnetic field, following the universal profile having the value at its core $B_c=10^{16}$ G.}\label{t}
\end{table*}
\end{center}
%\end{widetext}

%\begin{widetext}
\begin{center}
\begin{table*}[h!]
\begin{tabular}{|c|c|c|} 
\hline
\multicolumn{3}{|c|}{$M=1.4~M_{\odot}, t=100$ yrs} \\
\hline
Radial zone$\rightarrow$  & \textcolor{black}{$11.8-8.71$} km & \textcolor{black}{$\leq 8.71$} km \\
  \hline
  $\kappa_{e0,||}$ & \textcolor{black}{$10-1.5\times10^3$} & \textcolor{black}{$(1.5-7)\times 10^3$} \\
  \hline
  $\kappa_{e,||}(\Delta_p^{lower})$ & \textcolor{black}{$\sim 6$} & \textcolor{black}{$(2-7)\times 10^3$}  \\
\hline
 $\kappa_{e0,\perp}$ & \textcolor{black}{$0.001-0.4$} & \textcolor{black}{$0.4-1$} \\
 \hline
  $\kappa_{e,\perp}(\Delta_p^{lower})$ & \textcolor{black}{$0.0007-0.002$} & \textcolor{black}{$0.4-1$}  \\
  \hline\hline
  Radial zone$\rightarrow$  & \textcolor{black}{$11.8-6.45$} km & \textcolor{black}{$\leq 6.45$} km \\
  \hline
 $\kappa_{e0,||}$ & \textcolor{black}{$10-4\times10^3$} & \textcolor{black}{$(4-7)\times 10^3$}
 \\
 \hline
  $\kappa_{e,||}$($\Delta_p^{upper}$) & \textcolor{black}{$\sim 2$} & \textcolor{black}{$(4-7)\times 10^3$}\\
 \hline
 $\kappa_{e0,\perp}$ & \textcolor{black}{$0.001-0.8$} & \textcolor{black}{$0.8-1$} \\
 \hline
  $\kappa_{e,\perp}$($\Delta_p^{upper}$) & \textcolor{black}{$0.0002-0.004$} & \textcolor{black}{$0.8-1$}  \\
  \hline
\end{tabular}
\caption{Same as table 7, in presence of magnetic field, with the universal profile having the value at the core $B_c=10^{16}$ G.}\label{t}
\end{table*}
\end{center}
%\end{widetext}

\section{Discussion}\label{dis}
We study the behaviour of heat  capacity and thermal conductivity inside the magnetars having different mass limits $M\sim2.3M_\odot$ and $M\sim1.4M_\odot$ with their interior matter having $npe$ configuration governed by \textcolor{black}{DDME2} EoS. 
We analyse the effect of superfluidity/superconductivity, considering two extreme limits of the pairing energy gaps of the nucleons both in absence and presence of magnetic field. We conduct our analysis taking the universal profile of the magnetic field into consideration which generates the field value at the core of the magnetars about $10^{16}$ G. Different nucleons form pairs in different layers inside the magnetars having different strengths, once their respective critical temperatures are declined beyond certain critical limits, $e.g.$ the neutrons are in singlet state around the surface, but form triplet pairing in the deeper layers of the magnetars. The protons are only capable of existing in the singlet state in certain regions of the star, which is further dwindled if the effect of the magnetic field is taken into account. The nucleons remain in the paired state up to a much deeper layer inside the star from the surface for the upper boundaries of $\Delta_n$ and $\Delta_p$, as compared to their lower limits. In the massive star, all the nucleons become normal around the inner core. In the canonical mass star the neutrons remain superfluid till the inner core, while the proton pairing remains limited to specific region of the star. It should be noted that the electrons remain normal all throughout the star at any instant. 

In absence of magnetic field, the resultant heat capacity of all the particle species is diminished overall to some extent in the domains where the nucleons are superfluid/superconducting. When the effect of the magnetic field is implemented along with the nucleon pairing, in a certain region  of the star the heat capacity is increased relatively. This is observed more prominently for the upper boundaries of the $\Delta_n$ and $\Delta_p$ in comparison with their lower boundaries.

On the other hand, the thermal conductivity exhibits much diverse behaviour. Unlike the scenario of heat capacity, in absence of the magnetic field, the thermal conductivity of neutrons increases in the regions of the star where the proton superconductivity dominates over neutron superfluidity, but decreases in the area in which the strength of neutron pairing oversteps the proton pairing. The thermal conductivity of the electrons are reduced in the regions in which the protons are superconducting. Such enhancement or declination is much greater in case of the upper bounds of $\Delta_n$ and $\Delta_p$, as compared to their lower bounds. The electron thermal conductivity is reduced in presence of proton superconductivity. Needless to say, the reduction is higher in case of the upper bounds of the pairing energies of the nucleons.

Additionally, if the effect of magnetic field is incorporated, the neutron thermal conductivity is curtailed in some specific regions of the stars. The electron conductivity is splitted into two components, in the directions along and perpendicular to the magnetic field. Both the components are reduced even in normal matter configuration, by at least \textcolor{black}{eighteen} and \textcolor{black}{twenty four} orders of magnitude in case of the massive and canonical mass star at their inner core regions.In superfluid/superconducting matter, 
both of the components are minimized in certain regions near the surface due to the existence of proton superconductivity. This implies a slower rate of heat transfer inside the star due to the inclusion of magnetic field. Therefore, the magnetar may cool down at a slower rate.

We also inspect the thermal evolution of the two properties to elucidate the cooling phenomena of the magnetars studying the two quantities at an earlier time $t=100$ years and a latter time $t=1000$ years. The order of the heat capacity is lowered everywhere inside the stars with time, irrespective of whether the matter configuration is normal or superfluid/superconducting. In normal matter the thermal conductivity is decreased with time as expected. However, in presence of nucleon pairing, the deviation of the quantity from its normal counterpart becomes larger with time. For example, $\kappa_n$ attains larger value than its normal counterpart at certain regions of the star which gets even larger at $t=1000$ years than at $t=100$ years. The same scenario remains intact both in presence and absence of magnetic field, which possibly exhibits the cooling phenomenon occurring in the magnetars. The massive star undergoes faster cooling than the canonical mass star, due to its larger value of thermal conductivity.

In case of electron thermal conductivity, $\kappa_e$, its value remains the same for the upper and lower bounds of $\Delta_p$. However the presence of proton superconductivity causes one order of magnitude change in $\kappa_e$ compared to the case when protons are normal.
\textcolor{black}{Overall both $\kappa_n$ and $\kappa_e$ for the massive star are a little higher compared to the low mass star, although there is no significant change in their order of magnitude. This implies a shorter thermal relaxation period of the massive star ($t\lesssim100$ years) in comparison with the low mass star ($t\sim100-200$ years).} Larger conductivity of the massive star also indicates faster cooling. %Also both $\kappa_n$ and $\kappa_e$ show slight enhancement with time for both of the stars.
In presence of magnetic field $\kappa_e$ is anisotropic. The component parallel to the magnetic field ($\kappa_{e}^{||}$) is about \textcolor{black}{$\sim 4-5$ orders} of magnitude higher than the normal component ($\kappa_{e}^{\perp}$) in normal $npe$ matter. $\kappa_n$ is studied in presence of magnetic field when the nucleons inside the star form Cooper pairs. Magnetic field does not affect the magnitude of $\kappa_n$. In presence of magnetic field, the parallel and perpendicular components of $\kappa_e$ are not expected to show much deviation when proton superconductivity is taken into account.  
\textcolor{black}{In case of different EoSs the order of magnitude of the thermal properties are varied along with the superfluid/superconducting energy gap profile inside the star.  }

\section*{Acknowledgements}
\textcolor{black}{The authors would like to thank the anonymous referee for their constructive comments which enhance the quality of the manuscript.} The authors acknowledge the funding support from Science and Engineering Research Board, Department of Science and Technology, Government of India through Project No. CRG/2022/000069. 

\section*{Data Availability}

Data sharing is not applicable to this article as no data sets were generated during this study.

% Use the style of numbering in square brackets.
% If nothing is used, default style will be taken.
%\begin{enumerate}[a)]
%\item 
%\item 
%\item 
%\end{enumerate}  

%\begin{itemize}
%\item 
%\item 
%\item 
%\end{itemize}  

%\begin{description}
%\item[]
%\item[] 
%\item[] 
%\end{description}  

% Figure

% Uncomment and use as the case may be
%\begin{theorem} 
%\end{theorem}

% Uncomment and use as the case may be
%\begin{lemma} 
%\end{lemma}


\printcredits

\bibliographystyle{cas-model2-names}

\bibliography{thermal}

\end{document}